\documentclass[a4paper,fleqn,usenatbib]{mnras}

\usepackage[T1]{fontenc}
\usepackage{ae,aecompl}

\usepackage[dvips]{graphicx,xcolor}	
\usepackage{amsmath}	
\usepackage{amssymb}	
\usepackage{bm}           
\usepackage{threeparttable}
\usepackage{here}
\usepackage{newtxtext,newtxmath}

\newcommand{\lsim}{\, \mbox{\raisebox{-1.ex}
{$\stackrel{\textstyle<}{\textstyle\sim}$}}\,}
\newcommand{\vect}[1]{\!\!\!\mbox{ \,\boldmath $#1$}}
\newcommand{\KM}{\color{black}}
\newcommand{\HS}{\color{black}}
\newcommand{\GG}[1]{}

\title[KL-effect on Periastron Time Shift of Binary Pulsar]{ Post-Newtonian Kozai-Lidov Mechanism and its Effect on Cumulative Shift of Periastron Time of Binary Pulsar}

\author[H. Suzuki et al.]{
Haruka Suzuki$^{1}$\thanks{E-mail: suzuki@heap.phys.waseda.ac.jp (HS)},
Priti Gupta$^{2}$\thanks{E-mail: priti.gupta@tap.scphys.kyoto-u.ac.jp(PG)},
Hirotada Okawa$^{3}$\thanks{E-mail: h.okawa@aoni.waseda.jp(HO)},
and Kei-ichi Maeda$^{3,4}$ \thanks{E-mail: maeda@waseda.jp(KM)}
\\
$^{1}$Graduate School of Advanced Science and Engineering, Waseda University, Shinjuku, Tokyo 169-8555, Japan\\
$^{2}$Department of Physics, Kyoto University, Kyoto 606-8502, Japan\\
$^{3}$Waseda Institute for Advanced Study {\rm (WIAS)}, 1-6-1 Nishi Waseda, Shinjuku, Tokyo 169-8050, Japan \\
$^{4}$Department of Physics, Waseda University, Shinjuku, Tokyo 169-8555, Japan\\
}

\date{Accepted XXX. Received YYY; in original form ZZZ}

\pubyear{2020}

\begin{document}
\label{firstpage}
\pagerange{\pageref{firstpage}--\pageref{lastpage}}
\maketitle
\begin{abstract}
We study the Kozai-Lidov mechanism in a hierarchical triple system in detail by the direct integration of the first-order post Newtonian equations of motion. We analyse a variety of models with a pulsar to evaluate 
 the cumulative shift of the periastron time of a binary pulsar caused by the gravitational wave emission in a hierarchical triple system with Kozai-Lidov mechanism.
 We compare our results with those by the double-averaging method. 
 The deviation in the eccentricity, even if small,
 is important in the evaluation of the emission 
 of the gravitational waves.
We also calculate the cumulative shift of the periastron time by using obtained osculating orbital elements.
If Kozai-Lidov oscillations occur, 
the cumulative shift curve will bend differently from that of the isolated binary.
If such a bending is detected through the radio observation, 
it will be the first indirect observation of gravitational waves from a triple system.  
\end{abstract}




\section{Introduction}

Gravitational wave (GW) is one of the most interesting phenomena predicted by general relativity. 
It is the ripple on space-time caused by motions of massive objects like black holes.
Orbital motions of close binaries emit GW which extracts orbital energy and gradually shrinks the orbit. 
The shrinking binary orbits can be observed through radio signals if the binary includes a pulsar as its component \citep{Weisberg05}. 
Such a binary system with a pulsar is called a binary pulsar. 
A pulsar is a neutron star rotating fast and emitting radio signals with peaks whose period is quite precise. 
Due to this feature, it is possible to obtain various types of information from the observation of the radio signals from the pulsar; 
for example, we can know pulsar's rotational period, binary orbital period, and the information of binary orbital elements like semi-major axis and eccentricity \citep{Smarr76}. 
Hence, if it is observed for a long term, the time evolution of orbital shape due to GW emission can be followed. 

Such long-term observation of radio signals from a binary pulsar was in fact conducted for the PSR B1913+16 system. 
This system was found in 1975 and has been called Hulse-Taylor binary \citep{Hulse75}. 
It is one of the most famous binary pulsars. 
This binary has a quite eccentric and close orbit: its eccentricity and semi-major axis are 0.617 and 0.013 AU, respectively, and its orbital period is 7.75 hours \citep{Taylor76}. 
Because of these features, the orbital energy is extracted from this system by GW emission and it results in ongoing shrink of the orbit and decrease of the orbital period. 
This decrease of the period has been detected over 30 years with radio observation. 
The period shift effect clearly appeared in the cumulative shift of the periastron time (CSPT). 
The observed CSPT curve was explained quite well by the theoretical prediction of GW emission in general relativity \citep{Weisberg05, Weisberg10}. 
This observation was the first indirect evidence of the existence of GW. 

Numerous binary pulsars other than Hulse-Taylor binary have been found (see e.g. \citet{Lorimer08}). 
Some pulsars were reported as a part of triple systems. 
For example, the PSR B1620-26 system \citep{Thorsett99} and the PSR J0337+1715 system \citep{Ransom14} are triple systems. 
These triple systems are constructed with a close binary including a pulsar and another object orbiting around the binary. 
The triple systems that can be divided into an inner binary and outer orbiting companion are called as hierarchical triple systems.
Triple systems sometimes exhibit completely different orbital motions even if they have hierarchical structures. 
One of the most remarkable phenomena in hierarchical triple systems is the Kozai-Lidov (KL) mechanism \citep{Kozai62, Lidov62}. 
It is one of the most important orbital resonances that is mainly characterised by the secular changes of the eccentricity of the inner binary and the relative inclination between inner and outer orbits. 
These values oscillate exchanging their values with each other in secular timescale, that is, when the eccentricity increases, the inclination decreases, and vice versa, with timescale longer than both orbital periods. 
The eccentricity excitation in the inner binary is quite important for various astrophysical phenomena.
For example, the large eccentricity can enhance GW emission in the binary and finally cause the merger of black holes \citep{Blaes02, Miller02, Liu17}. 
In addition, the tidal force can also be enhanced with the excited eccentricity and the tidal disruptions of stars by supermassive black holes can be caused \citep{Ivanov05, Chen09, Chen11, Wegg11, Li15}. 
In the context of planetary science, the formation of hot Jupiters \citep{Naoz12, Petrovich15, Anderson16} or ultra-short-period planets \citep{Oberst17} are also said to be caused by KL mechanism. 
Recently, the GW emission from the hierarchical triple systems with KL mechanism has attracted attention of researchers. Some authors discussed about the waveform of GW from a binary in a hierarchical triple system and its observability \citep{Hoang19, Randall18, Gupta19}. 
If such systems exist and include pulsars as components of the binaries, the radio signal from the pulsar should also be detected.
The CSPT curve described from the signal will tell how the third companion and GW emission affect the evolution of the binary.

In this paper, 
we first analyse the KL mechanism in relativistic systems in detail 
and compare the orbital evolution by the direct integration of the equations of motion
and that by the well-known double-averaging method
\footnote{
{\KM
In this paper, "double averaging" denotes the commonly-used averaging of the dynamical equations 
for the orbital parameters
over two mean anomalies assuming multipole expansion of interaction terms of the potential.
Note that some authors use secular equations given by
averaged Hamiltonian without multipole expansion assuming the interaction term is small \citep{Saillenfest17, Li18}.
}
}.
We then 
investigate how CSPT curve changes with GW emission in hierarchical triple systems with KL mechanism. 
We treat general hierarchical triple systems in this paper, expanding the discussion in our previous letter \citep{Suzuki19}, which treated only one example.
If the CSPT curves predicted in this paper are detected through radio observation, it will be the first indirect observation of GW from a triple system. 
The paper is organised as follows: we summarise the important features of KL mechanism in \S\ref{sec:KL}. 
We describe our models in \S\ref{sec:models} and explain our methods in \S\ref{sec:method}. 
The results and discussions are in \S\ref{sec:result}. 
The conclusion follows in \S\ref{sec:conclusion}. 

\section{Kozai-Lidov Mechanism}
\label{sec:KL}

\begin{figure}
		\centering
		\includegraphics[width=5cm]{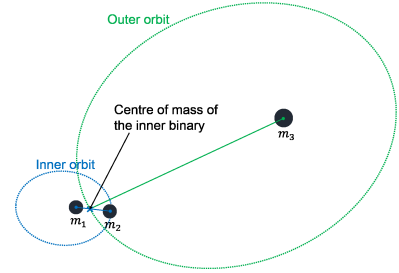}
		\caption{
		            The hierarchical triple system is constructed from inner and outer binaries.
					The inner binary consists of objects whose masses are $m_1$ and $m_2$, 
					and the outer one is the pair of the inner binary and the third body with mass $m_3$. 
					The outer  semi-major axis $a_{\rm out}$ is much larger than the inner one $a_{\rm in}$. 
				}
		\label{fig:3body}
	\end{figure}

Hierarchical triple systems are three-body systems in which the motions of components can be divided into two Keplerian elliptic orbits called inner and outer orbits due to highly hierarchical configuration such that the outer semi-major axis is much longer than the inner one (see  Fig.~\ref{fig:3body}).
We denote the masses of the components of inner binary by $m_1$ and $m_2$, 
and that of the tertiary companion by $m_3$.
Each orbit in the hierarchical triple system is described with six orbital elements.
In this paper, so called Kepler elements are used as the orbital elements; 
the semi-major axis $a$, the eccentricity $e$, the inclination $i$, the argument of periastron $\omega$, 
the longitude of ascending node $\Omega$, and the mean anomaly ${\cal M}$.
It is well-known that these elements are constant in a two-body system, except the mean anomaly, which corresponds to the phase in an elliptic orbit.
In the system that consists of three or more objects, in general, the trajectory of each component is
 not a closed elliptical orbit even in Newtonian dynamics.
However, when the Hamiltonian of the total system is given by the sum of two-body Hamiltonians with perturbative interactions like a hierarchical triple system, each trajectory can be approximated by an elliptical orbit but its shape gradually changes in time.
In such a case, the orbital elements of the {\it osculating orbit}, which is obtained by 
the instantaneous position and velocity, are used to describe the trajectory (see e.g. \citet{SSD2000}).
In this paper, the osculating orbital elements of inner and outer orbits are represented with the subscripts 'in' and 'out', respectively.
As for the outer orbit, we pursue the centre of mass of the inner binary rotating around the tertiary companion (see Fig.~\ref{fig:3body}).

Kozai-Lidov (KL) mechanism is one of the orbital resonances seen in hierarchical triple systems, which is discovered by \citet{Kozai62} and \citet{Lidov62}
\footnote{
Note that the framework of the fundamental formulation of this mechanism had been
already established by Von Zeipel in 1910 \citep{vonZeipel10, Ito19}. 
We shall call it Kozai-Lidov mechanism, however, because it is commonly used.
}.
In the system where KL-mechanism occurs, the eccentricity of inner orbit $e_\mathrm{in}$ and relative inclination
$I$ between inner and outer orbits  oscillate in secular timescale.
In this section, we shortly summarise some important features of KL-mechanism in Newtonian and post-Newtonian dynamics.
The basic features of KL-mechanism are well described with quadrupole-level approximation for a restricted triple system (see  e.g. \citet{Shevchenko17}), in which one of the components of the inner binary is assumed as a test particle.
We keep the lowest quadrupole order of the perturbed interaction terms in the Hamiltonian expanded in terms of the ratio of the semi-major axes.
The detailed explanation of this treatment is given in Appendix \ref{sec:KL_app}.

Not all of our models in this paper are the case of this restricted triple system.
For example, some models have the inner binary constructed with two neutron stars.
As shown in \S\ref{sec:method}, we will not use the double averaging method in our analysis 
but we directly integrate the equations of motion.
Hence the deviation from the test-particle limit is automatically taken into account.
Here we just introduce the basic features of KL-mechanism obtained from the test-particle treatment in order to analyse our results.
Note that  the detailed analysis for non-restricted hierarchical triple system was given in \citet{Naoz13a, Naoz13b}.
In \S\ref{sec:result}, we will revisit this point and will discuss the deviation seen in our simulation results from theoretical prediction with test-particle limit approximation.

\subsection{KL oscillations in Newtonian Dynamics}
\label{subsec:KL-Newton}
First we 
{\HS summarise important characteristics of KL-mechanism in a restricted triple system calculated in Newtonian mechanics.}
KL-mechanism is an orbital resonance in hierarchical triple systems characterised by the oscillation of the eccentricity of inner orbit $e_\mathrm{in}$ and the relative inclination  $I$ between inner and outer orbits on a secular timescale. 
We call this characteristic oscillation of $e_\mathrm{in}$ and $I$ as KL-oscillation.
The amplitude and timescale of KL-oscillation are determined by the conserved quantities in the restricted hierarchical triple system.
From the quadrupole-order restricted triple treatment, two conserved quantities are obtained: 
    \begin{eqnarray}
        &&\theta \equiv \sqrt{1-e_\mathrm{in}^2} \cos{I} , 
        \label{eq:theta}\\
        &&C_\mathrm{KL} \equiv e_\mathrm{in}^2\left( 1-\frac{5}{2} \sin^2I \sin^2\omega_\mathrm{in} \right).
        \label{eq:C_KL}
    \end{eqnarray}
When these values satisfy appropriate conditions, the KL-oscillation occurs.
{\HS 
The KL-oscillations are classified into two types depending on the sign of $C_\mathrm{KL}$.
KL-oscillation with $C_\mathrm{KL}\geq 0$ is called the ``rotation" type  because the periastron of the inner orbit rotates 
when the KL-oscillation proceeds, that is, the argument of periastron $\omega_\mathrm{in}$ 
increases monotonically.
On the other hand, KL-oscillation with $C_\mathrm{KL}\leq 0$ is called the ``libration" type 
because the argument of periastron $\omega_\mathrm{in}$ oscillates (librates) around $\pi/2$ or $3\pi/2$ {\KM with the KL-oscillation}. 
The {\KM possible} ranges of conserved values $(\theta^2, \, C_\mathrm{KL})$ for both rotation and libration types are 
depicted  in Fig. 1 in \cite{Antognini15}.
}
The amplitude and timescale of the KL-oscillation depend on the type of oscillations even if  the system size (masses and semi-major axes) is the same.
For the amplitude, the difference is clearly seen in the exact formulae of maximum 
and minimum eccentricities shown in Appendix \ref{subsec:maxmin_app}.
The timescale of the KL-oscillation $T_\mathrm{KL}$ is roughly estimated as
    \begin{equation}
      T_\mathrm{KL} \sim  
         \left( \frac{Gm_{\rm in}}{a_\mathrm{in}^3} \right)^\frac{1}{2}
         \, \frac{a_\mathrm{out}^3}{Gm_3} \,
            (1-e_\mathrm{out}^2)^\frac{3}{2}\,,
        \label{eq:roughT_KL}
    \end{equation}
where $G$ is the gravitational constant and $m_{\rm in} = m_1+m_2$ is the total mass of the inner binary.
This timescale depends only on the system size and the eccentricity of the outer orbit, 
but the exact oscillation period also depends on the conserved quantities of the system
 (see Appendix \ref{subsec:maxmin_app} for the reason).
In \S\ref{sec:result}, we confirm it by comparing our simulation results with different conserved quantities.

\subsection{Post-Newtonian Correction}
\label{subsec:GR}

In the restricted hierarchical triple system with quadrupole-level approximation, 
the GR correction is usually discussed by adding a simple correction term to the perturbation potential,
which is derived by double-averaging of the first order post-Newtonian (1PN) Hamiltonian of two-body relative motion (the detail is given in Appendix \ref{subsec:GR_app}).
Note that \citet{Will14a, Will14b} pointed out that this approach for 
the GR corrections  is not always appropriate.
Strictly speaking, for secular calculation due to the risk of the violation of energy conservation, 
we have to consider the effect of ``cross terms" between the Newtonian perturbations
 and the post-Newtonian precession effect.
In this section, however, we consider the GR correction without cross terms for 
interpretation of our numerical results (see also Appendix \ref{subsec:GR_app}).
In our simulation, as shown in \S\ref{sec:method}, 
the equations of motion are directly integrated. Hence the effect of the cross terms is automatically 
taken into account.	

The restricted triple systems with the GR correction have two conserved values as
 in the Newtonian dynamics.
$\theta$ does not change from Newtonian one, but $C_\mathrm{KL}$ is modified as
    \begin{equation}
        C_\mathrm{KL}^{(\mathrm{GR})} 
            = C_\mathrm{KL}(e, i, \omega)+\epsilon^\mathrm{(1PN)}
                \left(
                        \frac{ 1 }{ \sqrt{1-e_\mathrm{in}^2} } -1
                \right)
    \end{equation}
where
    \begin{equation}
        \epsilon^\mathrm{(1PN)} = 4\frac{r_\mathrm{g,in}}{a_\mathrm{in}}\frac{m_{\rm in}}{m_3}\left( \frac{a_\mathrm{out}}{a_\mathrm{in}} \right)^3 (1-e_\mathrm{out}^2)^{\frac{3}{2}}\,,
        \label{eq:epsilon}
    \end{equation}
 which is a dimensionless constant  
describing the strength of GR effect with  $r_\mathrm{g,in} = Gm_{\rm in}/c^2$.
Note that $C_\mathrm{KL}^{(\mathrm{GR})}$ is the same as $ C_\mathrm{KL}$ for circular orbit.

The classification conditions of KL-oscillations are $C_\mathrm{KL}^{(\mathrm{GR})} \geq 0$  for ``rotation" type while $C_\mathrm{KL}^{(\mathrm{GR})} \leq 0$ for the ``libration" type, respectively. 
The amplitude and timescale of KL-oscillation with the GR correction depend on the conserved quantities and vary from those in Newtonian analysis.
In \S\ref{sec:result}, we compare the Newtonian and GR results.

Generally, it is known that relativistic effects suppress the KL-oscillations.
There exists a critical value  $\epsilon^\mathrm{(1PN)}_\mathrm{cr}=3(1-e_\mathrm{in}^2)^{3/2}$,
which is found when the maximum and minimum eccentricities of the inner orbit
become equal\footnote{ This happens just for the libration type (see Fig. \ref{fig:emin_emax_NvsPN} in Appendix \ref{subsec:GR_app}). Hence the constraint (\ref{eq:GRKL}) may not be applied for the rotation type. 
However, even if the condition (\ref{eq:GRKL}) is not satisfied, the KL timescale becomes very long and then such a range is not so much interesting for observation.
}. Beyond the critical value ($\epsilon^\mathrm{(1PN)}>\epsilon^\mathrm{(1PN)}_\mathrm{cr}$),
the KL-oscillation does not occur (see e.g. \citet{Blaes02, Anderson17} for detail analysis).
The condition for the stable KL-oscillations ($\epsilon^\mathrm{(1PN)}<\epsilon^\mathrm{(1PN)}_\mathrm{cr}$)
 is rewritten as
    \begin{equation}
        \frac{ r_\mathrm{g,in} }{ a_\mathrm{in} }\frac{m_{\rm in}}{m_3}\left( \frac{a_\mathrm{out}}{a\mathrm{in}} \right)^3 
        \frac{ (1-e_\mathrm{out}^2)^{3/2} }{ (1-e_\mathrm{in}^2)^{3/2} } < \frac{3}{4}\,.
        \label{eq:GRKL}
    \end{equation}

\section{Models}
\label{sec:models}

We study GW emission effects on CSPT (cumulative shift of periastron time) 
of binary pulsars in hierarchical triple systems with the KL-oscillations.
As discussed in our previous letter paper \citep{Suzuki19}, 
this effect could be found in long-time observation of radio pulses from the pulsar.
We have shown only one model with initially circular inner binary as an example. 
In this paper, we analyse a broad range of parameters.
We first obtain constraints on parameters by imposing stability of the system and observable timescale 
and we then analyse several models in the allowed parameter range.

Before discussing the constraints, we first classify 
hierarchical triple systems 
 into three classes according to their mass ratio:
\begin{eqnarray*}
&{\rm Class~[1]}& m_{\rm in} \ll m_3\,,\\
&{\rm Class~[2]}& m_{\rm in} \sim m_3\,,\\
&{\rm Class~[3]}& m_{\rm in} \gg m_3\,.
\end{eqnarray*}
In Class [1], KL-oscillations are expected to occur, i.e., 
the inclination and eccentricity of inner orbit oscillates 
exchanging their values with each other \citep{VanLandingham16, Randall18, Hoang19}.
For Class [2], we may also see the KL-oscillations \citep{Blaes02, Wen03, Thompson11, Liu18} as 
in Class [1] as long as 
$a_\mathrm{out}\gg a_\mathrm{in}$. 
If $a_\mathrm{out}$ is not large enough as compared to $a_\mathrm{in}$, such a system does not have a sufficient hierarchy and then the interaction between the inner and outer orbits becomes strong. 
As a result, both orbital elements will change extremely with time and
the orbit will become chaotic. 
It may become unstable.

\begin{table*}
		\begin{tabular}{c||cc|ccc|c|cc|c}
			\hline
			 Model & inner binary&tertiary companion& $ m_1[\mathrm{M}_\odot]$  & $ m_2[\mathrm{M}_\odot]$  & $ m_3[\mathrm{M}_\odot]$ &class& $a_\mathrm{in}(0)$[AU] & $a_\mathrm{out}(0)$[AU] &$\epsilon^\mathrm{(1PN)}$ \\
			\hline
			 PNN  &     P-NS&NS&    1.4       &         1.4      &        1.4      &[1]&        0.01         &       0.2     &   0.177     \\
			 PNB  &      P-NS&BH&   1.4       &         1.4      &        30       &[2]&        0.01         &       0.5     &   0.129      \\
			 PNIB &     P-NS&IMBH&    1.4       &         1.4      &      $10^3$    &[2] &        0.01         &       2.5     &   0.484      \\
	    	 PNSB &     P-NS&SMBH&    1.4       &         1.4      &      $10^6$    &[2] &                  0.01         &      10.0     &   0.0310      \\ 
			\hline
			 PBB  &      P-BH&BH&    30       &         1.4      &        30      &[1] &         0.1         &       1.0     &   0.0130    \\
			 PBIB &       P-BH&IMBH&   30       &         1.4      &      $10^3$   &[2]  &         0.1         &       7.0     &   0.134      \\   
	         PBSB &       P-BH&SMBH&   30       &         1.4      &      $10^6$   &[2]  &                 0.1         &      40.0     &   0.249      \\   
			\hline
			 PIBIB&    P-IMBH&IMBH&   $10^3$      &         1.4      &      $10^3$   &[1]  &         0.1         &       1.2     &   0.684      \\
        	 PIBSB&    P-IMBH&SMBH&   $10^3$      &         1.4      &      $10^6$   &[2]  &                 0.1         &      10.0     &   0.396      \\
			\hline
		\end{tabular}
		\caption{
            		Model parameters:
            		$m_1$, $m_2$ and $m_3$ are the masses of components.
            		We fix the second object with mass $m_2=1.4 \, \mathrm{M}_\odot$ as a pulsar. 
            		$a_\mathrm{in}(0)$ and $a_\mathrm{out}(0)$ are the initial values of the 
            		semi-major axes of the inner and outer orbits, respectively.
            		$\epsilon^\mathrm{(1PN)}$ is the strength of the relativistic effect
            		defined by Eq. \eqref{eq:epsilon} for a restricted hierarchical triple system.
            		P, NS, BH, IMBH and SMBH mean a pulsar, neutron star, black hole, intermediate mass black hole and supermassive black hole,
            		 respectively.}
		\label{tab:model}
	\end{table*}
In Class [3], when the outer object can be treated as a test particle ($a_\mathrm{out}\gg a_\mathrm{in}$), 
the inner orbit is not affected so much by the tertiary object,
while the orbital elements of the outer orbit may change with time.
However, it is known that the eccentricity of the outer orbit does not change with time at least in the quadrupole order approximation.
 Instead we may expect the oscillation between the relative inclination $I$
 and the longitude of
ascending node of the outer orbit $\Omega_{\rm out}$
    in secular timescale \citep{Naoz17}.
Since we are interested in CSPT with the KL oscillations, 
i.e., CSPT via the time change of the pulsar's eccentricity,
we discuss only Classes [1] and [2].

In order to see CSPT through radio signals, each model should contain a pulsar as a component of the inner binary.
As a companion of the pulsar in the inner binary, 
in order to find large GW emissions from the inner binary and to neglect the tidal dissipation effect,
we may choose a compact object with a similar or larger mass than that of the pulsar, i.e., 
a neutron star (NS) or a black hole (BH).
If the companion is a non-compact object like a main sequence star, 
a strong tidal force from the pulsar
deforms the companion star and 
the orbital energy is dissipated by  friction in the star.
Since such dissipation by the tidal force may affect the periastron shift in addition to the GW emission, 
CSPT becomes more complicated, which is beyond the scope of this paper.
Hence we analyse three types of model for inner binaries; 
\begin{eqnarray*}
&{\rm P\mathchar`-NS}~binary & ({\rm pulsar+NS}) ,  \\
&{\rm P\mathchar`-BH}~binary & ({\rm pulsar}+ {\rm BH})  \\
&{\rm P\mathchar`-IMBH}~binary & ({\rm pulsar}+ {\rm intermediate~mass~BH})
\,.
\end{eqnarray*} 
$m_1$ and $m_2$ are the masses of the companion and pulsar in the inner binary, respectively. We choose those concrete values given in  Table \ref{tab:model}.

	\begin{figure}
		\centering
		\includegraphics[width=8.3cm]{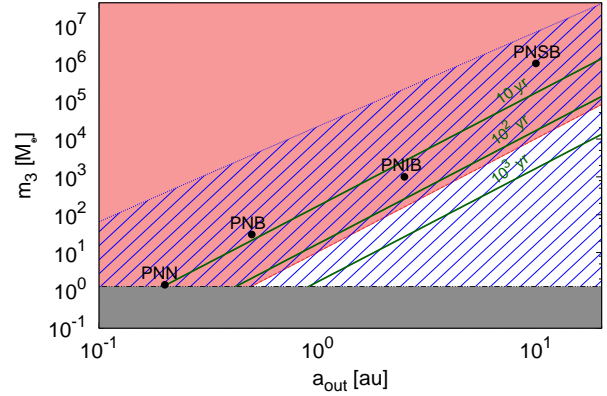}
		\caption{Stability constraints on the parameters of the outer orbit for stable KL-oscillations.
		The inner binary is a pulsar-neutron star system (P-NS binary), which  
		 parameters  are fixed as $m_1=1.4 \, \mathrm{M}_\odot$, $m_2=1.4 \, \mathrm{M}_\odot$ 
					and $a_\mathrm{in}=0.01$ AU.
					The black dashed line denotes the total mass of the inner binary $m_{\rm in}$.
					$m_3$ should be the same or larger than $m_{\rm in}$ for Class [1] and Class [2].
					In the blue thin-stripe region, a hierarchical triple system is stable. 
					 The condition for the KL-oscillations  not to be suppressed by the  post-Newtonian relativistic effect
					is given by the magenta-coloured region.
					The overlapped region gives a stable KL oscillations.
				    The dark-green lines show the timescales of KL-oscillations ($T_{\rm KL}=10, 10^2,$ and $10^3$ yrs),
				    which should be shorter than  our lifetime ($ < 100$ yrs) for observation.
					Our models given in Table~\ref{tab:model} are shown by the black dots.
				}
		\label{fig:allowedregion_NN}
	\end{figure}
	\begin{figure}
		\centering
		\includegraphics[width=8.3cm]{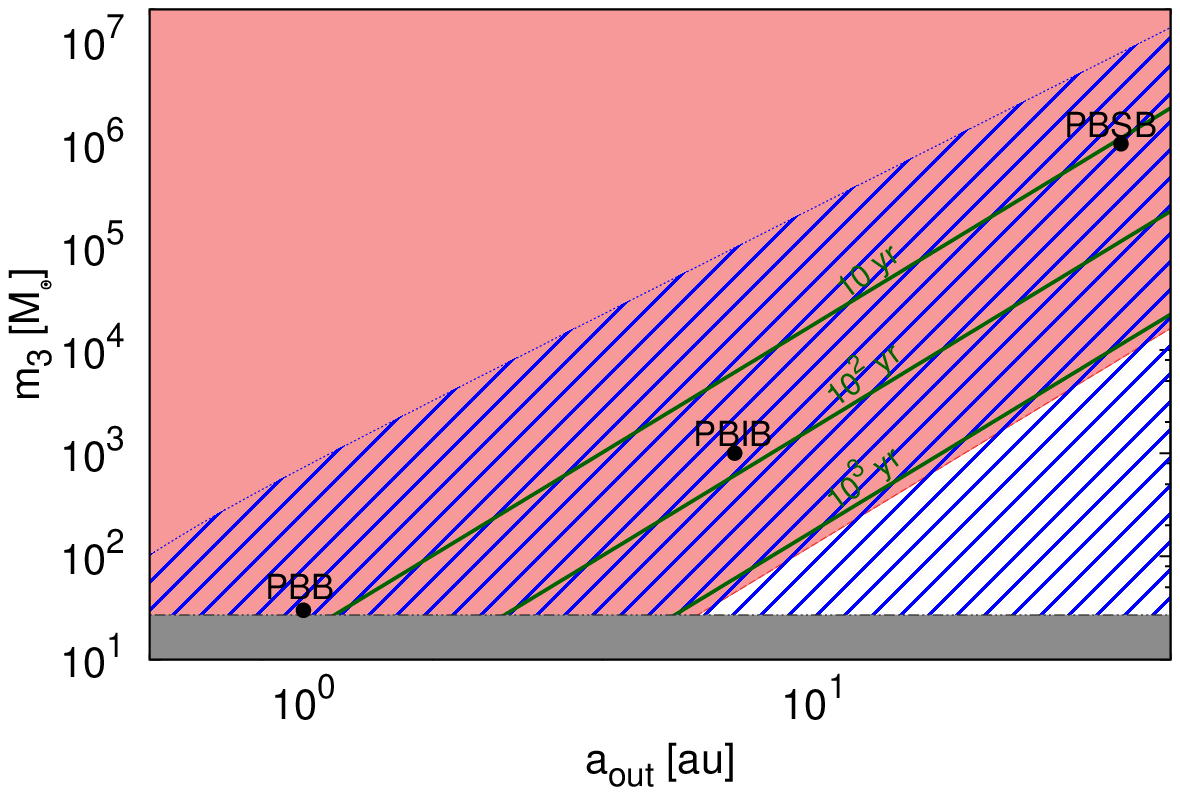}
		\caption{The same figure as Fig. \ref{fig:allowedregion_NN}, 
		but the inner binary is a pulsar-black hole system (P-BH binary), which  
		 parameters  are fixed as $m_1=30 \, \mathrm{M}_\odot$, $m_2=1.4 \, \mathrm{M}_\odot$ 
					and $a_\mathrm{in}=0.01$ AU.
		}
		\label{fig:allowedregion_NB}
	\end{figure}
		\begin{figure}
		\centering
		\includegraphics[width=8.3cm]{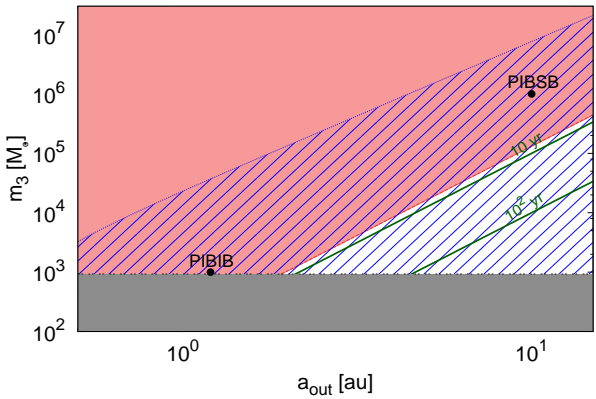}
		\caption{The same figure as Fig. \ref{fig:allowedregion_NN}, 
		but the inner binary is a pulsar-intermediate-mass black hole system (P-IMBH binary), which  
		 parameters  are fixed as $m_1=10^3 \, \mathrm{M}_\odot$, $m_2=1.4 \, \mathrm{M}_\odot$ 
					and $a_\mathrm{in}=0.1$ AU.
									}
		\label{fig:allowedregion_NIB}
	\end{figure}

There exist some conditions for the parameters of the outer orbit in order for the inner binary to 
exhibit stable KL-oscillations.
We show those constraints in Fig. \ref{fig:allowedregion_NN}-\ref{fig:allowedregion_NIB} 
in terms  of the semi-major axis of the outer orbit $a_\mathrm{out}$ and the mass of the third body $m_3$ by fixing parameters of the inner binary. 
The dashed black line shows the constraint for  the outer binary mass $m_3$, which should almost be the same or larger than the mass of the inner binary $m_{\rm in}$. 
The second condition 
 is stability of the hierarchical triple systems, i.e., the so-called ``chaotic boundary". 
As given in \cite{Mardling01}, the following condition should be satisfied so that the hierarchical structure of the system does not break at least in the initial state:
	\begin{equation}
		\frac{ a_\mathrm{out} }{ a_\mathrm{in} } >\frac{2.8}{ 1-e_\mathrm{out} } 
															\left[ \left( 1+\frac{m_3}{m_{\rm in}} \right) 
																	\frac{ 1+e_\mathrm{out} }{ (1-e_\mathrm{out})^\frac{1}{2} } \right]^\frac{2}{5}
																	\,.
		\label{eq:chaotic boundary}
	\end{equation}	
The stability condition \eqref{eq:chaotic boundary} is shown by the blue thin-stripe region.
The third condition is given by Eq.~\eqref{eq:GRKL}, which ensures KL-oscillation occurs even in a relativistic system.
We depict this condition by setting $e_\mathrm{in}=e_\mathrm{out}=0$  
because it does not change so much even for non-zero eccentricities.
This relativistic constraint is given by the magenta-coloured region.
In order to observe the effect of KL-oscillation on CSPT, the timescale of KL-oscillation should be short enough, compared with our lifetime.
As mentioned in \S\ref{subsec:KL-Newton}, the timescale of KL-oscillation is roughly estimated by Eq.~\eqref{eq:roughT_KL}.
We show some contour lines of $T_\mathrm{KL}$ by the dark-green lines ($T_\mathrm{KL}=10, 10^2$ and $10^3$ years). 

When the tertiary companion has the parameters both in 
 the blue thin-stripe region and the magenta-coloured region in Figs. \ref{fig:allowedregion_NN}-\ref{fig:allowedregion_NIB}, the KL-oscillation will occur with appropriate timescale. 
We also show our model parameters by the black dots with the model names in Fig.~\ref{fig:allowedregion_NN}-\ref{fig:allowedregion_NIB}.
We analyse nine models:
for P-NS inner binary, we  discuss four models; PNN, PNB, PNIB and PNSB, in which 
the tertiary companion is a neutron star (NS), black hole (BH), intermediate mass black hole (IMBH), and supermassive black hole (SMBH), respectively.
For P-BH inner binary, we consider three cases: PBB, PBIB and PBSB, in which 
the tertiary companion is a BH, IMBH and SMBH, respectively.
We also analyse the model PIBIB and PIBSB
; both systems have a P-IMBH inner binary, and an IMBH or SMBH as a tertiary companion.
We choose the masses of a pulsar (or NS), BH, IMBH and SMBH as
$1.4 \mathrm{M}_\odot$, $30\mathrm{M}_\odot$, $10^3\mathrm{M}_\odot$ and $10^6\mathrm{M}_\odot$, respectively.
The model parameters  are summarised in Table \ref{tab:model}.


Here we remark the Lense-Thirring precession effect.
This is one of the spin-orbit coupling effects appearing in 1.5 post-Newtonian order
correction \citep{Barker75}.
Recent studies have shown that the Lense-Thirring precession caused by the rapid rotation of an outer 
supermassive black hole in a hierarchical triple system changes the evolution of the KL-oscillation \citep{Fang19a, Fang19b, Liu19}. 
As in \citet{Liu19}, $T_\mathrm{LT}$ is evaluated by
    \begin{equation}
        T_\mathrm{LT} = \frac{2 c^3 a_\mathrm{out}^3 (1-e_\mathrm{out}^2)^{3/2} }{ \chi_3 G^2 m_3^2 (4+3m_{\rm in}/m_3)},
    \end{equation}
where $\chi_3 \leq 1 $ is the rotation parameter of the third object in the hierarchical triple system.
By using Eq.~\eqref{eq:roughT_KL}, $T_\mathrm{LT} \gg T_\mathrm{KL}$ gives the 
condition to neglect the Lense-Thirring effect, i.e., 
    \begin{equation}
        \left( \frac{a_\mathrm{in}}{\mathrm{AU}} \right)^\frac{3}{2} 
            \gg 10^{-12} \left( \frac{m_3}{ \mathrm{M}_\odot } \right)
                         \left( \frac{m_{\rm in}}{ \mathrm{M}_\odot } \right)^\frac{1}{2} .
    \end{equation}
We imposed $\chi_3 = 1$ in above estimation.
Since all models in Table~\ref{tab:model} satisfy this condition,
 we can neglect the Lense-Thirring effect in our calculation.

\section{Basic Equations}
\label{sec:method}

For the models explained in \S\ref{sec:models}, 
we directly integrate the equations of motion for their orbital evolution.
Then we analyse the behaviour of KL oscillations and 
evaluate the cumulative shift of periastron time (CSPT) of the inner binary.
\subsection{Equations of Motion and Initial Conditions}
\label{subsec:EOM}
\subsubsection{Equations of motion for three body system}
In order to solve relativistic motions of our three-body system composed of 
 compact objects, we use the first-order post-Newtonian equations of motion, 
 which are called as the Einstein-Infeld-Hoffmann (EIH) equations (\cite{EIH38}):
	\begin{eqnarray}
	&&\hskip -.5cm
	\frac{ \mathrm{d} \bm{v}_{k}}{\mathrm{d} t}
			=-G\sum_{n\neq k} m_{n}\frac{\bm{x}_{k} - \bm{x}_{n}}{|\bm{x}_{k} - \bm{x}_{n}|^{3}}
				 \Big[ 
						1-4 \frac{G}{c^2}\sum_{n'\neq k} \frac{m_{n'}}{|\bm{x}_{k} - \bm{x}_{n'}|}
				 \nonumber \\
            &&\hskip -.5cm~~
                  		-\frac{G}{c^2}\sum_{n'\neq n} \frac{m_{n'}}{|\bm{x}_{n} - \bm{x}_{n'}|} 
						\left \{
                             		   1-\frac{(\bm{x}_{k} - \bm{x}_{n}) \cdot (\bm{x}_{n} - \bm{x}_{n'})}
                                     {2|\bm{x}_{n} - \bm{x}_{n'}|^{2}} 
						\right \} 
				\nonumber \\
            &&\hskip -.5cm~~
						+\left( \frac{|\bm{v}_{k}|}{c} \right)^{2} + 2\left( \frac{|\bm{v}_{n}|}{c} \right)^{2} 
						-4 \frac{\bm{v}_{k} \cdot \bm{v}_{n}}{c^2}
                  		-\frac{3}{2} 
                  		\left\{ 
								   \frac{(\bm{x}_{k} - \bm{x}_{n})}{|\bm{x}_{k} - \bm{x}_{n}|} \cdot \frac{\bm{v}_{n}}{c} 
						\right\}^{2}   
                 \Big]  
				\nonumber \\
           &&\hskip -.5cm~~
                 -\frac{G}{c^2} \sum_{n \neq k} \frac{m_{n}(\bm{v}_{k}-\bm{v}_{n})}{|\bm{x}_{k} - \bm{x}_{n}|^{3}}
                		(\bm{x}_{k} - \bm{x}_{n}) \cdot (3\bm{v}_{n}-4\bm{v}_{k}) 
				\nonumber \\
           &&\hskip -.5cm~~
				-\frac{7}{2} \frac{G^{2}}{c^2} \sum_{n \neq k}\frac{m_{n}}{|\bm{x}_{k} - \bm{x}_{n}|}
                		\sum_{n'\neq n} \frac{m_{n'} (\bm{x}_{n} - \bm{x}_{n'})} {|\bm{x}_{n} - \bm{x}_{n'}|^{3}}\,, 
		\label{eq:EIH}
     \end{eqnarray} 
where  $m_k$, $\bm{v}_k$, $\bm{x}_k$ are the mass, velocity and position of the $k$-th component of the system
($k=1,2$ and $3$).
Note that this equation could be derived from the Lagrangian given by \citet{LD17}.
In our study, Eq.~(\ref{eq:EIH}) is numerically integrated by using the 6th order implicit Runge-Kutta method.
The coefficients of 6th-order Runge-Kutta are obtained from \citet{Butcher64}.
The back reaction of GW emission to the orbital evolution can be treated 
by including the 2.5 order post-Newtonian terms.
However, since the back reaction  
in a few KL-oscillation timescale is so small, it does not change our result.
Hence we consider only the first order of the post-Newtonian 
equations for the orbital evolution.

\subsubsection{Initial Conditions}

	\begin{table*}
		\begin{tabular}{cccccccc||l}
			\hline
			model & $\epsilon^\mathrm{(1PN)}$ &Type & $e_\mathrm{in}$  & $i_\mathrm{in}$[deg] & $\omega_\mathrm{in}$[deg] & $C_\mathrm{KL}^\mathrm{(GR)}$ &   $\theta^2$ &\Big{|}~$\omega_\mathrm{in}$[deg](Newtonian)\\
			\hline
    		PNN & 0.177 & ICL  & 0.01 &    60    &      60      &  $-3.18 \times 10^{-5}$ &   0.250      &\Big{|}~~~~~~~~~~57.0\\   
   				&  	    & ICR  & 0.01 &    60    &      30      &  $ 6.20 \times 10^{-5}$ &   0.250      &\Big{|}~~~~~~~~~~26.8\\
   				&	    & IEL  & 0.6  &    53    &      90      &  -0.170 & 0.232   &\Big{|}~~~~~~~~~~73.9\\
   		 		&	    & IER  & 0.6  &    45    &      60      &  0.0667 & 0.320    &\Big{|}~~~~~~~~~~53.8\\
			\hline
			PNB & 0.129 & ICL  &  0.01 &    60    &      60      &  $ -4.42 \times 10^{-5}$ &   0.250      &\Big{|}~~~~~~~~~~57.8\\   
   				&       & ICR  & 0.01 &    60    &      30      &  $ 5.96 \times 10^{-5}$ &   0.250      &\Big{|}~~~~~~~~~~27.7 \\
   				&       & IEL  & 0.6  &    53    &      90      &  -0.182 & 0.232   &\Big{|}~~~~~~~~~~76.3\\
   		 		&	    & IER  & 0.6  &    45    &      60      &  0.0548 & 0.320    &\Big{|}~~~~~~~~~~55.4\\
			\hline
			PNIB & 0.484 & ICL  & 0.01 &    60    &      60      &  $-1.64 \times 10^{-5}$ &   0.250      &\Big{|}~~~~~~~~~~52.0\\   
   				 &  	 & ICR  & 0.01 &    60    &      30      &  $ 7.73 \times 10^{-5}$ &   0.250      &\Big{|}~~~~~~~~~~20.4\\
   				 &	     & IEL  & 0.6  &    53    &      90      &  -0.0931 & 0.232   &\Big{|}~~~~~~~~~~62.7\\
   		 		 &	     & IER  & 0.6  &    45    &      60      &  0.143 & 0.320    &\Big{|}~~~~~~~~~~43.9\\
			\hline
		    PNSB & 0.0310 & ICL  & 0.01 &    60    &      60      &  $-3.91 \times 10^{-5}$ &          0.250      &\Big{|}~~~~~~~~~~59.5\\   
   				 &  	 & ICR  & 0.01 &    60    &      30      &  $ 5.47 \times 10^{-5}$ &   0.250      &\Big{|}~~~~~~~~~~29.5\\
   				 &	     & IEL  & 0.6  &    53    &      90      &  -0.206 & 0.232   &\Big{|}~~~~~~~~~~83.3\\
   		 		 &	     & IER  & 0.6  &    45    &      60      &  0.0302 & 0.320    &\Big{|}~~~~~~~~~~58.9\\
			\hline
			PBB & 0.0130 & ICL  & 0.01 &    60    &      60      &  $-4.00 \times 10^{-5}$ &   0.250      &\Big{|}~~~~~~~~~~59.8\\   
   			    &  	     & ICR  & 0.01 &    60    &      30      &  $ 5.38 \times 10^{-5}$ &   0.250      &\Big{|}~~~~~~~~~~29.8\\
   			    &	     & IEL  & 0.6  &    53    &      90      &  -0.211 & 0.232   &\Big{|}~~~~~~~~~~85.7\\
   		 		&	     & IER  & 0.6  &    45    &      60      &  0.0257 & 0.320   
   		 		&\Big{|}~~~~~~~~~~59.5\\
			\hline
			PBIB & 0.177 & ICL  & 0.01 &    60    &      60      &  $-3.39 \times 10^{-5}$ &   0.250      &\Big{|}~~~~~~~~~~57.7\\   
   				 &  	 & ICR  & 0.01 &    60    &      30      &  $ 5.98 \times 10^{-5}$ &   0.250      &\Big{|}~~~~~~~~~~27.6\\
   				 &	     & IEL  & 0.6  &    53    &      90      &  -0.181 & 0.232   &\Big{|}~~~~~~~~~~76.0\\
   		 		 &	     & IER  & 0.6  &    45    &      60      &  0.0559 & 0.320    &\Big{|}~~~~~~~~~~55.3\\
			\hline
	         PBSB & 0.0249 & ICL  & 0.01 &    60    &      60      &  $-3.94 \times 10^{-5}$ &   0.250      &\Big{|}~~~~~~~~~~59.6\\   
   				 &  	 & ICR  & 0.01 &    60    &      30      &  $ 5.44 \times 10^{-5}$ &   0.250      &\Big{|}~~~~~~~~~~29.6\\
   				 &	     & IEL  & 0.6  &    53    &      90      &  -0.208 & 0.232   &\Big{|}~~~~~~~~~~84.0\\
   		 		 &	     & IER  & 0.6  &    45    &      60      &  0.0287 & 0.320    &\Big{|}~~~~~~~~~~59.1\\
			\hline
			PIBIB & 0.684 & ICL  & 0.01 &    60    &      60      &  $-6.41 \times 10^{-5}$ &   0.250      &\Big{|}~~~~~~~~~~48.8\\   
   				  &  	  & ICR  & 0.01 &    60    &      30      &  $ 8.73 \times 10^{-5}$ &   0.250      &\Big{|}~~~~~~~~~~15.1\\
   				  &	      & IEL  & 0.6  &    53    &      90      &  -0.0430 & 0.232   &\Big{|}~~~~~~~~~~56.9\\
   		 		  &	      & IER  & 0.6  &    45    &      60      &  0.194 & 0.320    &\Big{|}~~~~~~~~~~37.5\\
			\hline
		     PIBSB & 0.396 & ICL  & 0.01 &    60    &      60      &  $-2.08 \times 10^{-5}$ &   0.250      &\Big{|}~~~~~~~~~~53.4\\   
   				  &  	  & ICR  & 0.01 &    60    &      30      &  $ 7.29 \times 10^{-5}$ &   0.250      &\Big{|}~~~~~~~~~~22.3\\
   				  &	      & IEL  & 0.6  &    53    &      90      &  -0.115 & 0.232   &\Big{|}~~~~~~~~~~65.5\\
   		 		  &	      & IER  & 0.6  &    45    &      60      &  0.122 & 0.320    &\Big{|}~~~~~~~~~~46.7\\
			\hline
	    \end{tabular}
		\caption{
		            	The important parameters in initial conditions for KL-oscillations for post-Newtonian calculations.
		            	We analyze four sets of initial parameters;
		            	`` initially circular libration'' (ICL), ``initially circular rotation'' (ICR),
		            	``initially eccentric libration''  (IEL) and ``initially eccentric rotation (IER).
		            	$e$, $i$, $\omega$ are  the eccentricity, the inclination, and the argument of the periastron, respectively.
		            	We also show two  conserved quantities, $C_\mathrm{KL}^\mathrm{(GR)}$ and $\theta^2$, in post-Newtonian dynamics.		            
		            	For ``Initially circular'', we set $e_\mathrm{in}=0.01$, while for ``initially eccentric'' we choose $e_\mathrm{in}=0.6$.
		            	The other parameters are determined to find 
		            	$C_\mathrm{KL}^\mathrm{(GR)}<0$ for libration and $C_\mathrm{KL}^\mathrm{(GR)}>0$ for rotation.
		            	For the outer orbit, $e_\mathrm{out}=0$ and $i_\mathrm{out}=0^\circ$ are used 
		            	and $\omega_\mathrm{out}$ cannot be defined.  
		            	About the parameters other than those shown in the table, 
		            	the longitude of the ascending node $\Omega$ is set  as $0$ for both inner and outer orbits, 
		            	and the mean anomaly ${\cal M}$ is set as $0^\circ$ and $20^\circ$ for inner and outer orbits. 
		            	To study the relativistic effect, we also perform the Newtonian calculation. We choose two conserved quantities as 
		            	$C_{\rm KL}=C_{\rm KL}^{\rm (GR)}$ and the same value of $\theta^2$ as the post-Newtonian one, which are obtained by setting the initial periastron argument as $\omega_\mathrm{in}$ given in the last column.
		             }
		\label{tab:ini_GR}
    \end{table*}	    
In order to set initial conditions for our simulation, we not only need the semi-major axis $a$ but also other parameters like the eccentricity $e$ and the inclination $i$.
These parameters fix the conserved quantities $\theta$ and $C_\mathrm{KL}^\mathrm{GR}$, 
which classify the type of KL-oscillation as ``libration'' or ``rotation'' (see  \S\ref{sec:KL}).
Hence we prepare four sets of initial parameters named as ``initially circular libration (ICL)'', ``initially circular rotation (ICR)'', ``initially eccentric libration (IEL) '' and ``initially eccentric rotation (IER)''.
For ``Initially circular'', we set $e_\mathrm{in}=0.01$, while for ``initially eccentric'' we choose $e_\mathrm{in}=0.6$.
The other parameters are determined to find $C_\mathrm{KL}^\mathrm{GR}<0$ for libration and $C_\mathrm{KL}^\mathrm{GR}>0$ for rotation.
The parameters of each type are summarised in Table \ref{tab:ini_GR} and are used for post-Newtonian calculations.

 To study the relativistic effect, we also perform the Newtonian calculation. We choose two conserved quantities as 
$C_{\rm KL}=C_{\rm KL}^{\rm (GR)}$ and the same value of $\theta^2$ as the post-Newtonian one, which are obtained by setting the initial periastron argument as $\omega_\mathrm{in}$ given in the last column in Table~\ref{tab:ini_GR}.

These initial orbital elements are converted into the position and velocity vectors, $\vect{x}_k$ and $\vect{v}_k$,
in Cartesian coordinates, whose origin is the centre of mass of whole system.
The $x$-$y$ plane of our coordinate system is chosen to be the initial outer orbital plane.
The detailed
conversion formula is given in Appendix \ref{subsec:initial} (See also e.g. \citet{SSD2000}).
By using Cartesian initial variables, the above EIH equations (\ref{eq:EIH}) are integrated numerically 
and the osculating orbital elements are evaluated at each time step.
The procedure to evaluate orbital elements from positions and velocities at each time step is 
also explained in Appendix \ref{subsec:post}.

The integrated inner orbit is not exactly a closed ellipse, 
but it fluctuates with small amplitudes because of the effect of the tertiary component.
As a result, the orbital parameters of the osculating orbit evaluated at each step are oscillating,
which seem to be artificial.
Hence we take an average of these elements for each inner cycle to extract the effective values at each cycle.  
We describe such averaged orbital elements with a bar, e.g., $\bar{a}_\mathrm{in}$ and $\bar{e}_\mathrm{in}$.
Those elements evolve in secular timescale due to the effect of the third body.

\subsection{Cumulative Shift of Periastron Time (CSPT)}
\label{subsec:Method_CSPT}
The orbital energy of inner binary, if it is close enough, is extracted little by little via the GW emission.
The energy dissipation makes the semi-major axis of the orbit shrink 
and then the period of the orbit becomes shorter and shorter.
As derived in \citet{Peters63}, the period change for each orbital cycle is \\
\begin{align}
		\dot{P}_\mathrm{in}
				&= -\frac{192\pi}{5} \left( \frac{P_\mathrm{in}}{2\pi} \right)^{-\frac{5}{3}} 
					 	\frac{G^2m_1 m_2}{c^5} \left( Gm_{\rm in} \right)^{-\frac{1}{3}} 
						\nonumber \\
				&\hskip .5cm \times		
						\frac{1}{\left( 1-\bar{e}_\mathrm{in}^2 \right)^\frac{7}{2}} 
						\left( 
							1+ \frac{73}{24}\bar{e}_\mathrm{in}^2 
							+ \frac{37}{96}\bar{e}_\mathrm{in}^4 
						\right) 
						, \label{eq:Pdot}
\end{align}
where $P_\mathrm{in}$ is the orbital period of the inner binary given by
	\begin{equation}
		P_\mathrm{in} = 2\pi \sqrt{\frac{\bar{a}^3_\mathrm{in}}{Gm_{\rm in}}} .
		\label{eq:P_in(a)}
	\end{equation} 
When the energy dissipation is evaluated for one binary cycle, the orbital elements can be treated as constant 
because  the back reaction of energy dissipation is small enough in such a timescale.
Here we use the averaged values, $\bar{e}$ and $\bar{a}$, instead of the osculating orbital elements, $e$ and $a$, to reflect the effective shape of the orbit for one cycle. 
When $\bar{e}$ and $\bar{a}$ evolve with secular timescale such as 
the KL-oscillation timescale, $\dot{P}_\mathrm{in}$ also changes with time.

This period shift can be seen by observing the cumulative shift of periastron time (CSPT)
 through radio signals from a binary pulsar just as 
 the observation of the Hulse-Taylor binary \citep{Weisberg05}.
In this paper, we expand the analysis to hierarchical three-body systems. 
The CSPT of the inner binary $\Delta_P$ is defined as
    \begin{equation}
	\Delta_P(T_N) =  T_N - P_{\mathrm{in}}(0)N\,,
	\label{eq:def_Delta_P}
	\end{equation} 
where $T_N$ is the $N$-th periastron passage time and $P_{\mathrm{in}}(0)$ is the initial orbital period of the inner binary.
From the definition of $T_N$, we obtain 
	\begin{equation}
		N = \int^{T_N}_0 \frac{1}{P_\mathrm{in}(t)} \mathrm{d} t\,,
		\label{eq:def_N}
	\end{equation}
where $P_\mathrm{in}(t)$ is the binary period at time $t$, which changes in time by the GW emission as
	\begin{equation}
		P_\mathrm{in}(t) =  P_{\mathrm{in}}(0) + \int^{t}_0 \dot{P}_\mathrm{in}(t') \mathrm{d}t' 
		\,.
		\label{eq:P_in(P_dot)} 
	\end{equation}
By substituting Eqs.~\eqref{eq:def_N} and \eqref{eq:P_in(P_dot)} into Eq.~\eqref{eq:def_Delta_P}, 
the CSPT $\Delta_P$ is described as
	\begin{eqnarray}
		\Delta_P(T_N) 
			&=&  T_N-\int^{T_N}_0 \mathrm{d}t \frac{P_{\mathrm{in}}(0) }{P_{\mathrm{in}}(0)+\int^{t}_0\dot{P}_\mathrm{in}(t') \mathrm{d}t' }
				    \nonumber \\
			&=&  \int^{T_N}_0 \mathrm{d}t 
					\frac{\int^{t}_0\dot{P}_\mathrm{in}(t') \mathrm{d}t'}{P_{\mathrm{in}}(0)+\int^{t}_0\dot{P}_\mathrm{in}(t') \mathrm{d}t'} 
					\,.
					\label{eq:derivation_Delta_P}
	\end{eqnarray}
Since the emission energy of GWs is quite small, we usually expect
	\begin{equation}
		\left| \int^{t}_0\dot{P}_\mathrm{in}(t') \mathrm{d}t' \right| \ll P_{\mathrm{in}}(0) .
		\label{eq:assumption_Pdot}
	\end{equation}
In fact, for Hulse-Taylor binary pulsar \citep{Weisberg05}, since we have
	\begin{eqnarray}
		P_\mathrm{b} &=& 0.32299\ \mathrm{day} ,\\
		\dot{P}_\mathrm{b} &=& -2.4184 \times 10^{-12} \ \mathrm{s/s}
		\,,
	\end{eqnarray}
the condition (\ref{eq:assumption_Pdot}) is true if $t \ll 3.7 \times 10^8 \ \mathrm{yrs}$.
Hence, when we are interested in the time-scale such that $T_N \ll 10^8 \ \mathrm{yrs}$, we approximate $\Delta_P$ as 
	\begin{equation}
		\Delta_P(T_N) \approx \frac{1}{P_{\mathrm{in}}(0)} \int^{T_N}_0 \mathrm{d}t \int^{t}_0 \mathrm{d}t' \dot{P}_\mathrm{in}(t') .
		\label{eq:Delta_P}
	\end{equation}
Note that if we assume $\dot{P}_\mathrm{in}(t)$ is almost constant, that is,
 $\dot{P}_\mathrm{in}(t) \approx \dot{P}_{\mathrm{in}}(0)$,  
$\Delta_P$ is given by
	\begin{equation}
		\Delta_P(T_N) \approx \frac{\dot{P}_{\mathrm{in}}(0)}{2P_{\mathrm{in}}(0)} T_N^2 ,
		\label{eq:approx_Delta_P}
	\end{equation}
which was used in \citet{Weisberg05}.

However, in a hierarchical triple system with the KL oscillation, $\dot{P}_\mathrm{in}(t)$ is not constant but 
may change in time with the KL-oscillation timescale. 
Hence, in this study, we evaluate $\Delta_P$ by Eq.~(\ref{eq:Delta_P}) with Eq.~(\ref{eq:Pdot}).

Our analysis can be applied to a general stable three-body (or $N$-body) system with a binary pulsar as long as the condition Eq. \eqref{eq:assumption_Pdot} is satisfied.
Here we stress that the CSPT could be observed through radio signals from the pulsar as the accumulated effect.
Highly accurate observation of radio pulsars enables us to see this CSPT even for 
such weak GW emission that
the back reaction of GW emission on the orbital elements is negligibly small. 
The CSPT observation through the radio signals from a binary pulsar in a triple system
 may be the precursor of detection of gravitational waves from a triple system with the KL-oscillation \citep{Gupta19}.

\section{Results and Discussions}
\label{sec:result}
\subsection{Orbital Evolutions}
\label{subsec:orbit}
In our simulation results, the stable orbital evolutions are observed in all the models shown in Table~\ref{tab:model}.
We show the results of PNN model and PNIB model as representative.
The mass hierarchy in PNN model is the smallest in all the models and it is expected that the deviation from test-particle approximation used in \S\ref{sec:KL} is the largest.
In PNIB model, on the other hand, $\epsilon^\mathrm{(1PN)}$ is the second largest as seen in Table \ref{tab:model} and relativistic effect in this model may become important.

    \begin{figure}
        \centering
        \begin{minipage}{7.5cm}
            \includegraphics[width=7.5cm]{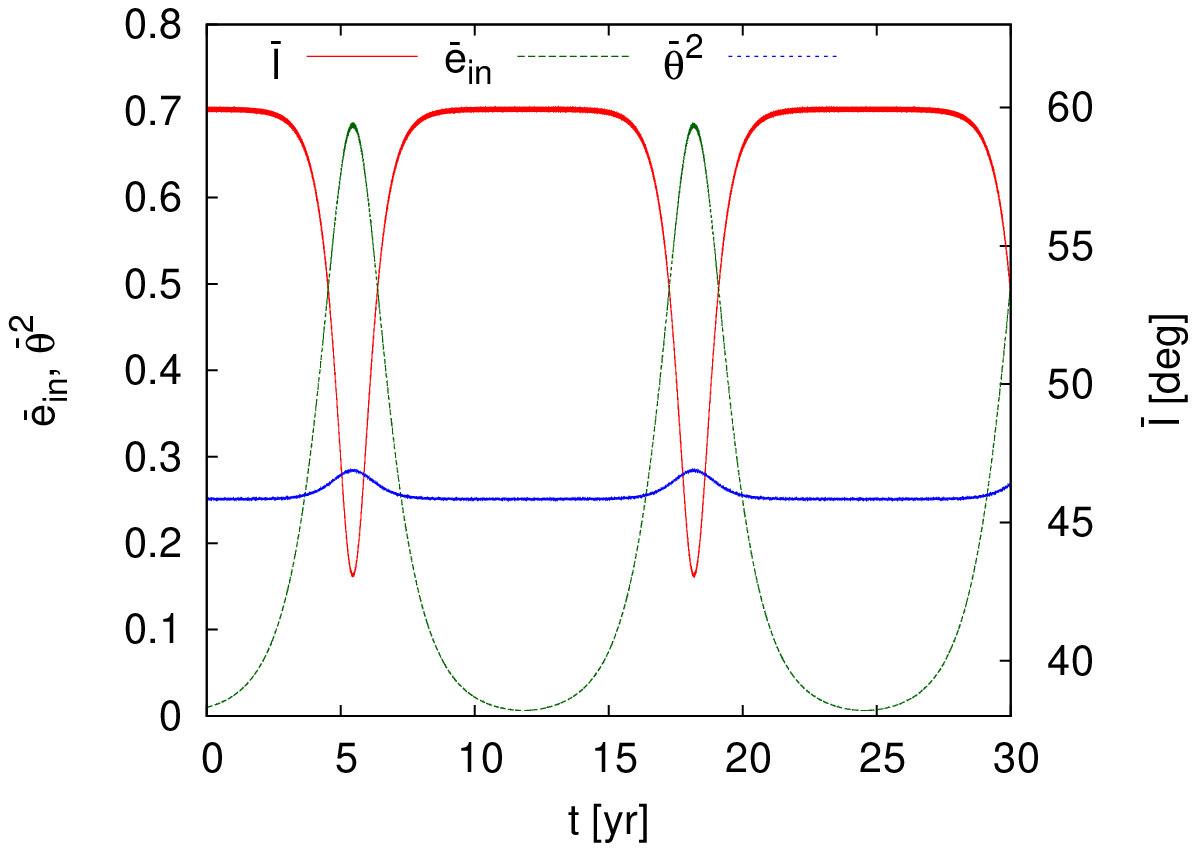}
        \end{minipage}\\
        \begin{minipage}{7.5cm}
            \includegraphics[width=7.5cm]{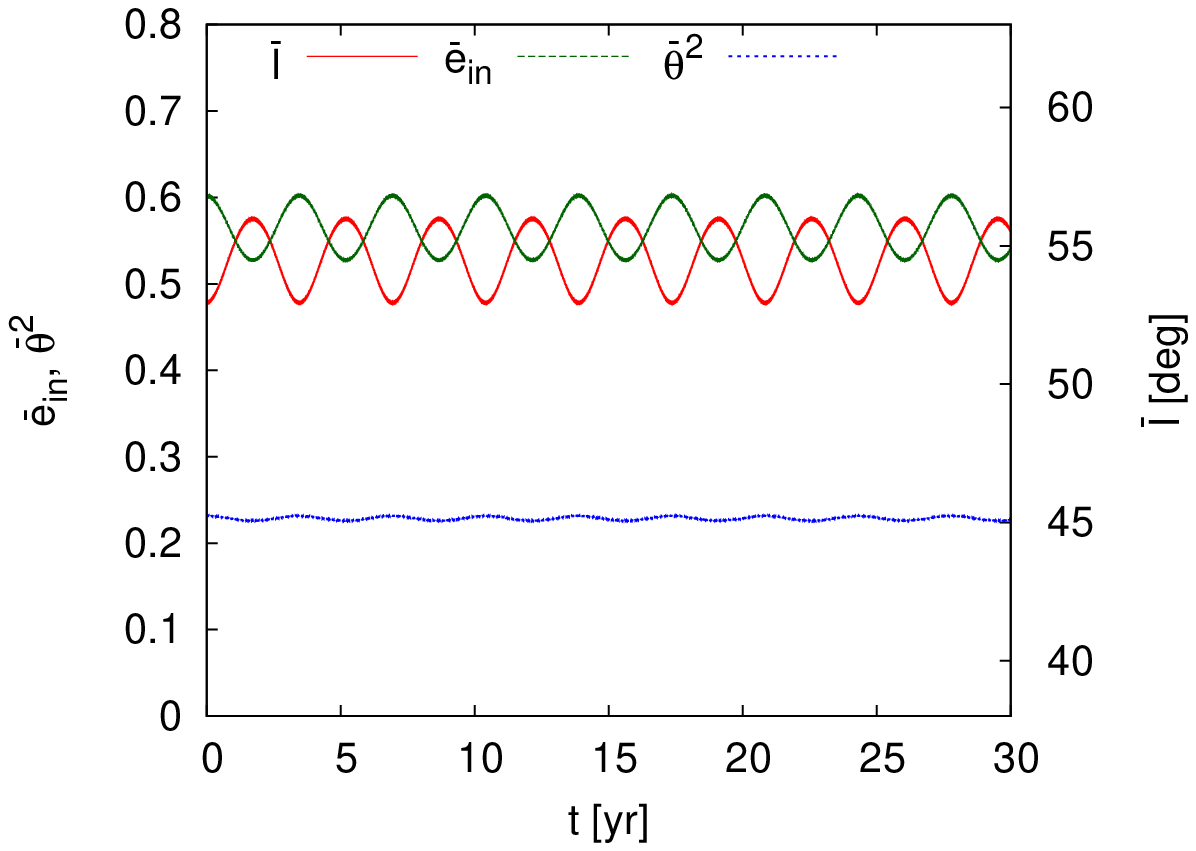}
        \end{minipage} 
        \caption{
                    Time evolution of the averaged inner eccentricity $\bar{e}_\mathrm{in}$ (green line), 
                    relative inclination $\bar{I}$ (red line) and KL-conserved value $\bar{\theta}^2$ (blue line) 
                    for the ``libration'' type KL oscillations in the PNN model.
                    Top and bottom panels correspond to the results of ICL and IEL types, respectively.
                 }
        \label{fig:PNN_KL_l}
    \end{figure}
    
      \begin{figure}
        \centering
               \begin{minipage}{7.5cm}
            \includegraphics[width=7.5cm]{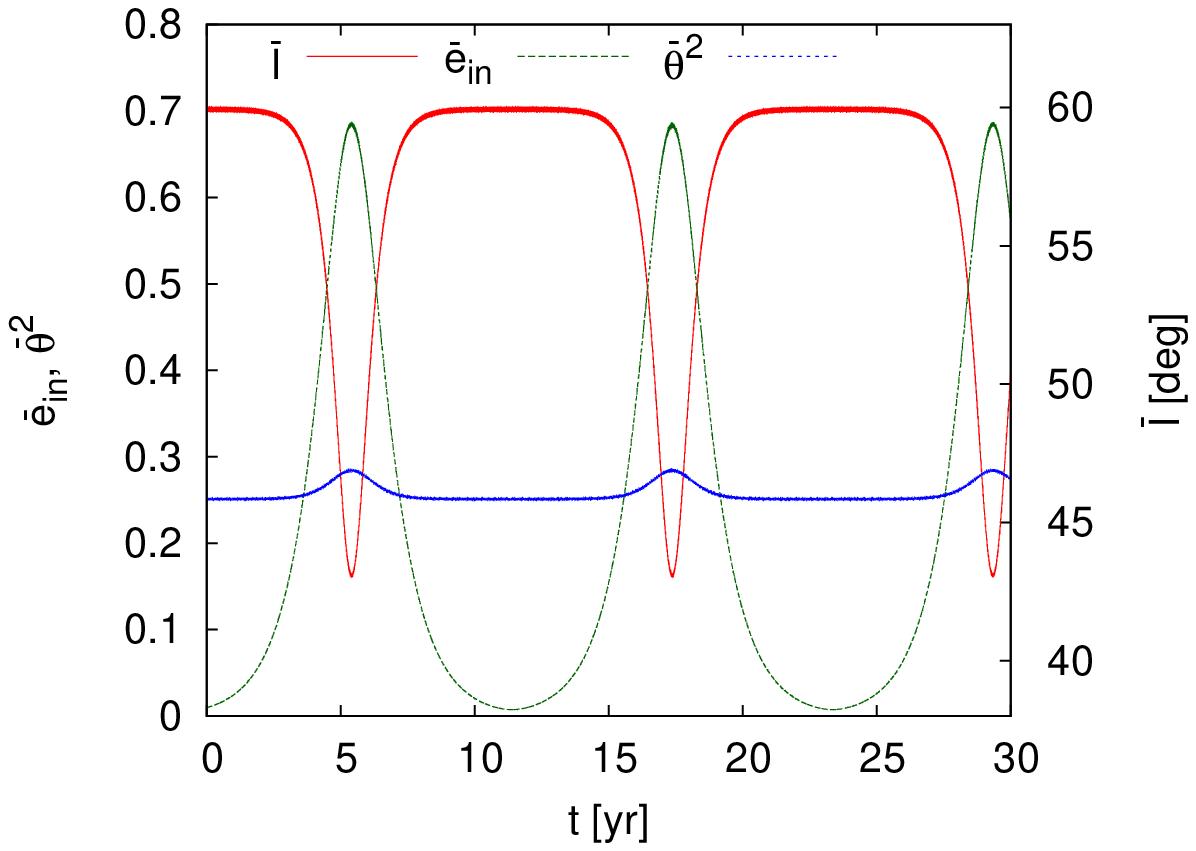}
        \end{minipage}\\
        \begin{minipage}{7.5cm}
            \includegraphics[width=7.5cm]{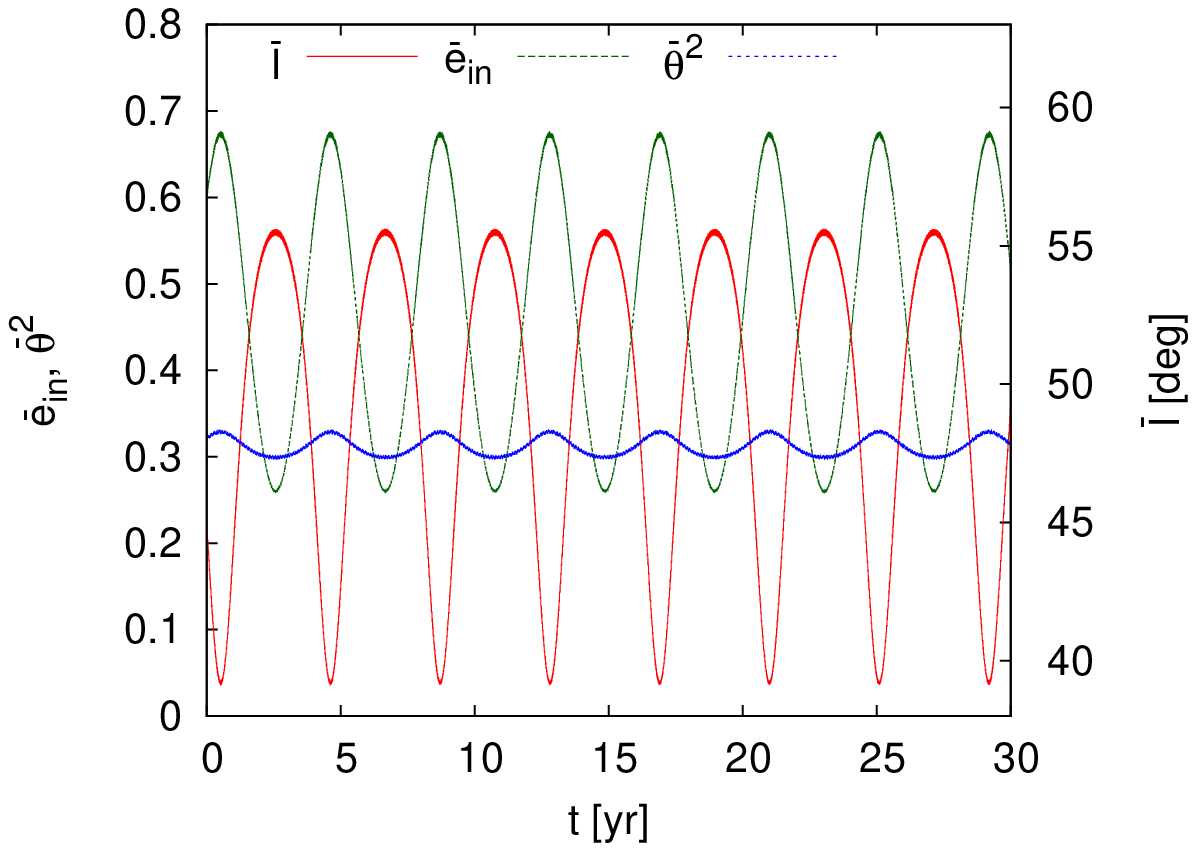}
        \end{minipage}
        \caption{
                    The same figure as Fig. \ref{fig:PNN_KL_l} for the ``rotation''  type KL oscillations in the PNN model.
                    Top and bottom panels are the results of ICR and IER types, respectively.
                 }
        \label{fig:PNN_KL_r}
    \end{figure}

    \begin{figure}
        \centering
        \begin{minipage}{7cm}
            \includegraphics[width=7cm]{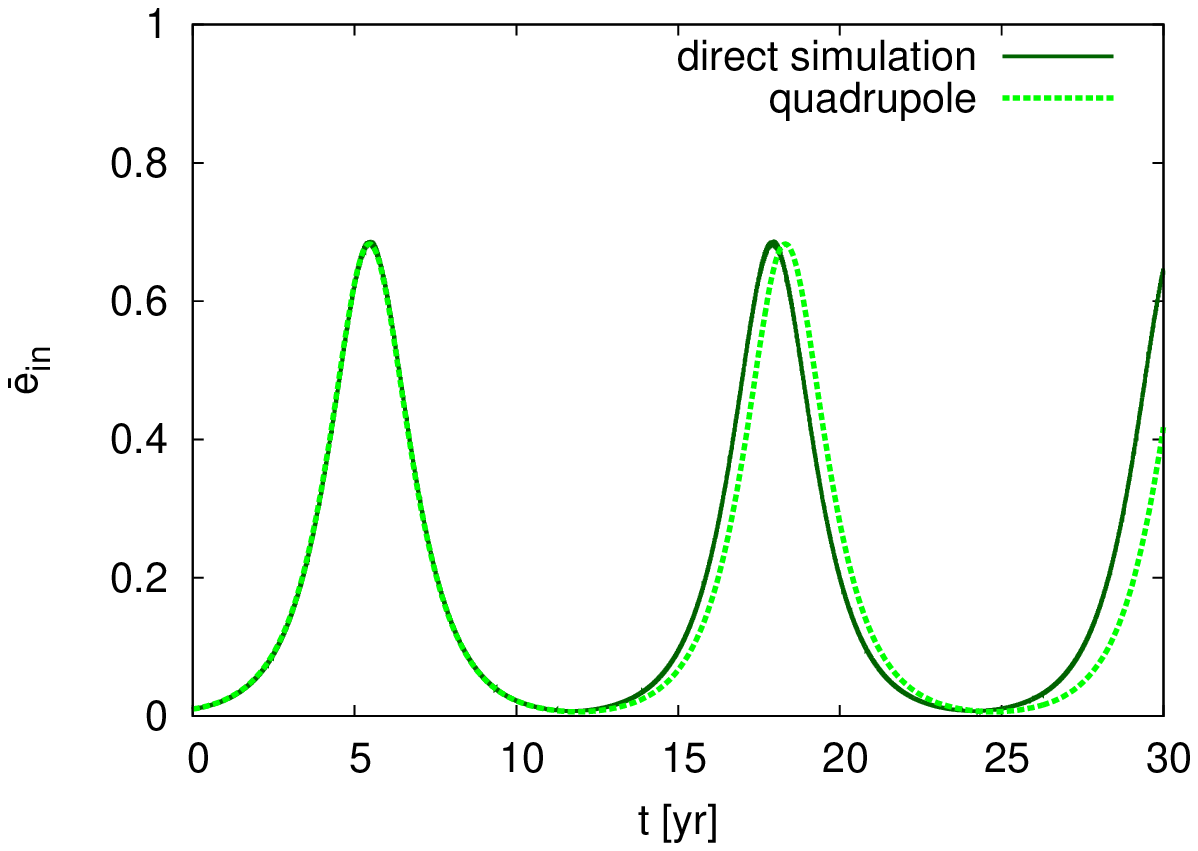}
        \end{minipage}\\
        \begin{minipage}{7cm}
            \includegraphics[width=7cm]{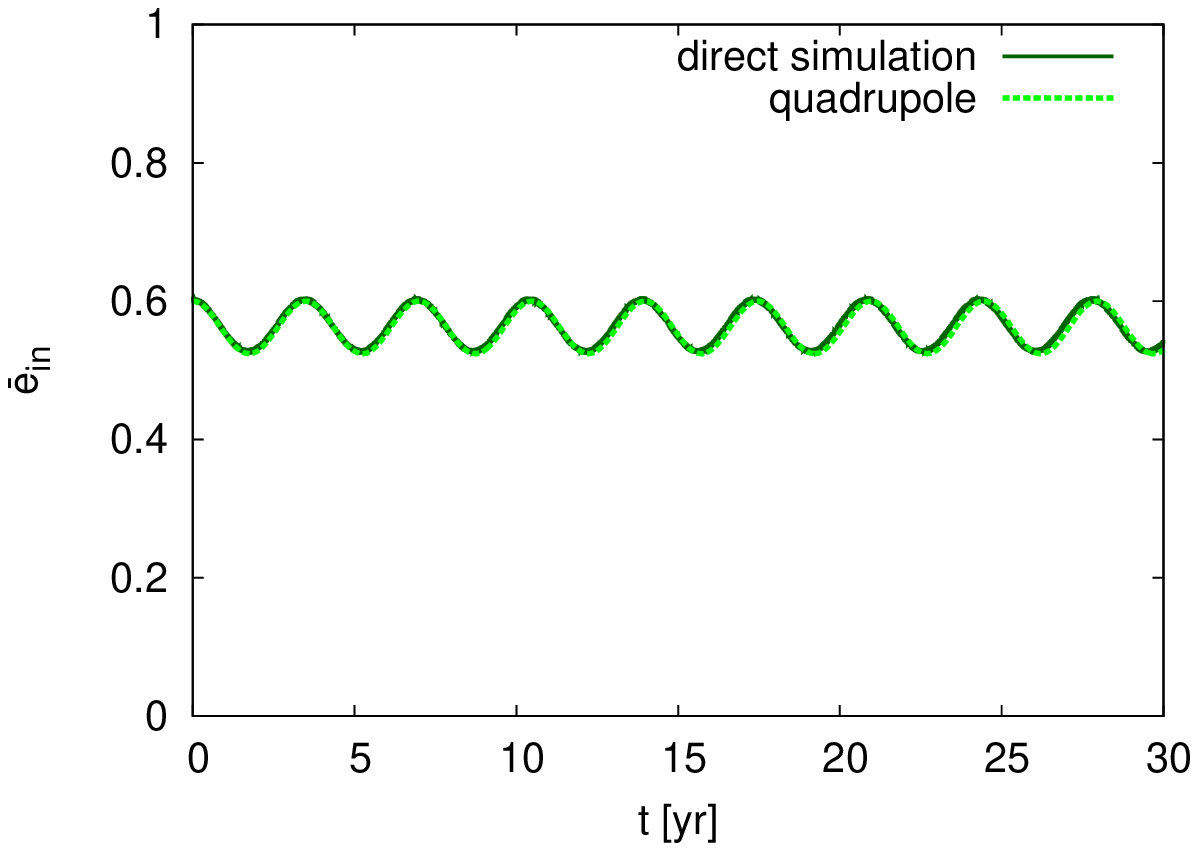}
        \end{minipage} 
        \caption{
                    Comparison between two evolution lines of the averaged inner eccentricity $\bar{e}_\mathrm{in}$ 
                    for the libration type of KL oscillations in the PNN model.
                    Top and bottom panels show the results of ICL and IEL types, respectively.
                    The solid line describes the evolution obtained from direct simulation while the dashed line denotes the result obtained by double-averaged calculation.
                 }
        \label{fig:PNN_e_l}
    \end{figure}
     \begin{figure}
        \centering
               \begin{minipage}{7cm}
            \includegraphics[width=7cm]{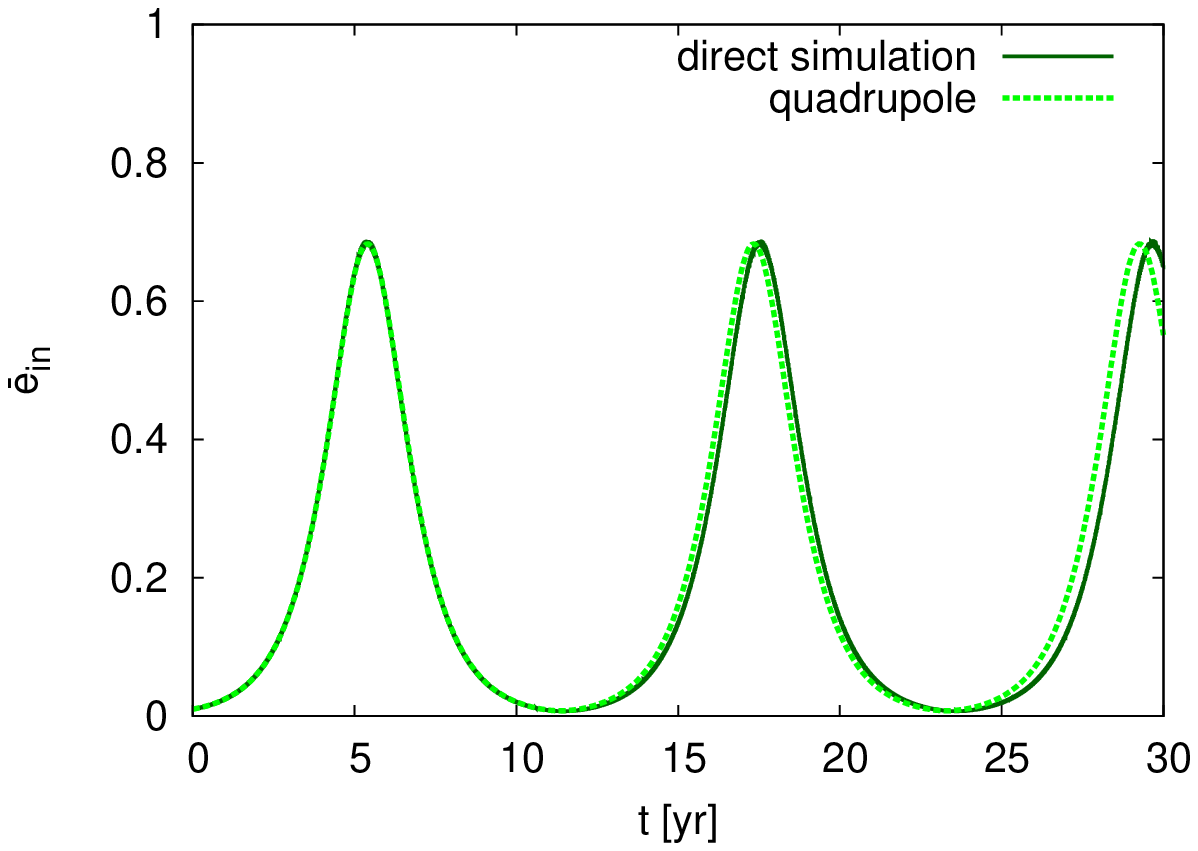}
        \end{minipage}\\
        \begin{minipage}{7cm}
            \includegraphics[width=7cm]{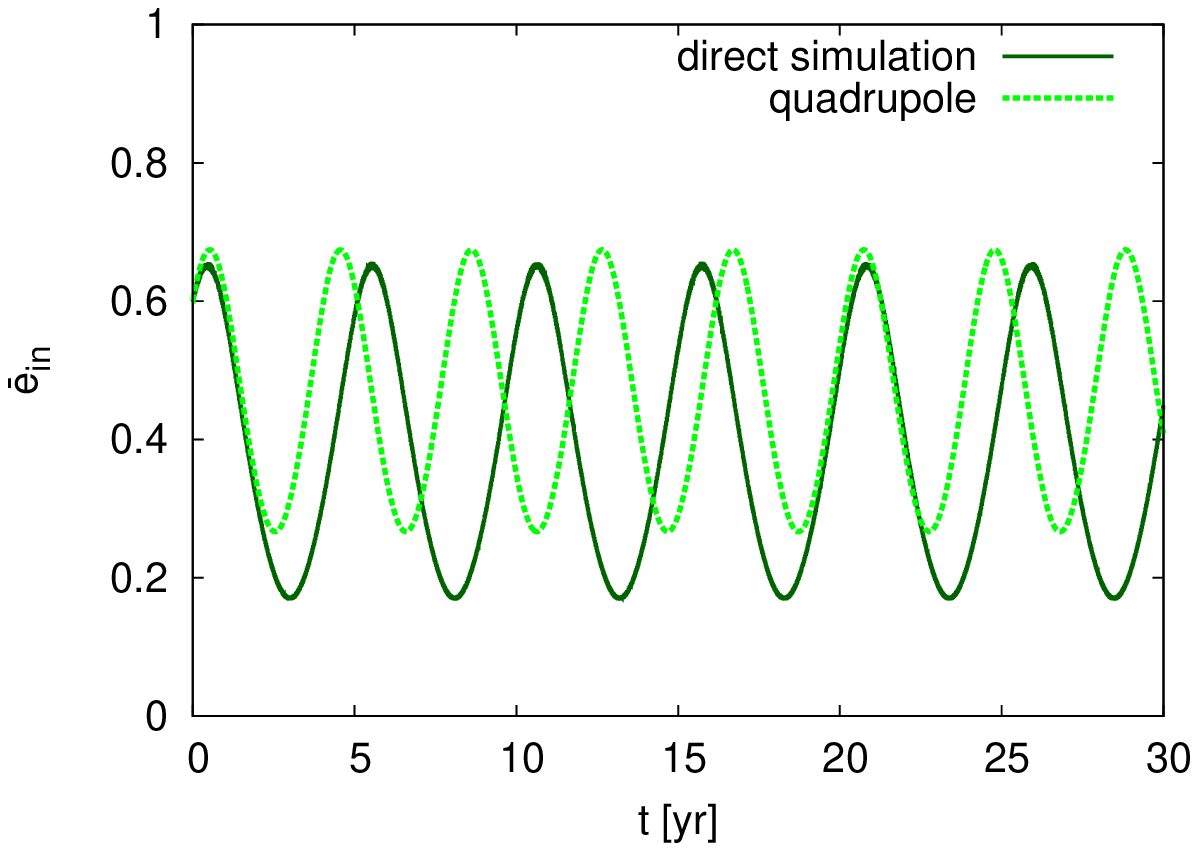}
        \end{minipage} 
        \caption{
                    The same figure as Fig.~\ref{fig:PNN_e_l} for the ``rotation'' type KL-oscillations in PNN model.
                    The top and bottom panels show the results of ICR and IER types, respectively.
                }
        \label{fig:PNN_e_r}
    \end{figure}

\subsubsection{PNN model}
\underline{Evolution of Orbital Parameters}\\[1em]
Figs.~\ref{fig:PNN_KL_l} and \ref{fig:PNN_KL_r} show the time evolution of the averaged inner eccentricity $\bar{e}_\mathrm{in}$, relative inclination $\bar{I}$ and KL-conserved value $\bar{\theta}^2$ of the PNN model. 
Fig.~\ref{fig:PNN_KL_l} shows the result of libration type KL-oscillations, while Fig.~\ref{fig:PNN_KL_r} exhibits those of rotation type KL-oscillations.
In each figure, top and bottom panels correspond to the results of the initially circular and eccentric cases, respectively, whose parameters are given in Table~\ref{tab:ini_GR}. 
In Figs.~\ref{fig:PNN_KL_l} and \ref{fig:PNN_KL_r}, the KL-oscillation is observed in all panels with different amplitudes and timescales:
The initially eccentric cases (bottom panels) have smaller amplitude and shorter timescale than those of initially circular cases (top panels).
This is because in the initially circular case, the eccentricity oscillates 
between zero and some finite value, while in the initially eccentric case, it oscillates 
between two finite values around the initial value.
The same behaviour is also found from the figures of the eccentricity in the double-averaging method.  $C_{\rm KL}$ (or $C_{\rm KL}^{\rm (GR)}$) is 
very small in the initially circular case, while it is 
not so small in the initially eccentric case (see  Fig. \ref{fig:emin_emax_N} 
in Appendix \ref{sec:KL_app} for the Newtonian case).

However, in all panels of both figures, $\theta^2$ is not exactly constant but oscillates with the same period as that of the KL-oscillation although it should be constant in the analysis with test-particle quadrupole approximation in \S\ref{sec:KL}.
This is because all masses of the components in the system are the same in PNN model and the hierarchy assumed in \S\ref{sec:KL} is not enough in this model, that is, the test-particle approximation does not work exactly in this model.
This small deviation from the test-particle limit is consistent with 
the discussion given in \cite{Naoz13a}. \\

\noindent
\underline{Direct Integration v.s. Double-averaging Method}\\[1em]
In Figs.~\ref{fig:PNN_e_l} and \ref{fig:PNN_e_r}, we show the evolution of inner eccentricity obtained in our direct integration (dark-green solid line) as well as 
that calculated with Lagrange planetary equation Eq.~\eqref{eq:lagrange1}-\eqref{eq:lagrange5} (light-green dashed line).
The latter one corresponds to the result obtained by double-averaging 
under the quadrupole approximation\footnote{Double-averaging equations~\eqref{eq:lagrange1}-\eqref{eq:lagrange5} are integrated by the fourth order implicit Runge-Kutta method {\KM using  W4 method (\cite{Okawa18,Fujisawa18}), which is 
an improved version of the Newton-Raphson method,} as an internal nonlinear solver}.
Each panel in Figs.~\ref{fig:PNN_e_l} and \ref{fig:PNN_e_r} is the same evolution as that shown in the corresponding panel in Figs.~\ref{fig:PNN_KL_l} and \ref{fig:PNN_KL_r}. 

{\KM At first glance, the difference is very small except for the IER type (the bottom panel in Fig.  \ref{fig:PNN_e_r} ),
but 
we find some difference between the solid line and dashed lines in all panels as shown below. }
The timescale of KL-oscillation obtained from direct integration is smaller than that calculated in double-averaging method 
{\HS in the panels of Fig.~\ref{fig:PNN_e_l}, but it is larger in the panels in Fig.~\ref{fig:PNN_e_r}}.
The deviation in timescale is much more obvious {\HS in the bottom panel of Fig.~\ref{fig:PNN_e_r}}.
For the amplitude, the tendency of the difference is not the same in all panels.
In the results of ICL and ICR types (top panels in Fig.~\ref{fig:PNN_e_l} and \ref{fig:PNN_e_r}), the amplitude of KL-oscillation is {\HS larger} in our direct simulation than that obtained with double-averaged calculation {\HS with quadrupole expansion}.
Both curves in these panels have the same minimum values but the maximum values are 
 enhanced  in {\HS dark}-green lines.
In the result of IER type (bottom panel of Fig.~\ref{fig:PNN_e_r}), 
the enhancement of the amplitude in the {\HS direct simulation}
is observed as seen in ICL and ICR types, but both maximum and minimum values in light-green line are different from those of dark-green line:
both maximum and minimum values are larger in light-green line than those of dark-green line. 
The result of IEL type, (top panel of Fig.~\ref{fig:PNN_e_l}), on the other hand, 
{\HS the amplitude of the dark-green line is almost the same as that of the light-green line.} 
These differences are summarised in Table \ref{tab:KLtimescale_PNN}.

{\KM One may wonder the double averaging method can be improved 
if we take into account the higher-multipole interaction terms. 
Here we just comment about the calculation with the double averaging method up to the octupole-order expansion.
We have also performed numerical calculation including the octupole-order expansion \citep{Ford20, Naoz13a, Naoz13b}.
In the models with PN-binary, the results obtained from octupole-order equations are 
completely the same as quadrupole ones because the octupole terms are always proportional to the mass difference ($m_1-m_2$).
Hence, to see the effect of the octupole-order terms, we 
analyze the models with different-mass binaries (e.g. model PBB).
We show the results in Appendix~\ref{sec:octupole}.
The octupole-order terms seem to improve the results obtained by quadrupole ones, but 
it is not always the case (see Appendix~\ref{sec:octupole} for the detail).
However, both quadrupole- and octupole-order double-averaged calculations 
do not exactly reproduce the evolution obtained by direct integration (see Appendix~\ref{sec:octupole}). 
}

We remark that these differences between eccentricity evolution obtained from direct integration and {\HS those by double-averaging methods} may be crucial when 
we evaluate the GW emission for the systems with finite masses, that is, one may overestimate or underestimate the maximum or minimum value of the 
eccentricity when we use the double-averaging method.
The amplitude and frequency of the gravitational waves are strongly sensitive to the eccentricity, especially for the highly eccentric orbit like $e>0.9$. 
It may be important to calculate the evolution of such an orbit by direct integration.
\begin{table}
\begin{tabular}{c|c|c|c|c|c}
\hline
Type & $\theta^2$   & $C_{\rm KL}^{\rm (GR)}$         &
$e_{\rm min}$&  $e_{\rm max}$  &  $T_{\rm KL}$[yrs]  \\
\hline
ICL&   $0.25$ &  $-3.18\times 10^{-5}$   &  $0.00728$  &   $0.687$     &$12.465$   \\
\cline{4-6}
&    &   &   $0.00613$  &   $0.683$      &$12.858$   \\
\hline
ICR&  $0.25$ &  $6.20\times 10^{-5}$    & $0.00689$  &  $0.687$        &$12.117$  \\
\cline{4-6}
&    &   &   $0.00768$  &  $0.683$       &$11.937$  \\
\hline
 IEL & $0.232$   &  $-0.170$  &  $0.525$  &   $0.605$        &$3.473$   \\
\cline{4-6}
 &   &  &   $0.524$  &   $0.600$        &$3.477$   \\
\hline
IER &  $0.32$  &  $0.0667$  &  $0.170$ &  $0.654$       &$5.083$ 
\\
\cline{4-6}
 &   &   &    $0.267$ &  $0.675$        &$4.046$ 
\\
\hline
\hline
	\end{tabular}
		\caption{Comparison between the results by the direct integration and those by the double averaging method for the PNN model with $\epsilon^{\rm (1PN)}=0.177$.
		We show the maximum and minimum eccentricities, and the KL oscillation period $T_{\rm KL}$. 
		The first rows give the results by the direct integration, while the second rows show the 
		results  by the double averaging method.}
			\label{tab:KLtimescale_PNN}
\end{table}
    \begin{figure}
        \centering
        \begin{minipage}{7.5cm}
            \includegraphics[width=7.5cm]{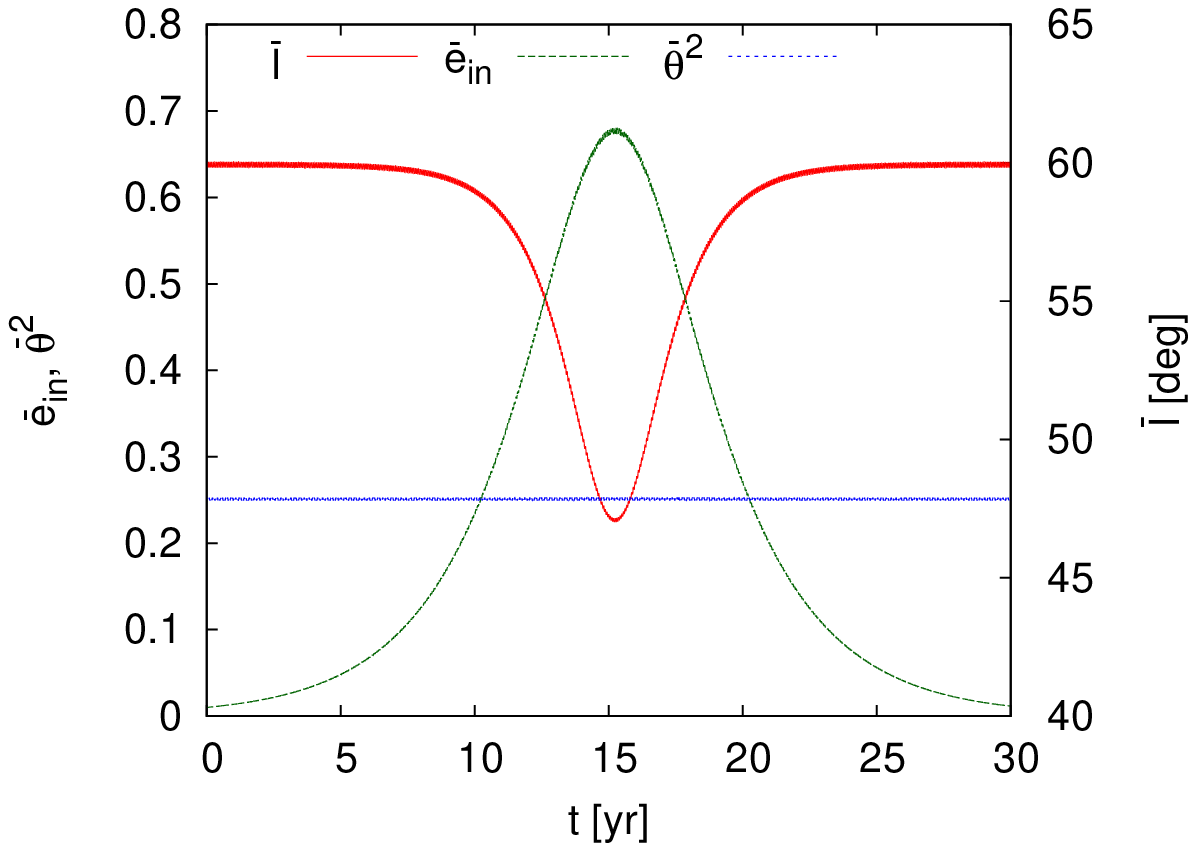}
        \end{minipage}\\
        \begin{minipage}{7.5cm}
            \includegraphics[width=7.5cm]{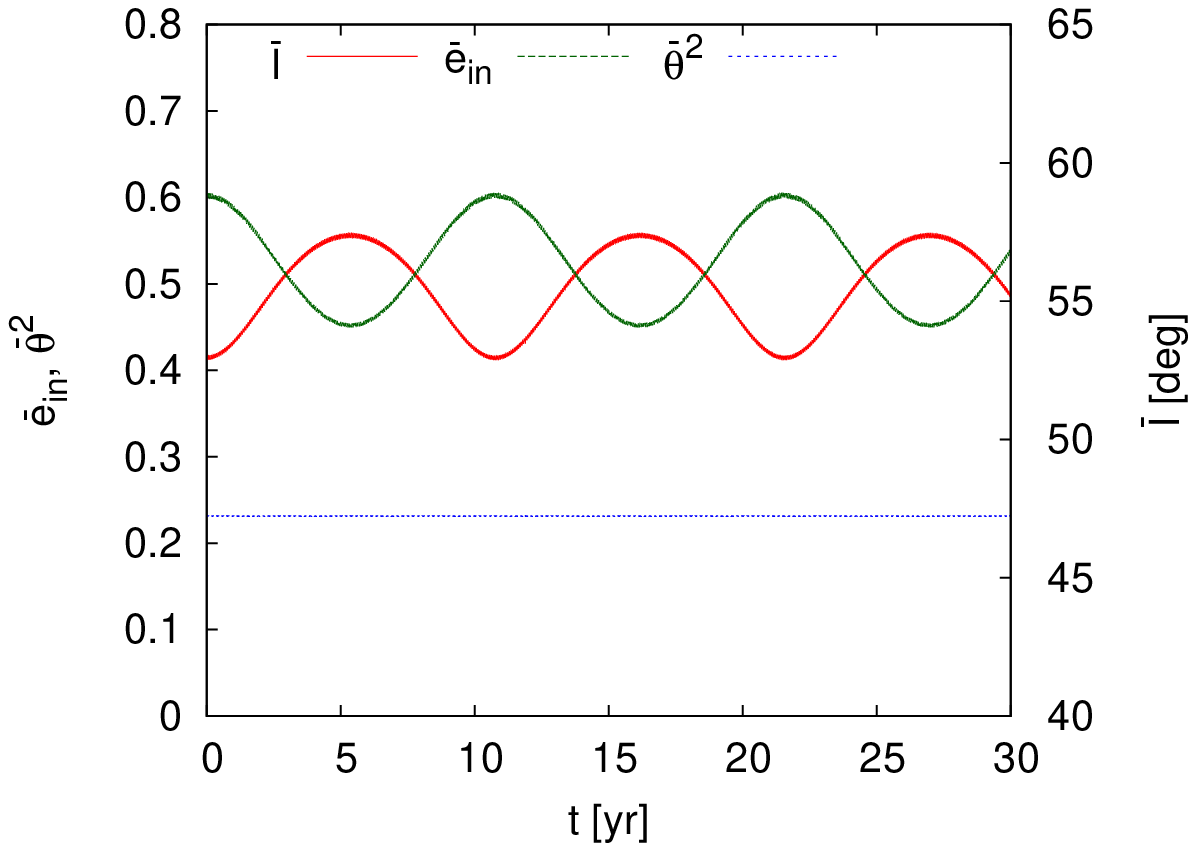}
                   \end{minipage}
        \caption{
                    The same figure as Fig. \ref{fig:PNN_KL_l} for PNIB model.
                    Top and bottom panels correspond to the results of ICL and IEL types, respectively.
                }
        \label{fig:PNIB_KL_l}
    \end{figure}
    \begin{figure}
        \centering
              \begin{minipage}{7.5cm}
            \includegraphics[width=7.5cm]{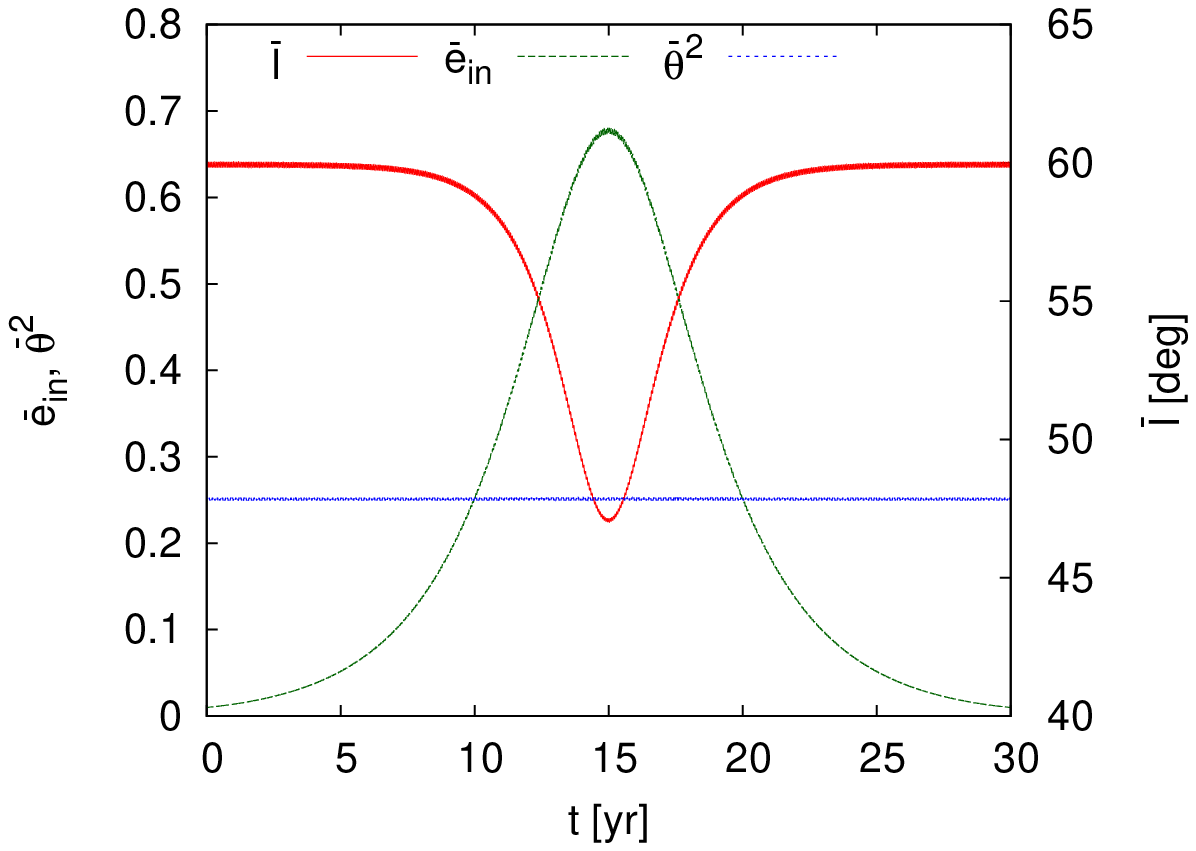}
        \end{minipage}\\
        \begin{minipage}{7.5cm}
            \includegraphics[width=7.5cm]{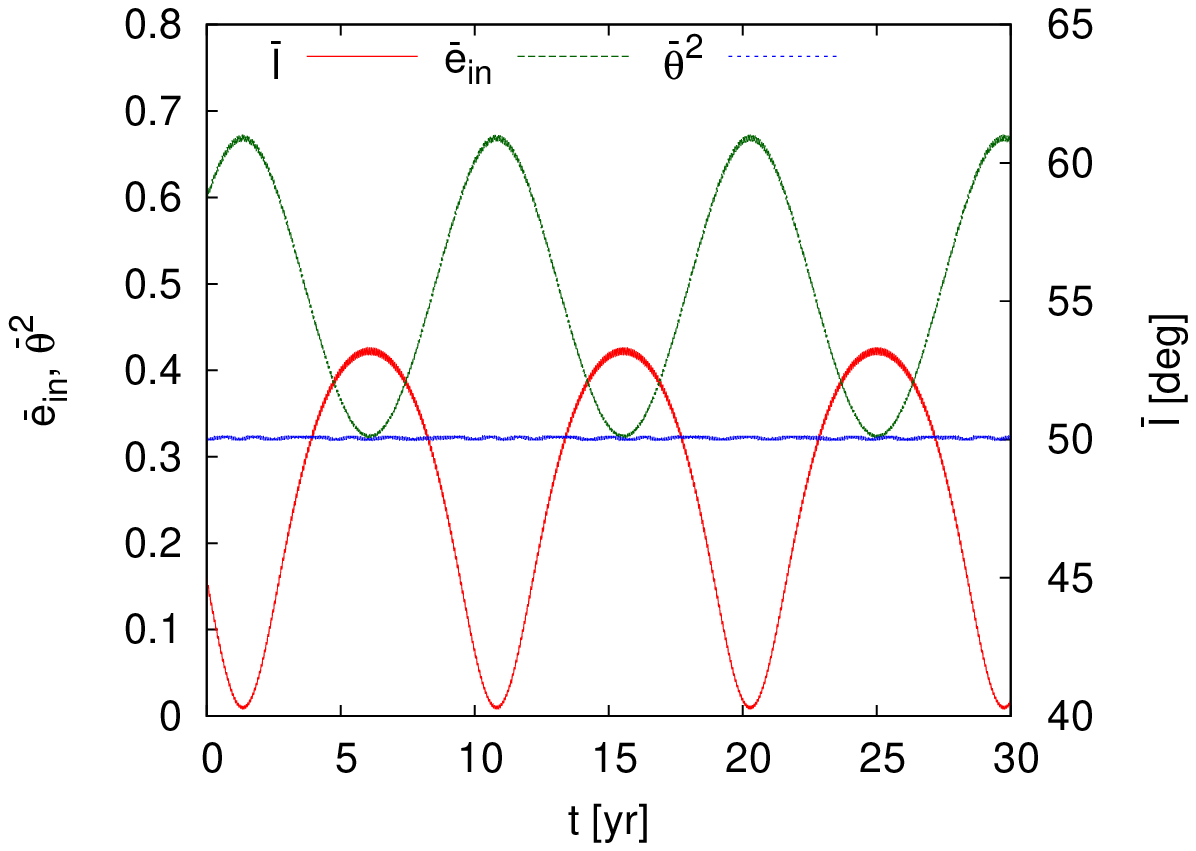}
        \end{minipage}
        \caption{
                    The same figure as Fig. \ref{fig:PNN_KL_r} for PNIB model.
                    Top and bottom panels correspond to the results of ICR and IER types, respectively.
                }
        \label{fig:PNIB_KL_r}
    \end{figure}

\subsubsection{PNIB model}
\underline{Evolution of Orbital Parameters}\\[1em]
Figs.~\ref{fig:PNIB_KL_l} and \ref{fig:PNIB_KL_r} are the same figures as Figs.~\ref{fig:PNN_KL_l} and \ref{fig:PNN_KL_r} but for PNIB model. 
Figs.~\ref{fig:PNIB_KL_l} and \ref{fig:PNIB_KL_r} reflect the features of libration and rotation types of KL-oscillations, respectively.
In each figure, top and bottom panels are the results of initially circular and eccentric types, respectively.
As seen in Figs.~\ref{fig:PNN_KL_l} and \ref{fig:PNN_KL_r}, initially eccentric cases (bottom panel) have smaller amplitude and shorter timescale than those of initially circular cases (top panels), which is similar to the PNN model.
As for the KL oscillation period, it does not seem to depend on the oscillation 
types  in 
the initially circular case, while in the initially eccentric case, 
the rotation type (the bottom panel in Fig.~\ref{fig:PNIB_KL_r}) 
gives shorter oscillation time than that in the libration type (the bottom panel in Fig.~\ref{fig:PNIB_KL_l}).

$\bar{\theta}^2$ is almost constant in PNIB model unlike that in PNN model.
It is because the test-particle approximation is valid in PNIB model.
In fact, the deviation 
from the double-averaging method
is smaller than that of PNN model.\\

\noindent
\underline{Newtonian v.s. post-Newtonian}\\[1em]
PNIB model has the second largest value of 
$\epsilon^\mathrm{(1PN)}$ in Table~\ref{tab:model} and 
its relativistic effect is the strongest in our models except PIBIB model.
Since the main features are the same in both models, we shall discuss the PNIB model as
a representative of relativistic ones.

\begin{figure}
        \centering
        \begin{minipage}{7cm}
            \includegraphics[width=7cm]{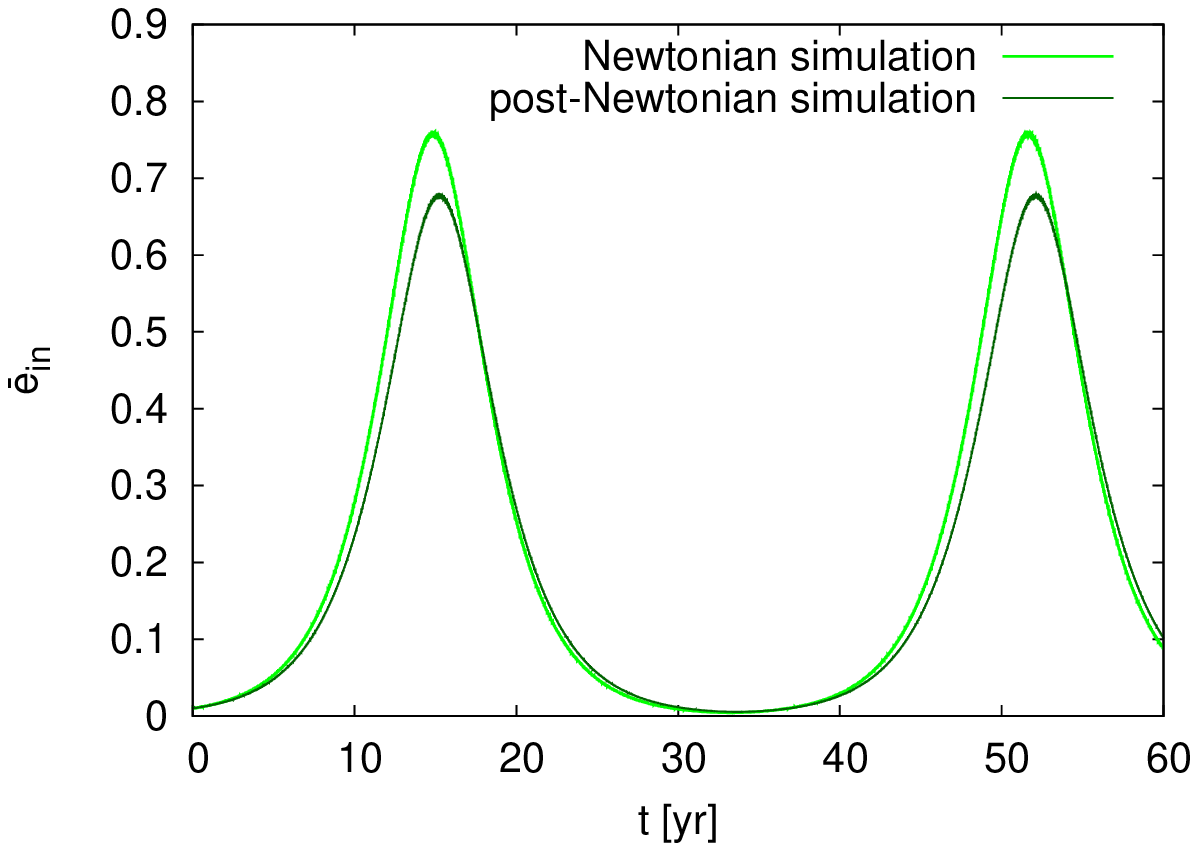}
        \end{minipage}\\
        \begin{minipage}{7cm}
            \includegraphics[width=7cm]{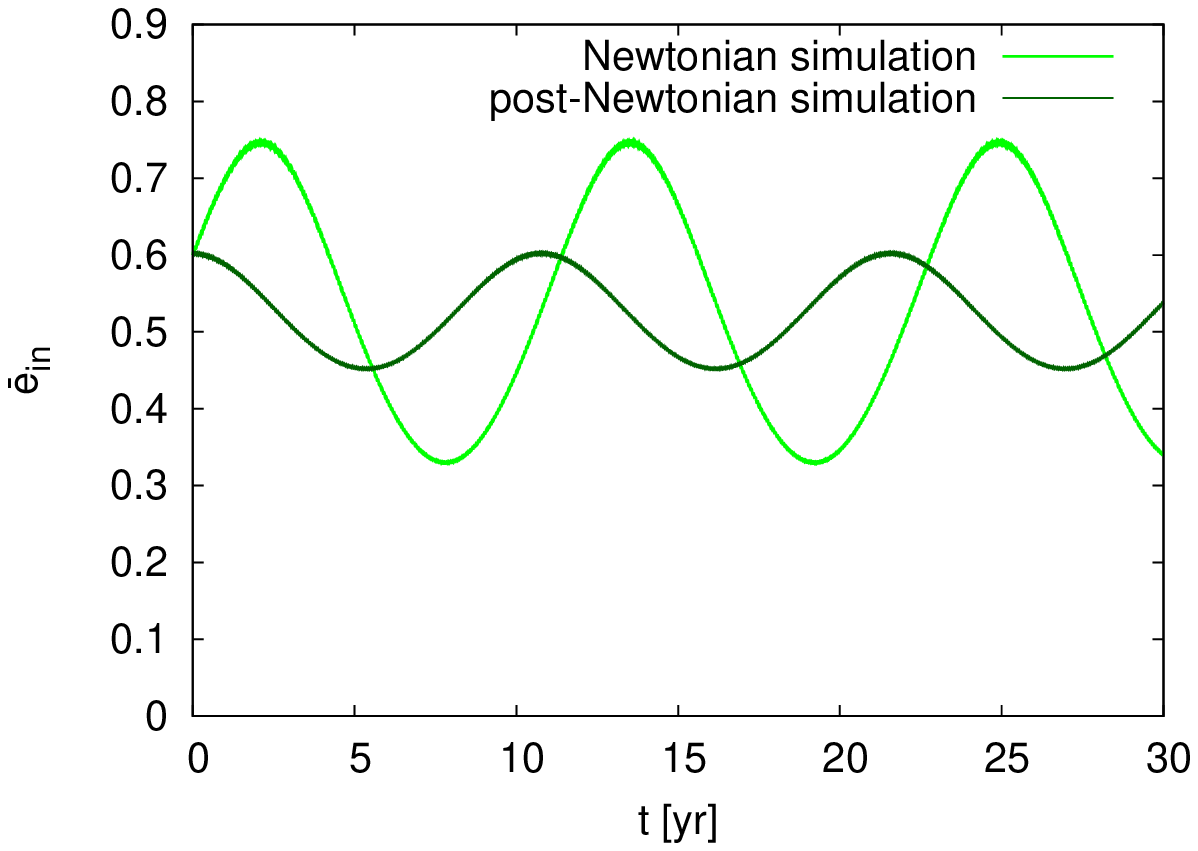}
        \end{minipage} 
        \caption{
                    Comparison between Newtonian and post-Newtonian evolution curves of the averaged inner eccentricity $\bar{e}_\mathrm{in}$ for the ``liblation'' type of KL oscillations in the PNIB model.
                    Top and bottom panels correspond to the results of ICL and IEL types, respectively. 
                    The light- and dark-green curves describe the results obtained from Newtonian and post-Newtonian direct simulations.
                }
        \label{fig:PNIB_e_l}
    \end{figure}

\begin{figure}
        \centering
        \begin{minipage}{7cm}
            \includegraphics[width=7cm]{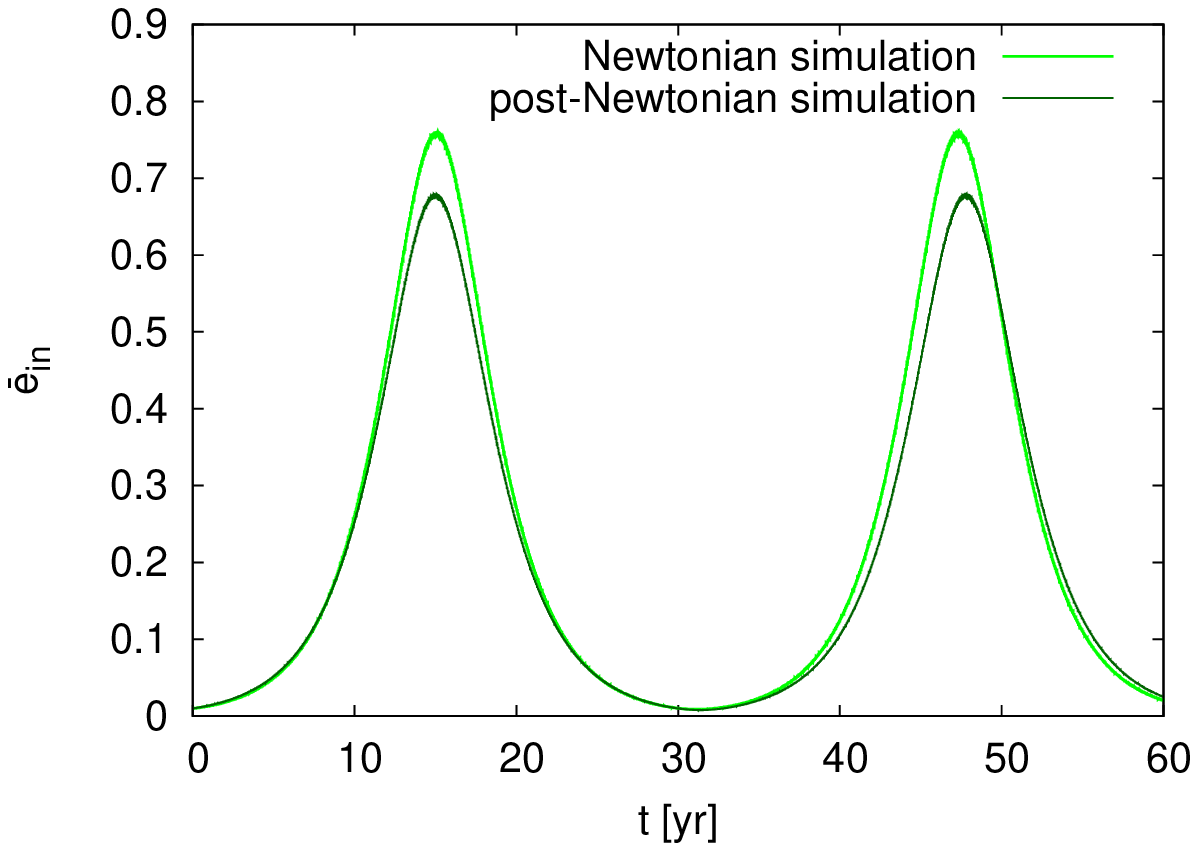}
        \end{minipage}\\
        \begin{minipage}{7cm}
            \includegraphics[width=7cm]{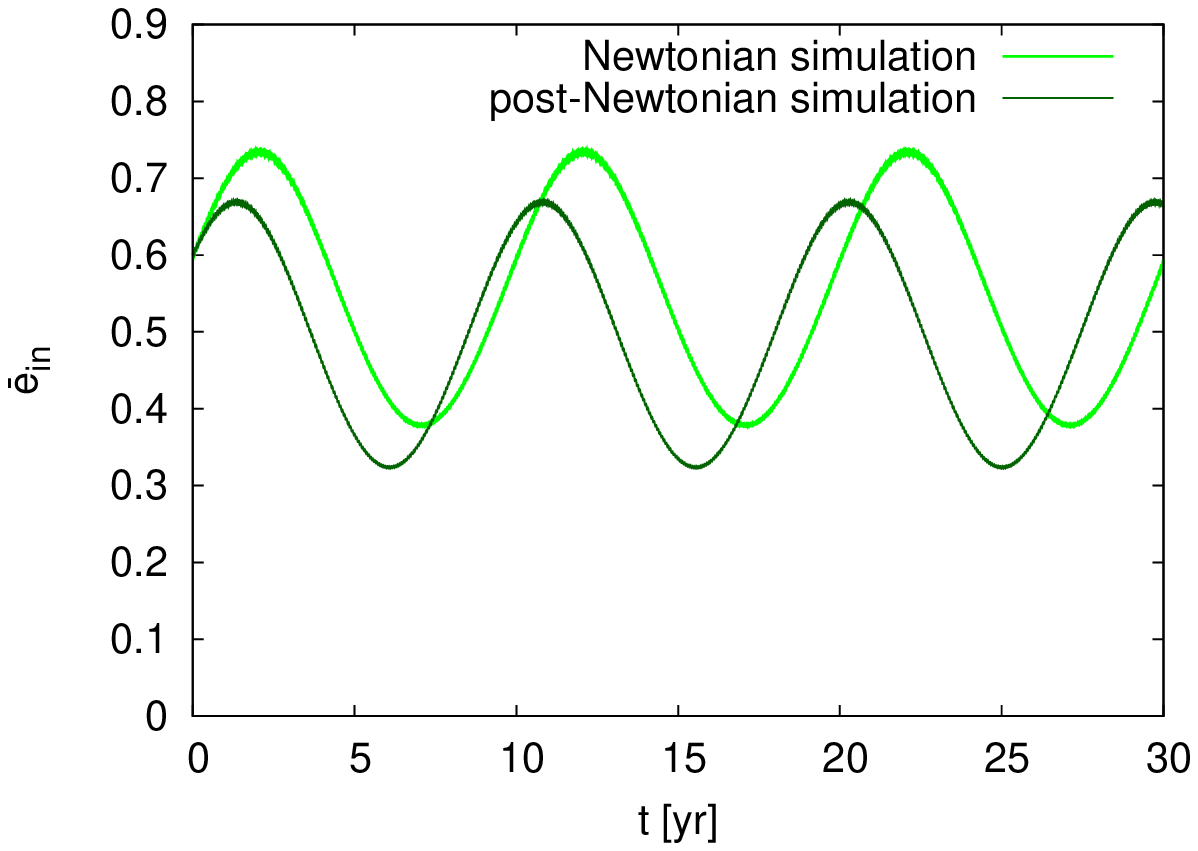}
        \end{minipage} \\
        \caption{
                    The same figure as Fig. \ref{fig:PNIB_e_l} for the ``rotation'' type KL-oscillations in the PNIB model.
                    Top and bottom panels correspond to the results of ICR and IER types, respectively.
                 }
        \label{fig:PNIB_e_r}
    \end{figure}

In Figs.~\ref{fig:PNIB_e_l} and \ref{fig:PNIB_e_r}, we show 
the evolution of the eccentricities obtained by Newtonian and post-Newtonian direct simulations.
Each figure exhibits the results of libration and rotation types of KL-oscillations.
The top and bottom panels in each figure correspond to  
the results of initially circular and eccentric types.
The Newtonian and post-Newtonian results are described by  the light- and dark-green curves, respectively.
The tendency of the difference between two curves is not the same in all panels.
In the results of ICL and ICR types (top panels of Figs.~\ref{fig:PNIB_e_l} and \ref{fig:PNIB_e_r}), the amplitude of KL-oscillation is smaller in post-Newtonian simulation than that obtained from Newtonian calculation. Both curves in those results have the same minimum values (about zero), but the maximum value is suppressed in post-Newtonian curve.
The KL-timescale is a little longer in post-Newtonian result in the initially circular types.
\begin{table}
\begin{tabular}{c|c|c|c|c|c|c|c}
\hline
Type & $\theta^2$& $C_{\rm KL}^{\rm (GR)}$  &          &
$e_{\rm min}$&  $e_{\rm max}$  &$T_{\rm KL}$[yrs]  \\
				\hline
ICL &  $0.25$  & $-1.64\times10^{-5}$ &N& $0.00438$ &   $0.761$    &$ 36.779$  \\
\cline{4-7}
 &    &    & 1PN &  $0.00550$ &   $0.680$     &$36.913$  \\
\hline
ICR& $0.25$ & $7.73\times10^{-5}$ &N  & $0.00869$ &   $0.761$  &$32.231$   \\
\cline{4-7}
 &    &    & 1PN  &  $0.00776$ &   $0.680$     &$32.60 $ \\
\hline
IEL  & $0.232$   & $-0.0931$ &N& $0.328$ &   $0.749$    &$ 11.419$  \\
\cline{4-7}
 &     &    & 1PN &  $0.450$ &   $0.605$     &$10.804$  \\
\hline
IER &$0.32$   & $0.143$ &N & $0.377$ &   $0.738$   &$9.992$  \\
\cline{4-7}
&      &   & 1PN &  $0.322$ &   $0.672$    &$9.504$ \\
\hline
\hline
	\end{tabular}
		\caption{The comparison between Newtonian and post-Newtonian results
		for the PNIB model.
		 $T_{\rm KL}$ denotes the KL-oscillation period. The first rows give the Newtonian results, while the second rows show the 
		results with 1st post-Newtonian correction ($\epsilon^{\rm (1PN)}=0.484$).}
			\label{tab:KLtimescale_PNIB}
\end{table}	
In the results of IEL and IER types (bottom panels of Figs.~\ref{fig:PNIB_e_l} and \ref{fig:PNIB_e_r}), on the other hand, the KL-timescale obtained in post-Newtonian calculation is shorter than that obtained from Newtonian one.
Interestingly, IEL (bottom panel of Fig.~\ref{fig:PNIB_e_l}) and IER (bottom panel of Fig.~\ref{fig:PNIB_e_r}) have different features in the amplitude.
In the result of IEL type, the amplitude obtained by post-Newtonian simulation are smaller than those of Newtonian result; unlike results of the ICL and ICR types, both maximum and minimum values are suppressed in this case.
On the other hand, in the result of IER type, the amplitudes of Newtonian and post-Newtonian results are almost the same but both maximum and minimum values of post-Newtonian result are shifted downward.

These complicated features can be understood basically by using the double-averaging method,
which is given in Appendix \ref{subsec:GR_app}.
As shown in Fig. \ref{fig:emin_emax_NvsPN}, 
the curve of the maximum-minimum eccentricity in terms of
$C_{\rm KL}$ in Newtonian dynamics is shifted to the right 
when the post-Newtonian correction term is taken into account. 
Here we have used $C_{\rm KL}^{\rm (GR)}$ instead of 
$C_{\rm KL}$ as the horizontal axis because it is conserved and classifies the oscillation types, libration or rotation.
Hence when we include the post-Newtonian correction term,
fixing two conserved quantities 
($\theta^2$ and $C_{\rm KL}^{\rm (GR)}=C_{\rm KL}$),
we find that the maximum value decreases and the minimum value increases for the libration type, while 
both maximum and minimum values decrease for the rotation type. 
As for the KL oscillation, the analysis by the double-averaging method explains the results by the direct integration (compare Figs.~\ref{fig:PNIB_e_l} and \ref{fig:PNIB_e_r}  with
Table \ref{tab:KLtimescale_PNIB}. ).


%
\subsubsection{Irregularity of KL oscillation period}
\label{irregular_period}
    \begin{figure}
        \centering
        \begin{minipage}{7cm}
            \includegraphics[width=7cm]{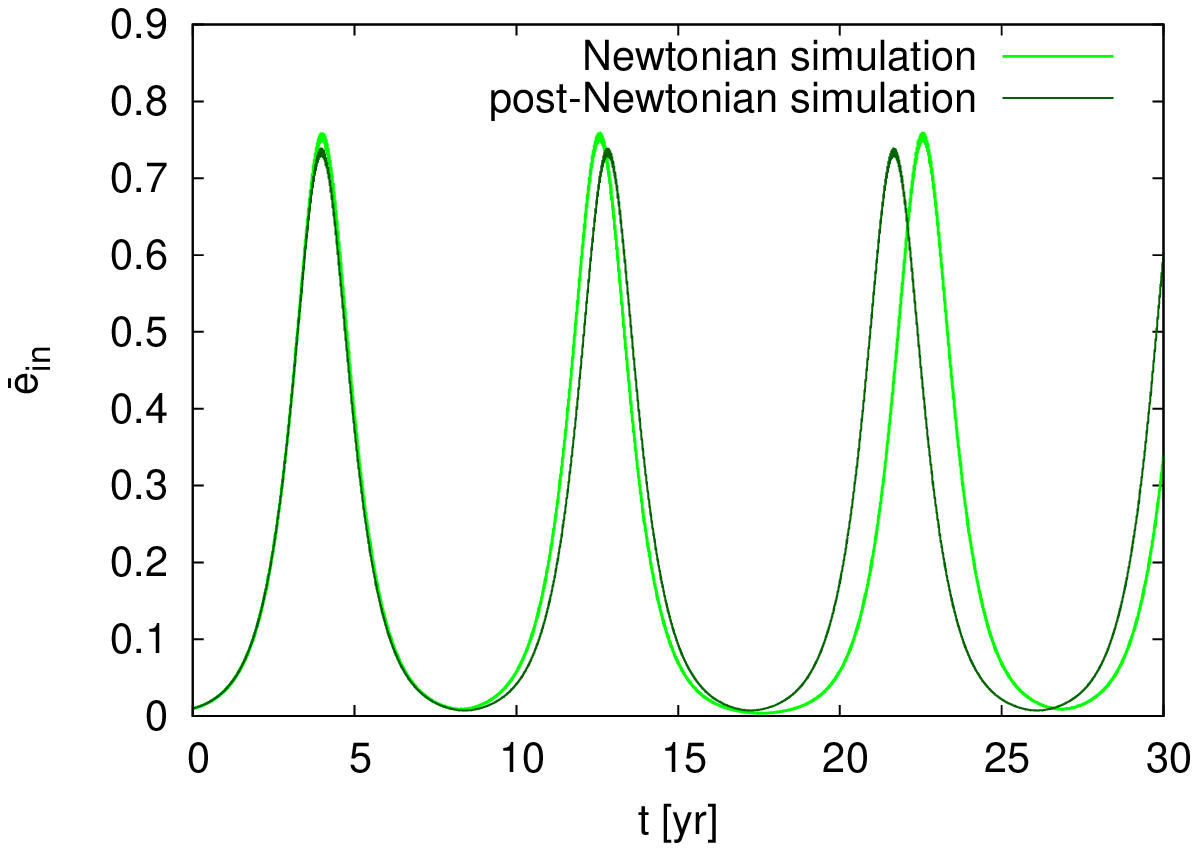}\\
           \includegraphics[width=7cm]{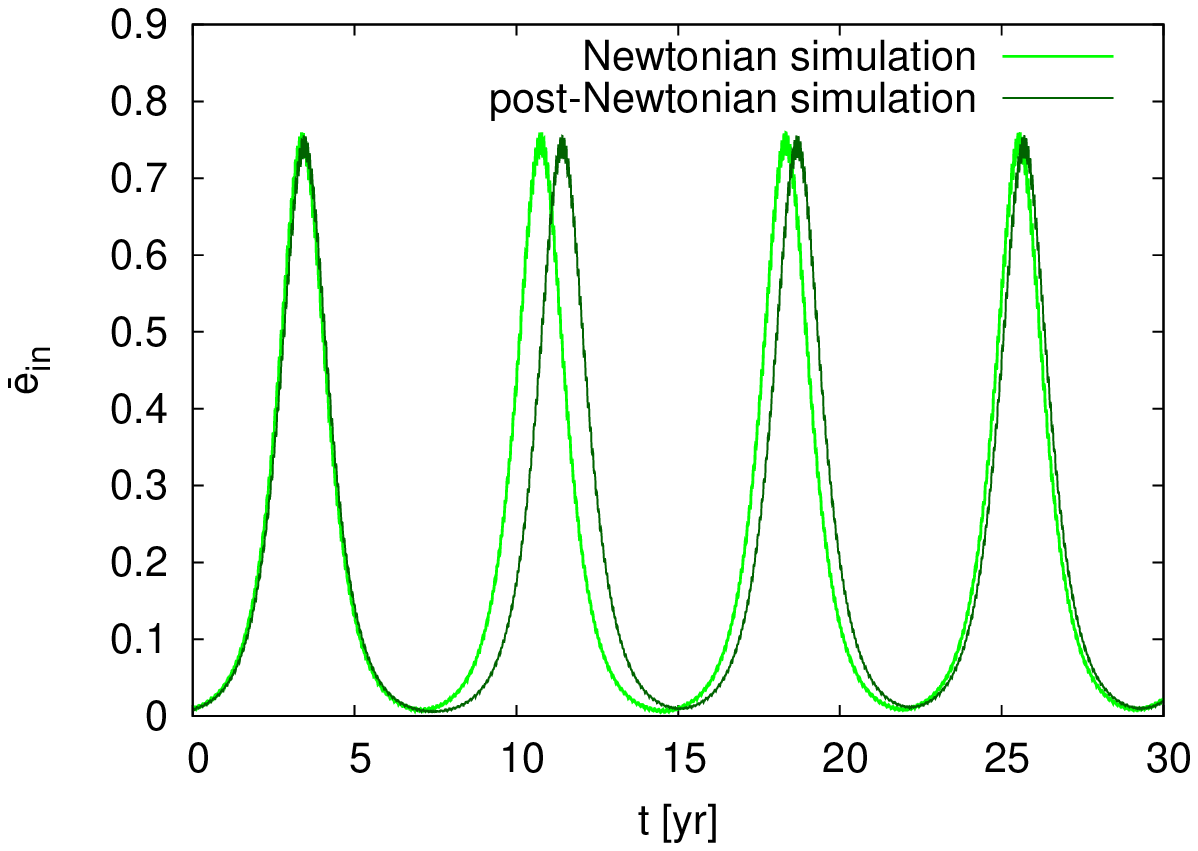}
       \end{minipage}\\
        \caption{
                    The same figures as Fig. \ref{fig:PNIB_e_l} for ICR type KL-oscillations in the PNB model
                    (top) and the PBB model (bottom). 
                     The period from one maximum to the next one is not regular 
                     for the Newtonian case in the top figure and for the post-Newtonian case 
                     in the bottom figure.
                 }
        \label{fig:PNB_PBB_e_r}
    \end{figure}
    \begin{figure}
        \centering
        \includegraphics[width=7cm]{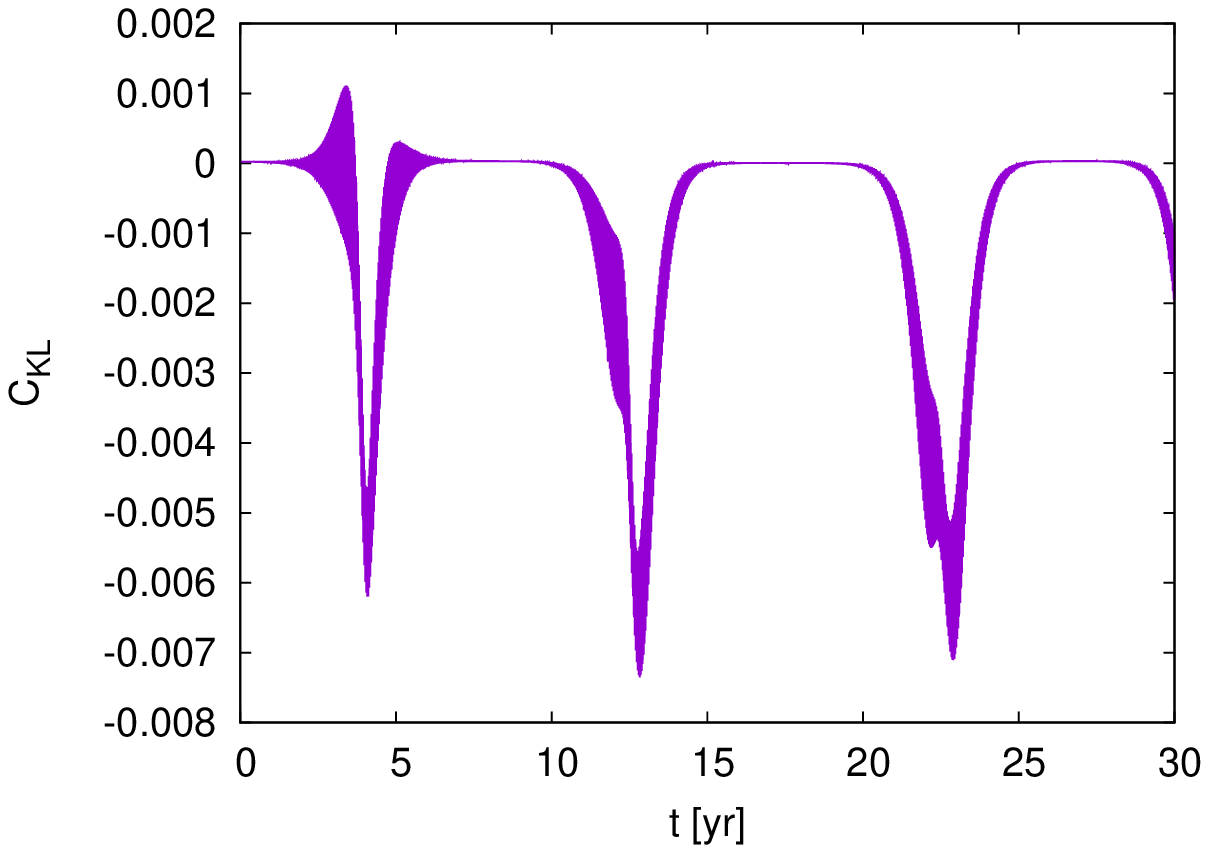}\\
         \includegraphics[width=7cm]{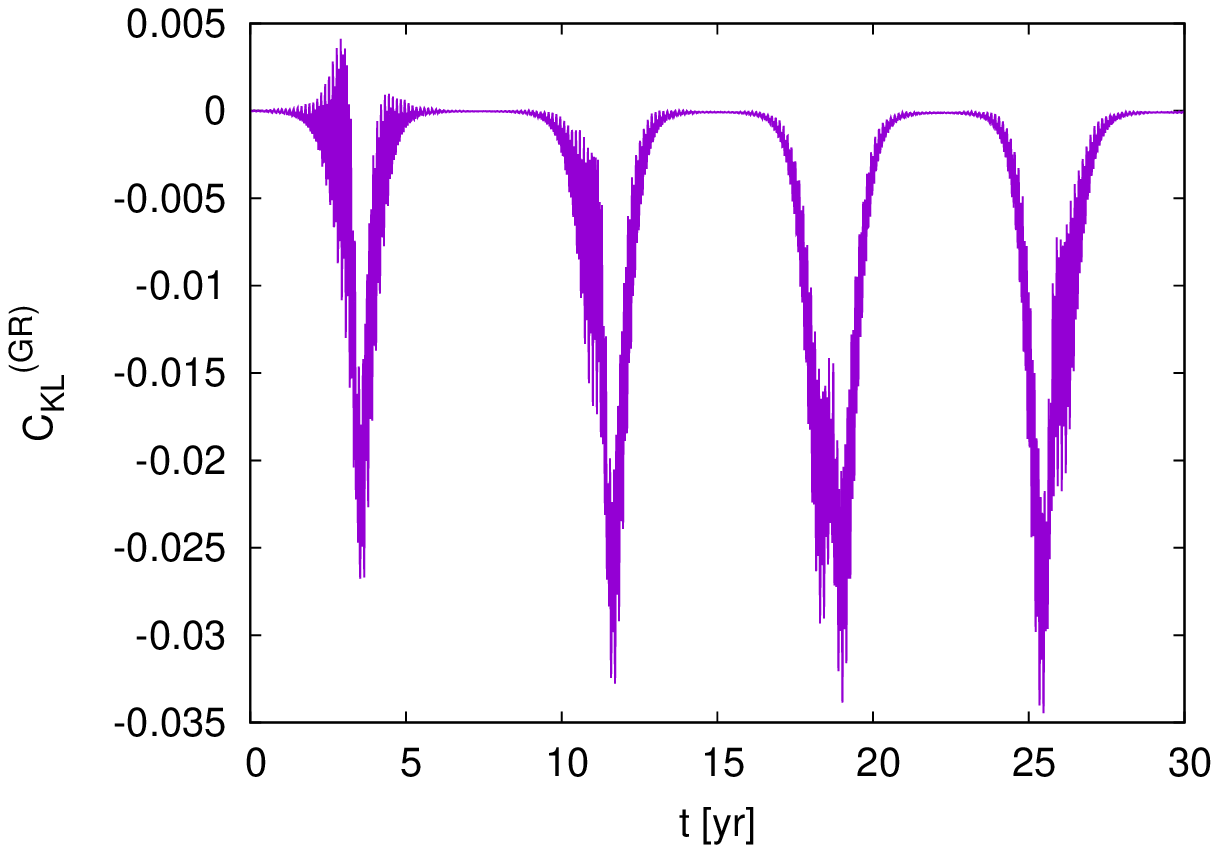}
      \caption{
                    Time evolutions of $C_\mathrm{KL}$ and $C_\mathrm{KL}^\mathrm{(GR)}$
                    for ICR type KL-oscillations in the PNB model (top)
                    and the PBB model (bottom), respectively.
                 }
        \label{fig:PNB_PBB_cKL_r}
    \end{figure}

As we showed above, the amplitude of KL oscillation and its period 
can be understood basically by the double-averaging method.
However we find that there appears an irregularity of the period
in some models.
For example, the KL-oscillations in ICR type of the PNB and PBB models 
show irregular periods  (see Fig.~\ref{fig:PNB_PBB_e_r}).
This irregular behaviour of the KL-oscillation period was already found in \cite{Antonini_2012}.
They calculated orbital evolutions of BH binaries around SMBH by using N-body integrator
and found the irregular periods and amplitudes in the KL-oscillation (Fig.~3 in their paper)
.

Since the calculations in \cite{Antonini_2012} and ours are performed by the direct integration,
one may naturally expect some deviation from the double averaging method,
in which the KL-oscillation period is regular.
However, since the deviation in our calculation is very small,
the double averaging method may provide almost correct results. 
Note that 
the amplitude and timescale of KL-oscillation 
are strongly dependent on two conserved quantities $\theta^2$ and $C_\mathrm{KL}^\mathrm{(GR)}$, 
but not so much 
on $\epsilon^\mathrm{(1PN)}$ except for the ICL oscillation type, in which
the relativistic effect is large because it changes the existence range 
of KL oscillation.
Hence we analyse the behaviour of the ``conserved" quantities 
in our simulations.
As for $\theta^2$, although it oscillates with the outer orbit period,
the averaged value is almost constant except around 
the time when the eccentricity
reaches the maximum value.
We then show 
the time evolution of $C_{\rm KL}$ and $C_{\rm KL}^{\rm (GR)}$ in 
the top and bottom of Fig.~\ref{fig:PNB_PBB_cKL_r}, respectively.
It is because the irregularity is clearer for Newtonian calculation 
in the PNB model, while it is so for the post-Newtonian calculation 
in the PBB model.
These figures show that $C_{\rm KL}$ or $C_{\rm KL}^{\rm (GR)}$ 
is not conserved when the eccentricity 
reaches the maximum value. 
However it becomes almost constant again 
when the eccentricity decreases.

In order to see the detail, in Table \ref{tab:irregular_period}, 
 we show the numerical values of the oscillation periods.
 The period $n$ ($n=1, 2, 3$) denotes the period from the $n$-th peak of the eccentricity
 to the $(n+1)$-th peak. 
We also show the constant ``conserved" values after the eccentricity passes through the maximum value in Table \ref{tab:irregular_period}.
We evaluate the KL oscillation periods by the double-averaging method with those values of 
 $C_{\rm KL}/C_{\rm KL}^{\rm (GR)}$, which are given in the third row of each period
 in Table \ref{tab:irregular_period}.
 We find that those periods are consistent with the numerical ones by the direct integration.
We believe that these small deviations of the ``conserved" values in each period
causes small irregularity of the KL oscillation period.
We still have a small difference from the numerical simulation, which may be 
because of large deviation of  $C_{\rm KL}/C_{\rm KL}^{\rm (GR)}$ near the maximum 
eccentricity.

	\begin{table}
		\begin{tabular}{c|cc|cc}
			\hline
		model&\multicolumn{2}{c}{PNB}&\multicolumn{2}{c}{PBB}
		    \\
		    \cline{2-5}
 		&Newtonian&1PN&Newtonian&1PN
 		\\
 		\hline
 		period 1 & 8.6 yrs & 8.8 yrs & 7.4 yrs & 8.0 yrs
 		\\
  $C_{\rm KL}$/$C_{\rm KL}^{\rm (GR)}$ & $3.2\times 10^{-5}$ & $5.5\times 10^{-5}$ & $2.6\times 10^{-5}$ & $-2.7\times 10^{-5}$
 		\\
 		\cline{2-5}
 	 &9.34 yrs&8.90 yrs&8.06 yrs&8.03 yrs
		\\
 		\hline
		\hline
 		period 2 & 10.0 yrs & 8.8 yrs & 7.6 yrs & 7.2 yrs
 		\\
 $C_{\rm KL}$/$C_{\rm KL}^{\rm (GR)}$ & $5.2\times 10^{-6}$  & $5.6\times 10^{-5}$ & $1.9\times 10^{-5}$ & $-7.9\times 10^{-5}$
 		\\
 		\cline{2-5}
 		 &10.77 yrs&8.88 yrs&8.26 yrs&7.31 yrs
		\\
 		\hline
	    \hline
 		period 3 & 8.6 yrs & 8.8 yrs & 7.2 yrs & 7.1 yrs
 		\\
 $C_{\rm KL}$/$C_{\rm KL}^{\rm (GR)}$ & $3.2\times 10^{-5}$ & $5.6\times 10^{-5}$ & $3.1\times 10^{-5}$ & $-1.1\times 10^{-4}$
 		\\
 		\cline{2-5}
 		 &9.34 yrs&8.88 yrs&7.94 yrs&7.09 yrs
		\\
 		\hline
		\hline
	    \end{tabular}
		\caption{The period of KL oscillations.
 The period $n$ ($n=1, 2, 3$) denotes the period from the $n$-th peak of the eccentricity
 to the $(n+1)$-th peak. $C_{\rm KL}/C_{\rm KL}^{\rm (GR)}$ is the ``conserved" value
 after the eccentricity passes through the maximum value.
 The periods calculated by the double-averaging method with the same values of 
 $C_{\rm KL}/C_{\rm KL}^{\rm (GR)}$ are given in the third rows of each period.} 
		\label{tab:irregular_period}
    \end{table}	    

\subsection{Cumulative Shift of Periastron Time (CSPT)}
\label{subsec:CSPT}
    \begin{figure}
        \centering
        \begin{minipage}{7cm}
            \includegraphics[width=7cm]{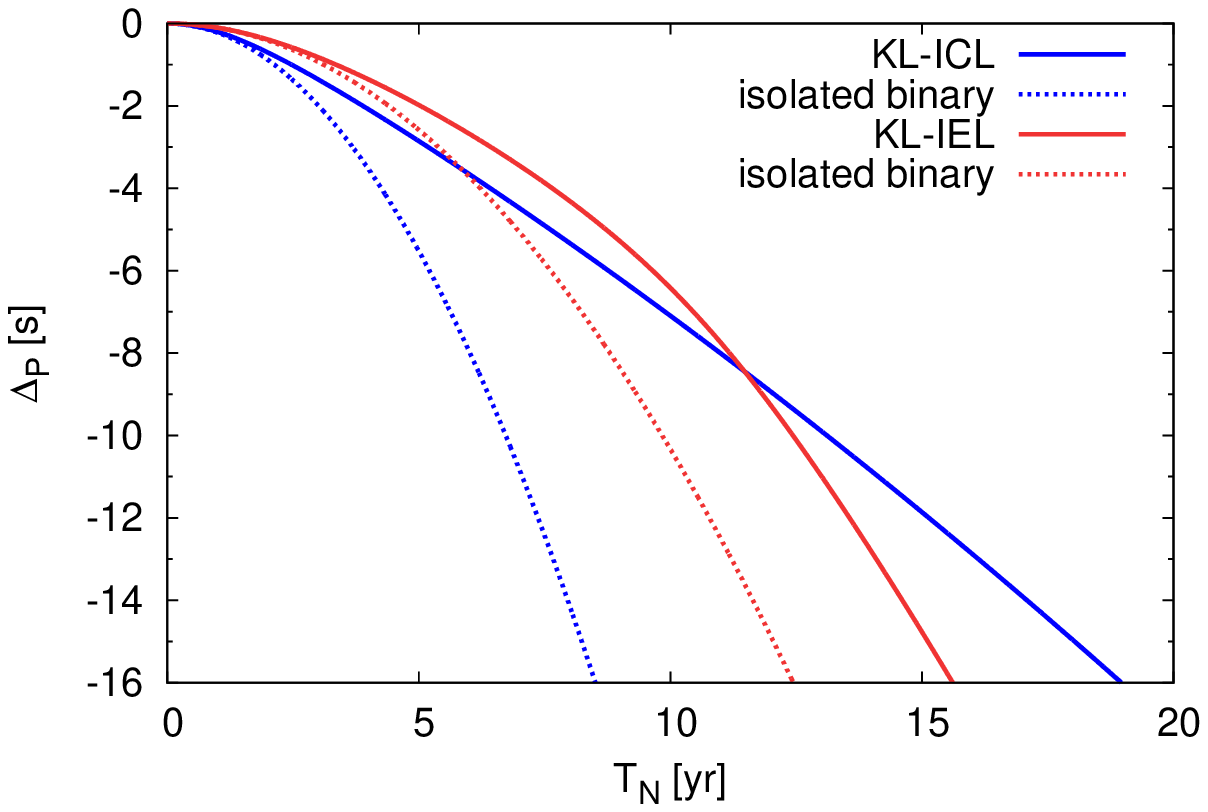}
        \end{minipage}\\
        \begin{minipage}{7cm}
            \includegraphics[width=7cm]{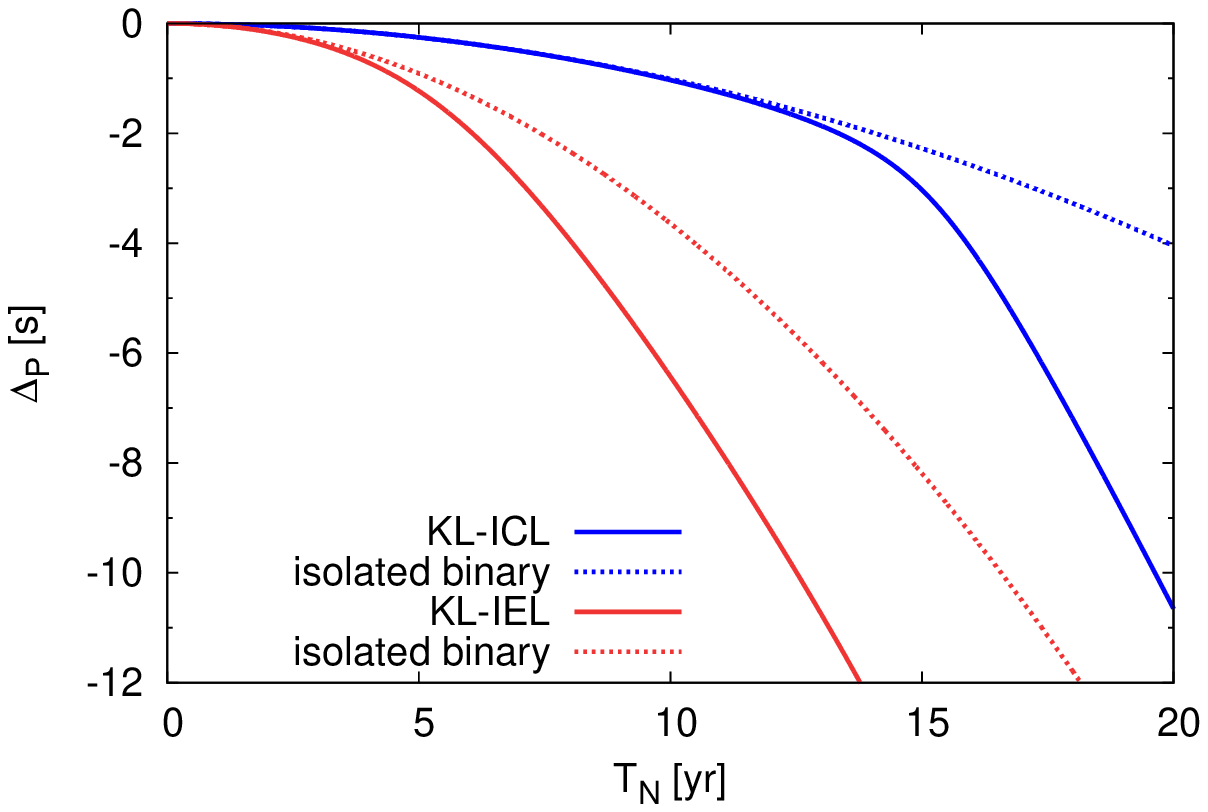}
        \end{minipage} 
        \caption{
                    The CSPT curve for libration type of PNIB model is shown.
                    Top and bottom panels are the results integrated
                    from the time of maximum and minimum eccentricities, respectively. 
                    The blue and red solid curves correspond to ICL and IEL types,
                    respectively. The dashed curves are those of isolated binaries whose parameters are the same as the initial values of the inner binaries of corresponding types.
                }
        \label{fig:CSPT_l}
    \end{figure}
    \begin{figure}
        \centering
        \begin{minipage}{7cm}
            \includegraphics[width=7cm]{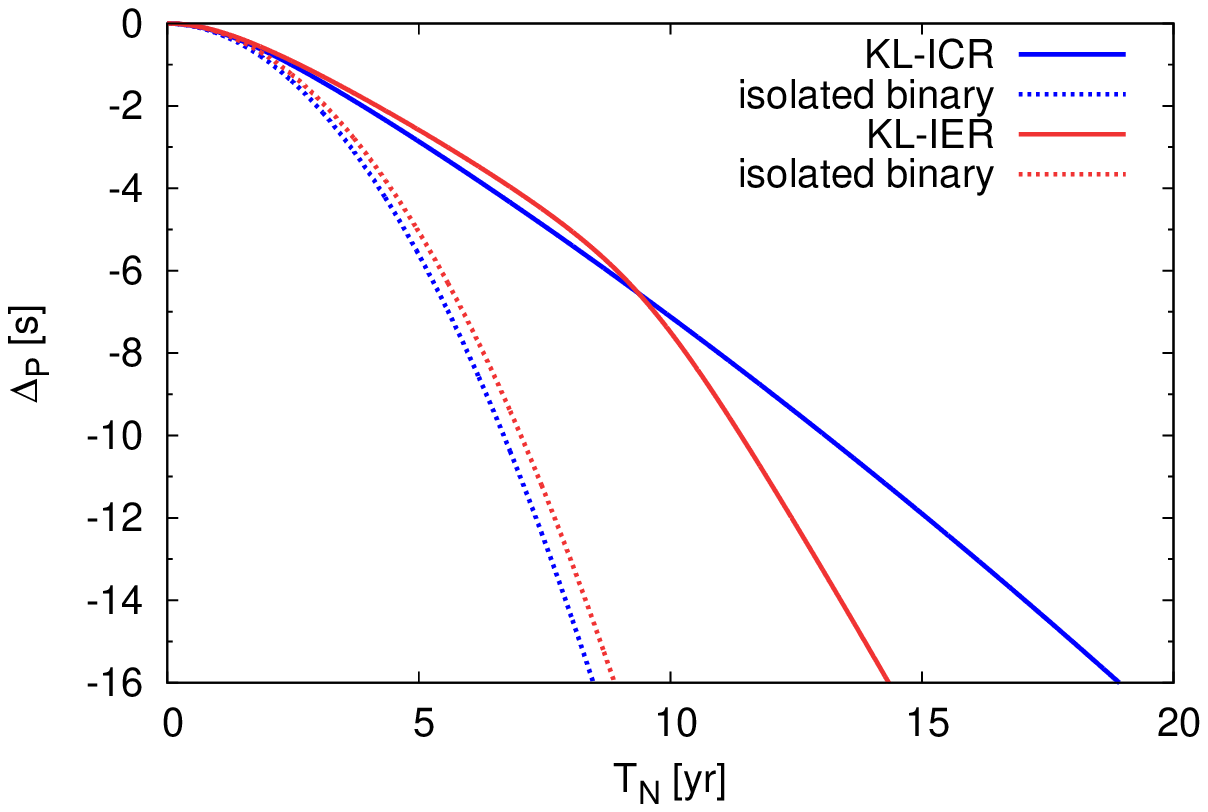}
        \end{minipage}\\
        \begin{minipage}{7cm}
            \includegraphics[width=7cm]{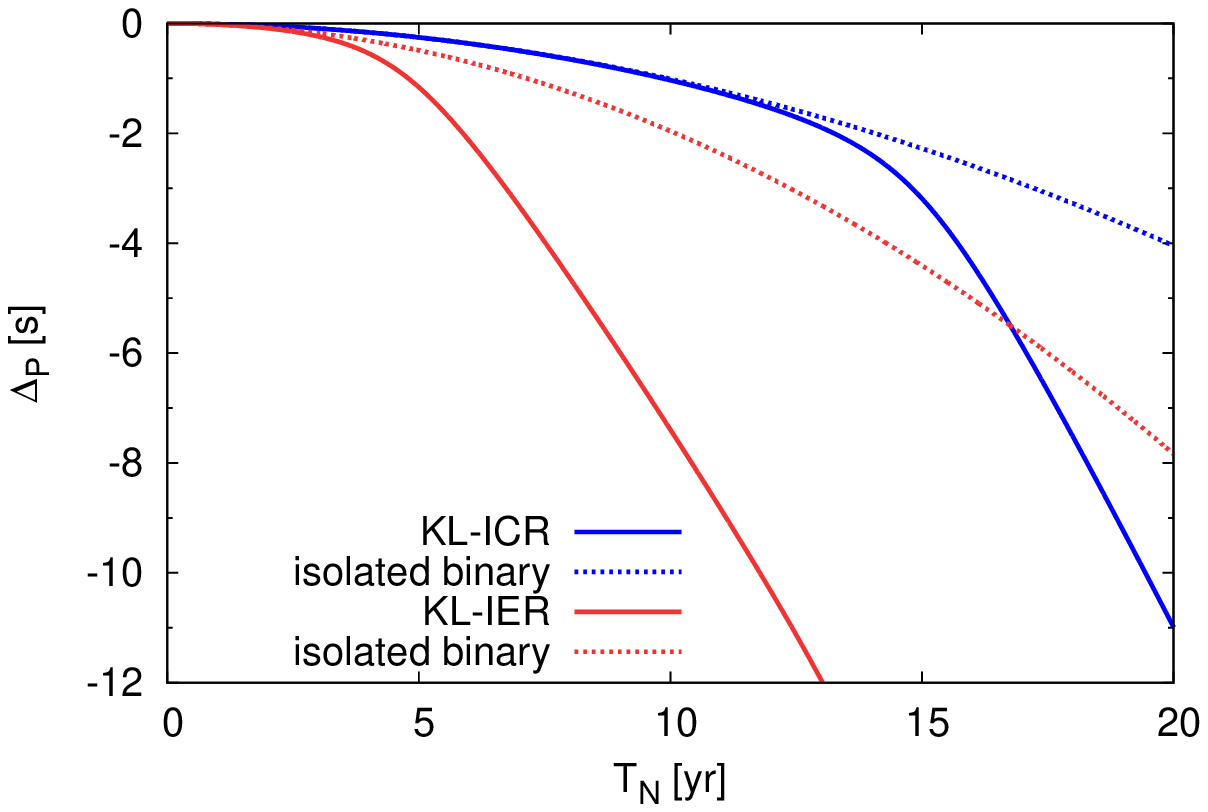}
        \end{minipage} 
        \caption{
                    The same figures as Fig.~\ref{fig:CSPT_l} but for rotation case is shown.
                    The blue and red solid curves are the results of ICR and IER types.
                }
        \label{fig:CSPT_r}
    \end{figure}

The KL-oscillations shown in \S\ref{subsec:orbit} affect the evolution of the CSPT $\Delta_P$ of binary pulsar in the hierarchical triple system.
As we showed in the previous letter \citet{Suzuki19}, 
if a hierarchical triple system shows
the KL oscillations in observation period, we expect the bending of CSPT curve.
It is because when the eccentricity becomes large, the amount of 
GW emission increases, and then the change of orbital period gets large.
Here we shall discuss how the bending of CSPT curve depends on the models or types of KL oscillations.

For each model in Table~\ref{tab:model}, we have 
calculated the time-evolution of CSPT as explained in \S\ref{sec:method}.
Since the behaviour of the CSPT curve of these models is 
{\HS similar, we show the results for PNIB model in figures as representative.}
Figs.~\ref{fig:CSPT_l} and \ref{fig:CSPT_r} show the results of libration and rotation types KL-oscillations, respectively.
In each panel, the red and blue solid curves show the results of initially circular and eccentric types, respectively.
The top panels show the CSPT curves calculated 
from the time when the maximum eccentricity is found
in each KL-oscillation type (at $t=15.21$yr, $t=0$yr, $t=14.96$ yr and $t=1.32$ yr for ICL, IEL, ICR and IER types, respectively), 
while the bottom panels exhibit those calculated from the time when minimum eccentricity is reached (at $t=0$yr, $t=5.40$yr, $t=0$ yr and $t=6.10$ yr  for ICL, IEL, ICR and IER types, respectively). 
It shows that the CSPT curves become completely different  
depending on the choice of the initial time of integration $T_N=0$ 
even for the same model.
For reference, we also show the CSPT curves of the isolated binary whose parameters are the same as the initial parameters of the inner binary in corresponding hierarchical triple models, by the red and blue dashed curves.

The CSPT curves of isolated binaries are approximated by the quadratic functions as  Eq.~\eqref{eq:approx_Delta_P}.
At first the CSPT curves of KL triple system 
coincide with the quadratic curves of corresponding isolated binaries, but when the eccentricity changes with KL-mechanism, 
the curves of the triple-system bend and the discrepancy from the binary curves
becomes large as already shown in \citet{Suzuki19}. 
This is because the period change of the inner binary due to GW emission ($\dot{P}_\mathrm{in}$) depends on the orbital eccentricity as given by Eq.~\eqref{eq:Pdot}.
Hence when the orbital eccentricity changes, 
 $\dot{P}_\mathrm{in}$ also changes, and then the CSPT curve 
 deviates largely from the quadratic curve.
 
In the top panels of Figs.~\ref{fig:CSPT_l} and \ref{fig:CSPT_r}, the solid curves at first coincide with the quadratic curves with eccentric orbits, but they switch to the less steeper curves as the eccentricities become smaller by KL-mechanism.
This feature results in the slower decrease of $\Delta_P$ in the triple system compared with that of the isolated eccentric binary.
The slope and bending timescale of red and blue solid curves are different from each other depending on the amplitude and KL-timescale.
While, in the bottom panels of Figs.~\ref{fig:CSPT_l} and \ref{fig:CSPT_r},  the switch from the circular curves to the eccentric steeper curves causes rapid decrease of $\Delta_P$ in the triple system curves than those of isolated circular binaries.

\begin{figure}
        \centering
        \begin{minipage}{7cm}
            \includegraphics[width=7cm]{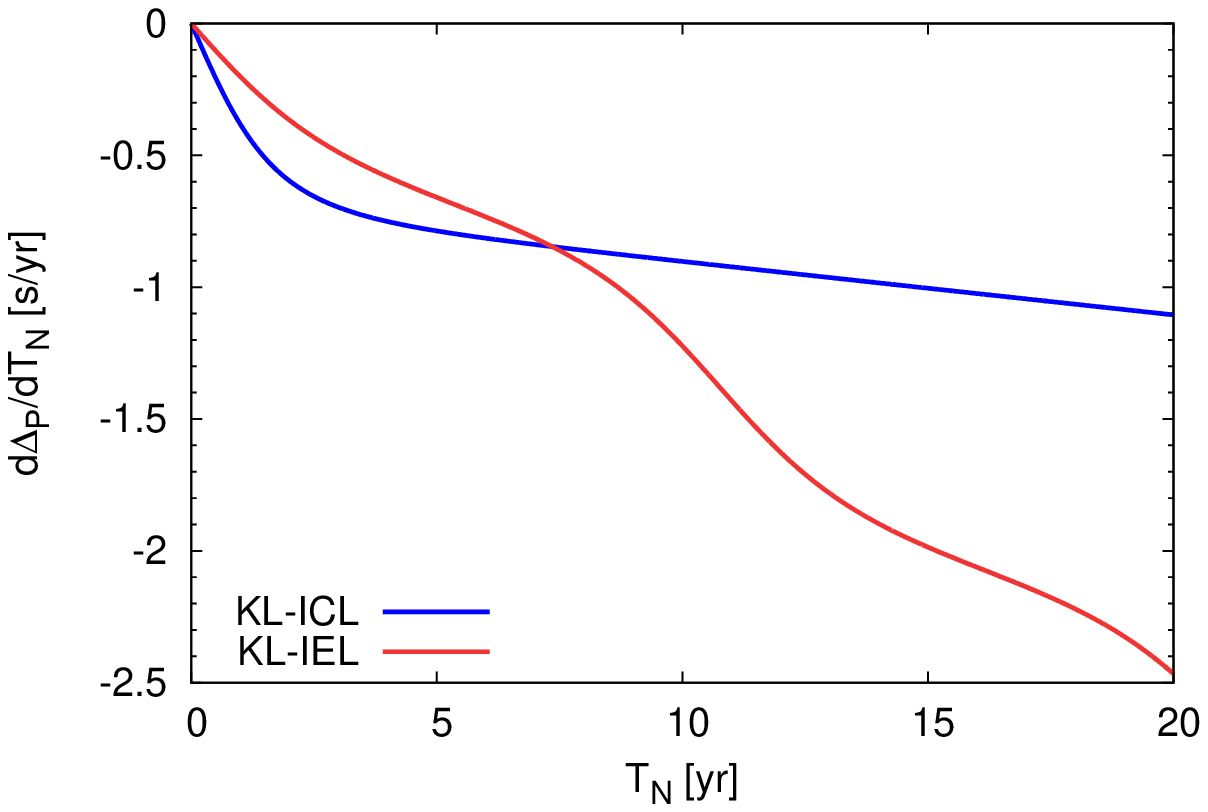}
        \end{minipage}\\
        \begin{minipage}{7cm}
            \includegraphics[width=7cm]{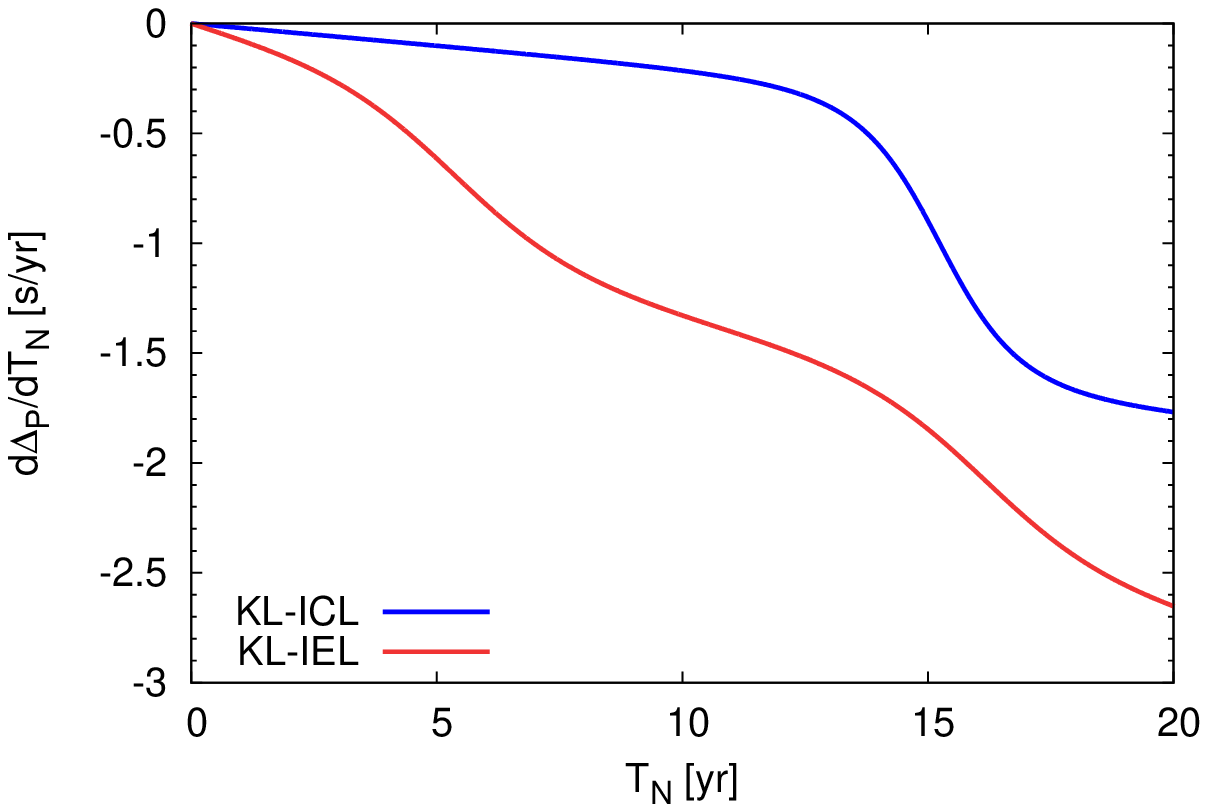}
        \end{minipage} 
        \caption{
                    Time derivative of CSPT $d\Delta_P/dt$ calculated for libration type of KL-oscillations in PNIB model.
                    Top and bottom panels are the results calculated from the time of maximum and minimum eccentricities, respectively. 
                    The blue and red solid curves are the results of ICL and IEL types.
                }
        \label{fig:CSPT_dot_l}
    \end{figure}

\begin{figure}
        \centering
        \begin{minipage}{7cm}
            \includegraphics[width=7cm]{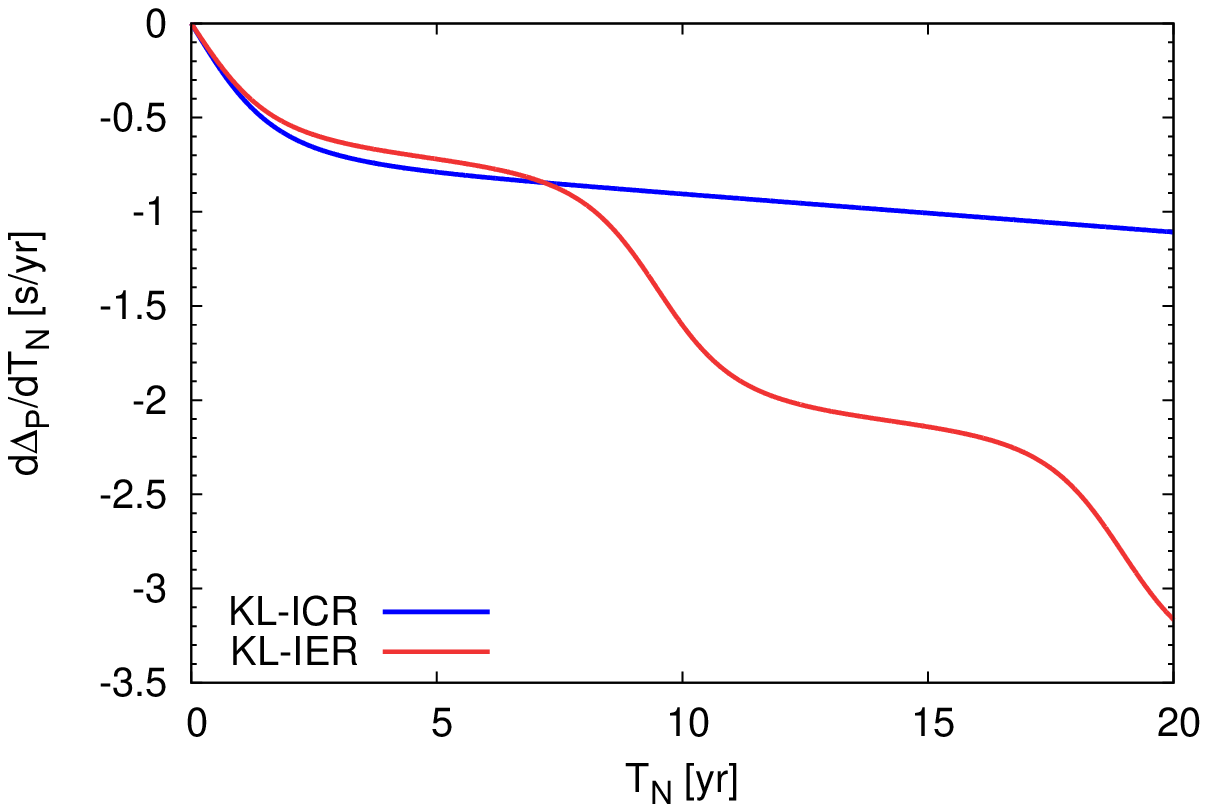}
        \end{minipage}\\
        \begin{minipage}{7cm}
            \includegraphics[width=7cm]{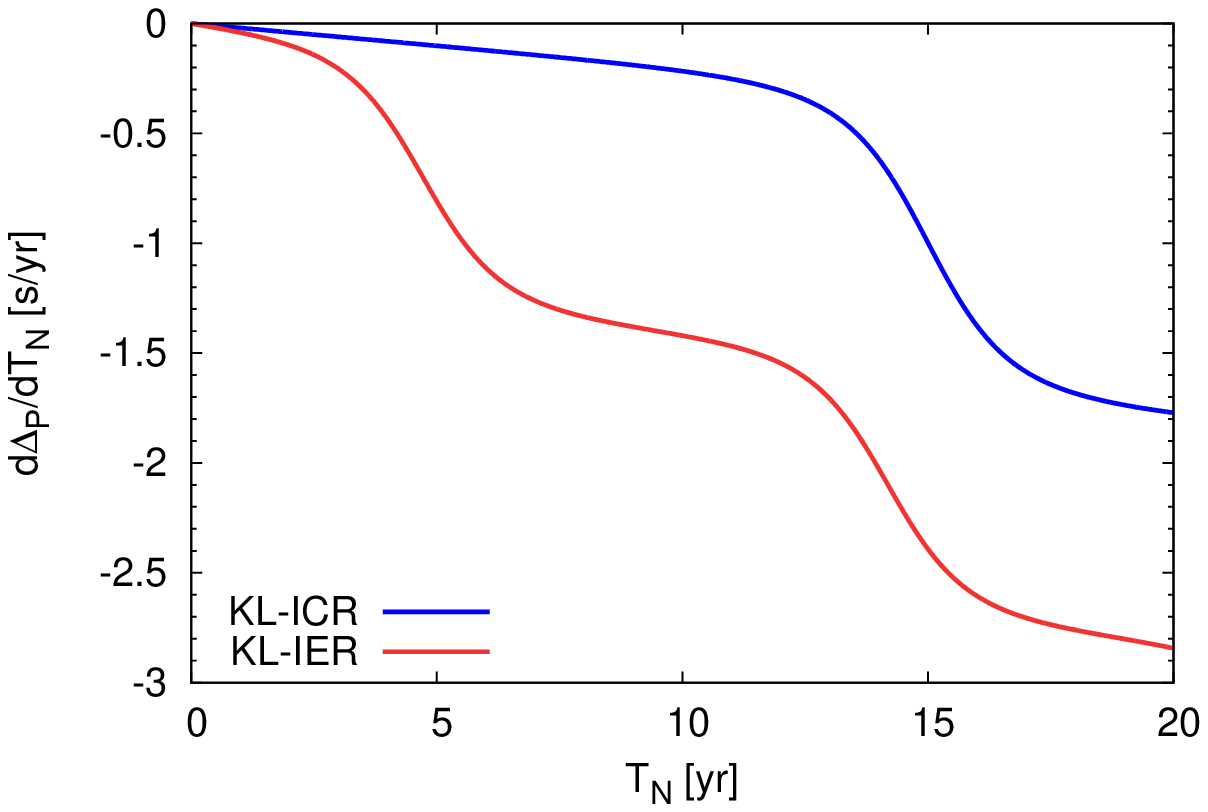}
        \end{minipage} 
        \caption{
                    The same figures as Fig.~\ref{fig:CSPT_l} but for rotation type of KL-oscillations in PNIB model are shown.
                }
        \label{fig:CSPT_dot_r}
    \end{figure}

This bending feature may be useful to see KL-oscillation from pulsar observation.
The shape of the CSPT curve has the information of the eccentricity 
and the KL-oscillation timescale in its slope change.
The bending  of the CSPT curve is clear when 
the curve is integrated from minimum eccentricity,
but the curve from the maximum eccentricity does not show 
clear bending.
However, the change of the CSPT curve becomes clearer 
if the time-derivative of $\Delta_P$ is plotted.  
In Figs.~\ref{fig:CSPT_dot_l} and \ref{fig:CSPT_dot_r}, the time-evolution of $d\Delta_P/dT_N$ for each KL-type is plotted.
Figs.~\ref{fig:CSPT_dot_l} and \ref{fig:CSPT_dot_r} show the results of libration and rotation types KL-oscillations, respectively.
In each panel, the red and blue curves show the results of 
initially circular and eccentric types.
The top and bottom panels in those figures show $d\Delta_P/dT_N$ curves calculated from the time when the maximum and minimum eccentricities are obtained, respectively. 
{\KM We find the clear slope change of the $d\Delta_P/dT_N$ curves even for the curve integrated from the maximum eccentricity.}

{\HS 
When the bending of the CSPT curve occurs, the slope of the $d\Delta_P/dT_N$ curve changes.}
{\KM Here we define the slope as
\begin{equation}
S(T_N)={d^2 {\Delta}_P\over dT_N^2}={\dot P_b(T_N)\over P_b(0)}
\end{equation}
from Eq. (\ref {eq:Delta_P}).
Hence we find that when eccentricity gets large, the slope becomes steep, i.e., 
the absolute value of the slope become large, and vise versa.}
{\HS 
Hence, the difference between the minimum and maximum values of slope ($S_\mathrm{min}$ and $S_\mathrm{max}$) 
indicates the amplitude of KL-oscillation.   
}

{\KM The magnitude of the slope depends not only on the eccentricity but also the model parameters.
If a system has smaller semi-major axis or larger masses for its inner binary, the GW emission rate is
larger and then the slope becomes steeper.
In Tables~\ref{tab:s_maxmin}, we summarise the minimum slope $S_\mathrm{min}$ and the maximum slope $S_\mathrm{max}$ as well as the Kozai-Lidov time scale $T_\mathrm{KL}$, which gives the time interval between minimum and maximum slopes, for 
all models we have calculated.}

{\KM For the Hulse-Taylor binary, we find the slope is $S_{\rm HT}\approx -8.57 \times 10^{-2}$.
Hence we expect that we can observe the change of the slope for most models except for 
the models with P-BH inner binary.
The models with P-IMBH inner binary
show largest slope.
However the models with P-BH inner binary give
smallest absolute values, for which we may need more precise observation to find the CSPT curve.
In each model, the difference between $S_\mathrm{min}$ and $S_\mathrm{max}$ in the IEL type KL-oscillation  
is smallest of all types.
This is because the amplitude of KL-oscillation in this type is smallest
 as seen in Figs.~\ref{fig:PNN_e_l} and \ref{fig:PNIB_e_l}.
}

\begin{table}
		\begin{tabular}{c||c||c||cc|}
			\hline
			 Model & Type & $T_\mathrm{KL}$ [yr] & $S_\mathrm{min}[\mathrm{s/yr}^2]$ & $S_\mathrm{max}[\mathrm{s/yr}^2]$  \\
			\hline
             PNN   & ICL  &        12.7          &    -0.475      &  $-2.02 \times 10^{-2}$ \\
                   & ICR  &        12.0          &    -0.476      &  $-2.02 \times 10^{-2}$ \\
                   & IEL  &        3.47          &    -0.215      &  -0.117 \\
                   & IER  &        4.08          &    -0.423      &  $-3.10 \times 10^{-2}$ \\
            \hline
             PNB   & ICL  &        9.18          &    -0.902      &  $-2.02 \times 10^{-2}$ \\
                   & ICR  &        8.84          &    -0.903      &  $-2.02 \times 10^{-2}$ \\
                   & IEL  &        2.48          &    -0.223      &  -0.167 \\
                   & IER  &        3.40          &    -0.480      &  $-2.79 \times 10^{-2}$ \\
            \hline
             PNIB  & ICL  &        36.9          &    -0.442      &  $-2.02 \times 10^{-2}$ \\
                   & ICR  &        32.6          &    -0.442      &  $-2.02 \times 10^{-2}$ \\
                   & IEL  &        10.8          &    -0.215      &  $-7.28 \times 10^{-2}$ \\
                   & IER  &        9.50          &    -0.406      &  $-3.91 \times 10^{-2}$ \\
            \hline
             PNSB  & ICL  &        1.98          &     -1.21      &  $-1.99 \times 10^{-2}$ \\
                   & ICR  &        1.97          &     -1.21      &  $-2.02 \times 10^{-2}$ \\
                   & IEL  &        0.588         &    -0.255      &  -0.179 \\
                   & IER  &        0.876         &    -0.527      &  $-2.34 \times 10^{-2}$ \\
            \hline
             PBB   & ICL  &       7.05     & $-2.70 \times 10^{-2}$ &  $-4.85 \times 10^{-4}$ \\
                   & ICR  &       7.99     & $-2.72 \times 10^{-2}$ &  $-4.88 \times 10^{-4}$  \\
                   & IEL  &       2.23     & $-6.33 \times 10^{-3}$ &  $-4.62 \times 10^{-3}$ \\
                   & IER  &       3.19     & $-1.38 \times 10^{-2}$ &  $-6.01 \times 10^{-4}$  \\
            \hline
             PBIB  & ICL  &       75.1     & $-2.13 \times 10^{-2}$ &  $-4.85 \times 10^{-4}$ \\
                   & ICR  &       69.9     & $-2.13 \times 10^{-2}$ &  $-4.86 \times 10^{-4}$  \\
                   & IEL  &       21.6     & $-5.40 \times 10^{-3}$ &  $-3.88 \times 10^{-3}$  \\
                   & IER  &       29.1     & $-1.20 \times 10^{-2}$ &  $-6.90 \times 10^{-4}$  \\
            \hline
             PBSB  & ICL  &       14.0     & $-2.82 \times 10^{-2}$ &  $-4.85 \times 10^{-4}$ \\
                   & ICR  &       14.5     & $-2.86 \times 10^{-2}$ &  $-4.87 \times 10^{-4}$  \\
                   & IEL  &       4.09     & $-6.28 \times 10^{-3}$ &  $-4.37 \times 10^{-3}$  \\
                   & IER  &       5.96     & $-1.41 \times 10^{-2}$ &  $-6.08 \times 10^{-4}$  \\
            \hline
             PIBIB & ICL  &        2.50          &    -6.04      &  -0.516 \\
                   & ICR  &        2.26          &    -6.01      &  -0.518  \\
                   & IEL  &        0.899         &    -6.34      &  -0.771  \\
                   & IER  &        0.491         &    -10.5      &  -1.20  \\
            \hline
             PIBSB & ICL  &        1.20          &     -11.5      &  -0.516 \\
                   & ICR  &        1.20          &     -11.5      &  -0.519 \\
                   & IEL  &        0.400         &     -7.69      &  -1.48 \\
                   & IER  &        0.339         &     -13.7      &  -0.915 \\
            \hline
		\end{tabular}
		\caption{
		          $T_\mathrm{KL}$, $S_\mathrm{min}$, $S_\mathrm{max}$ for all models are summarised.
            	}
		\label{tab:s_maxmin}
	\end{table}

It has already been pointed out that the KL-oscillation should be observed through the long-period radio observation of the orbital elements of the binary pulsar \citep{Gopakumar09, Portegies11}. 
In real observation, however, the observational data is sometimes missed due to some reasons; for example, in the observation of the Hulse-Taylor binary, the data was not obtained for a decade of 1990s because of the major upgrades of Arecibo telescope \citep{Hulse94}.
If this unseen period is completely overlapped with the time when eccentricity is changed from the initial value with KL-oscillation, it is difficult to recognise whether KL-oscillation occurs or not only from orbital element data.
Even in such case, we can conclude that KL-oscillation occurs in the system if 
the CSPT curve deviates from that of isolated binary in late phase.


Some readers may worry about the spin evolution of the pulsar caused by the spin-orbit coupling in 1.5 order post-Newtonian terms \citep{Barker75} because it may change the direction of the pulsar rotation axis and affect the radio observation, that is, the change of beaming direction of pulse signal may cause the disappearance of the pulsar. 
Following \citet{Liu17, Liu18}, the evolution of spin in relativistic KL-oscillation can be characterised with the "adiabaticity parameter" $\mathcal{A}$ defined as the ratio of the de-Sitter spin precession rate $\Omega_\mathrm{SL}$ to the orbital precession rate by KL-oscillation $\Omega_\mathrm{L}$.
The adiabaticity parameter $\mathcal{A}$ is described as
    \begin{equation}
        \mathcal{A} \equiv \left| \frac{\Omega_\mathrm{SL}}{\Omega_\mathrm{L}} \right|
                    \simeq  4\frac{r_\mathrm{g,in}}{a_\mathrm{in}}\frac{m_1+\mu_\mathrm{in}/3}{m_3}
                                \left( \frac{a_\mathrm{out}}{a_\mathrm{in}} \right)^3 (1-e_\mathrm{out}^2)^{\frac{3}{2}} ,
        \label{eq:adiabaticity}
    \end{equation}
where $\mu_\mathrm{in} = m_1m_2/(m_1+m_2)$ is the reduced mass of the inner binary.
This parameter is quite similar to $\epsilon^\mathrm{(1PN)}$ defined as Eq.~\eqref{eq:epsilon}.
Hence for  the system with KL-oscillation, which satisfies the condition Eq.~\eqref{eq:GRKL}, we find that
the adiabaticity parameter $\mathcal{A}$
satisfies 
    \begin{equation}
   \mathcal{A} \lsim \frac{m_1(3m_1+4m_2)}{(m_1+m_2)^2}(1-e_{\rm in}^2)^{\frac{3}{2}}\leq 3
   \,.
    \end{equation}
The adiabaticity parameters of our models are summarised in Table~\ref{tab:adiabaticity}.

In case of $\mathcal{A} \ll 1$, the spin evolution is classified as "non-adiabatic", that is, the orbital precession by KL-oscillation is much faster than the relativistic spin precession, and then the spin axis cannot 'catch up' with the precession of angular-momentum axis.
In such a situation, the spin axis of the pulsar is expected to be parallelly transported just as in the Newtonian case, and then the beaming direction of the radio signal is
expected not to change so much even when the inclination changes by KL-oscillation.
The PBB model corresponds to this case.
For the other models, $\mathcal{A}$ is still smaller than unity,
but not so much.
The spin axis of the pulsar in the system with such mid-range of $\mathcal{A}$ is perturbed around its initial direction as shown in \citet{Liu18}.
If the perturbation of the spin direction is large enough so that the beaming angle of the pulsar goes out from the observable range, the radio signal from the pulsar will disappear and will rarely re-appear due to its complicated evolution.
If the disappearance of a pulsar in triple system is observed, it will be an important example of the 1.5 post-Newtonian effect on the KL-oscillation.
The critical value of $\mathcal{A}$ that causes the disappearance of the signal should depend on the emission mechanism of the pulsar, the intensity of the radio signal, the distance to the system, and the opening angle of the radio telescope.
If the CSPT is observed for a whole period of KL-oscillation despite the precession of the spin direction of the pulsar, it means the pulsar is successively observed from some different directions and such observation may give new information about the pulsar.

	\begin{table}
	    \centering
		\begin{tabular}{cc}
			\hline
			 name & $\mathcal{A}$ \\
			\hline
			 PNN  &     0.103     \\
			 PNB  &     0.075      \\
			 PNIB &     0.282      \\
             PNSB &     0.0181      \\
			\hline
			 PBB  &     0.0125    \\
			 PBIB &     0.129      \\   
             PBSB &     0.0242      \\   
			\hline
			 PIBIB&     0.683      \\
			 PIBSB&     0.396      \\
			\hline
		\end{tabular}
		\caption{
            		The adiabaticity parameter $\mathcal{A}$ of each model is summarised.
		        }
		\label{tab:adiabaticity}
	\end{table}
{\HS 
The hierarchical three-body system which causes the bending of CSPT curve needs high inclination so that KL-oscillation occurs.
Such highly inclined triple systems may need to be formed by the dynamical interaction in dense environments like the globular clusters and the galactic nuclei \citep{Kulkarni93, Samsing_2014, Zevin19}. 
Hence we need population synthesis with large numerical simulation to estimate event rates of the observation of bending of CSPT.
We also need to  consider the distance to the system which should be close enough to observe the radio signals from it.
Though the population synthesis simulation considering all factors is beyond the scope of this paper, we can expect that the observation of the bending of the CSPT curve may be a rare event. 
}
However, as discussed in \citet{Suzuki19}, this interesting signal is important not only to confirm the existence of the third body but to provide 
a first indirect evidence of GW emission from the triple system with KL-oscillations.
GW emission makes the inner binary more compact and GW waveform from such compact triple system with KL-oscillation can be observed by future GW detectors \citep{Gupta19} like
LISA \citep{Amaro-Seoane17}, DECIGO \citep{Sato17}, and Big Bang Observer \citep{Harry06}.

{\HS 
For some binary pulsars, for example, PSR J1840-0643 \citep{Knispel13}, the possible existence of the tertiary companion has not been denied.
Observing such binary pulsars for a long {\KM period may lead to } 
discovery of a first indirect evidence of GW emission from the triple system with KL-oscillations.
}

\section{Conclusions}
\label{sec:conclusion}

In this paper, 
taking the 1st post-Newtonian relativistic correction into account,
we have studied the KL-oscillations in hierarchical triple systems with a pulsar and calculated  the cumulative shift of periastron time (CSPT).
The KL-mechanism is one of the orbital resonances that appear in the hierarchical triple systems characterised as the exchanging oscillation with the inner eccentricity and the relative inclination.
When the eccentricity of the binary pulsar is excited by KL-oscillation, it enhances GW emission from the binary and it changes the shape of the CSPT curve.
We have analysed the KL-oscillations in several models with a pulsar, 
and those effects on the CSPT curves.

We have first analysed the KL-oscillations for the models with different initial parameters.
We have classified those models into four types (ICL, IEL, ICR, and IER).
We have calculated their orbital evolution  by the direct integration of 1st post-Newtonian equations of motion.
The four KL-types have different amplitudes and timescales and, in addition, the non-test particle limit effect and the relativistic effect appear differently.
In the result of the model with weak mass hierarchy (e.g. PNN model), we find that KL-``conserved" value $\theta^2$ is not conserved but oscillating whereas it should be constant in double-averaged method with test-particle limit approximation. 
It has also been found that the amplitudes and timescales obtained in direct integration do not coincide with those in double-averaged method.
The tendency of these discrepancies is different in the four types of KL-oscillations.
The amplitudes and frequencies of the emitted gravitational waves are quite sensitive to the eccentricity, and these differences between eccentricity evolution in direct integration and that obtained from double-averaged method may be crucial when we evaluate the GW emission for the systems with finite masses, that is, one may overestimate or underestimate the maximum or minimum value of the eccentricity when we use the double-averaged method.

In the model with large $\epsilon^{(1\mathrm{PN})}$  (e.g. PNIB model), we could observe clear differences between the results obtained by Newtonian and post-Newtonian direct integrations. 
The post-Newtonian effects appear differently in the four types of KL-oscillations.
The complicated behaviours can be understood theoretically by using the double-averaging method with 1st-order post-Newtonian corrections. 
However, in some models (e.g. PNB and PBB models), we have 
observed KL-oscillation with irregular periods, which cannot be explained by double-averaging method with quadrupole-order approximation.
This may be because the KL-conserved quantities are not exactly constant in the direct integration.

The KL-oscillation effect appears in the CSPT curve as the bending of the curve.
The slope of the curve at each phase reflects the maximum or minimum eccentricity and the time between two bending points corresponds to the timescale of KL-oscillation.
The CSPT curves become completely different depending on the choice of the initial time of integration even for the same model. 
The bending of the CSPT curve is clear when the curve is integrated from minimum eccentricity, but the curve from the maximum eccentricity does not show clear bending.
In such case, the time derivative of the CSPT can be a good indicator for the bending of the CSPT curve.

The system that causes this interesting signal may be rare because such compact hierarchical triple systems with high inclination need to be formed by dynamical interaction in a dense environment like a globular cluster or the galactic center.
However,  once such systems are observed with the pulsar signal, it is very important because it is the first indirect observation of GW from triple systems.
In addition, it will be the precursor of the direct detection of the waveform by the future gravitational detectors like LISA, DECIGO and Big Bang Observer.
Some highly relativistic triple systems should show the spin precession of the pulsar caused by the 1.5 post-Newtonian effect from the outer orbit and it will change the beaming angle of the pulsar.
If the beaming angle of the pulsar is perturbed and goes out of the observable range, the radio signal from the pulsar will disappear and rarely appear again.
The disappearance of the signal from a pulsar in triple system will provide 
one of the important examples of the 1.5 post-Newtonian effect on the KL-oscillation. 
On the other hand, if the CSPT is observed for a whole period of KL-oscillation despite the precession of the spin direction of the pulsar, it corresponds to the successive observation of a pulsar from different directions and such observation may give new information about a pulsar.

\section*{Data availability}

The data underlying this article will be shared on reasonable request to the corresponding author.

\section*{Acknowledgements}
P.G. is supported by Japanese Government (MEXT)
Scholarship.
This work was supported in part by JSPS KAKENHI Grant Numbers JP20J12436, 
JP17H06359 and JP19K03857, and by Waseda University Grant for Special
Research Projects (Project number: 2019C-254 and 2019C-640).

\newpage



\bibliographystyle{mnras}
\bibliography{Full_paper_draft_ver1} 


\newpage

\appendix

\section{Detail Analysis of Kozai-Lidov Mechanism by Double Averaging Method}
\label{sec:KL_app}
\subsection{Newtonian Dynamics}
\label{subsec:maxmin_app}
Here we discuss the restricted hierarchical triple system.
We choose our reference plane to define the inclinations as the initial orbital plane of the outer orbit.
Since the outer inclination is conserved in the restricted triple system, we find that
$i_\mathrm{out}=0$ and then the inner inclination $i_\mathrm{in}$ is
the same as the relative inclination  $I$ between inner and outer orbits
\footnote[1]{
Note that in the non-restricted triple system case, the outer inclination will also evolve with time. 
In such case, the relative inclination is calculated as 
{\HS
$\cos I = \cos i_{\mathrm{in}} \cos i_{\mathrm{out}} 
 				   + \sin i_{\mathrm{in}} \sin i_{\mathrm{out}} 
 						\cos \left( \Omega_{\mathrm{in}} - \Omega_{\mathrm{out}} \right) .$
}
}.
 The secular time evolution of
the osculating orbital elements of the inner orbit  is described by the 
Lagrange planetary equations, which is decoupled from the orbital motion of the outer orbit
in the restricted hierarchical triple system as
    \begin{eqnarray}
        &&\frac{da}{dt}=0, 
        \label{eq:lagrange1}\\
        &&\frac{de}{dt}=
            -\frac{\sqrt{1-e^2}}{na^2 e}
            \frac{\partial V_S}{\partial \omega}, 
        \label{eq:lagrange2}\\
        &&\frac{di}{dt}=
            \frac{\cot i}{na^2 \sqrt{1-e^2}}
            \frac{\partial V_S}{ \partial \omega}, 
        \label{eq:lagrange3}\\
        &&\frac{d\omega}{dt}=
            \frac{\sqrt{1-e^2}}{ na^2 e}
            \frac{\partial V_S}{ \partial e}-
            \frac{\cot i}{ na^2 \sqrt{1-e^2}}
            \frac{\partial V_S}{\partial i}, 
        \label{eq:lagrange4}\\
        &&\frac{d\Omega}{dt}=
        \frac{1}{ na^2 \sqrt{1-e^2}\sin i}
        \frac{\partial V_S}{ \partial i}, 
        \label{eq:lagrange5}
    \end{eqnarray}
where $n$ is the mean motion of the inner orbit, which is defined by 
   \begin{equation}
        n = \sqrt{\frac{Gm}{a^3}}\,,
    \end{equation}
and $V_S$ is the double-averaged perturbation potential in the Hamiltonian 
of the motion of a test-particle in the triple system.
``Double-averaged" means that the corresponding term is 
averaged for both  periods of inner and outer orbits.
In this section, we drop  the subscript ``in" for the inner orbit variables just for brevity.

$V_S$ is obtained by expanding the perturbative interaction potential term in the Hamiltonian with $a/a_\mathrm{out}$ up to the quadrupole moment and performing its double-averaging procedure.
It is described by the orbital elements as
    \begin{equation}
        V_S = V_0 v_S(e, i, \omega),
    \end{equation}
where
    \begin{align}
        &V_0=\frac{G m_3 a^2}{16a_\mathrm{out}^3(1-e_\mathrm{out})^{3/2}}, \\
        &v_S=(2+3e^2)(3\cos^2i-1)+15e^2\cos{2\omega}\sin^2i.
    \end{align}
Introducing the following three variables:
    \begin{align}
        &\eta \equiv \sqrt{1-e^2}, \\
        &\mu \equiv \cos{i}, \\
        &\tau \equiv \frac{V_0}{na^2}t\,,
    \end{align}
where $\tau$ is the dimension-free time parameter measured by
 the typical oscillation timescale $na^2/V_0$,
we find that 
 the basic equations \eqref{eq:lagrange2}-\eqref{eq:lagrange4} are rewritten   as
    \begin{eqnarray}
        \frac{d\eta}{d\tau} &=& \frac{\partial v_S}{\partial \omega}, 
        \label{eq:eta}\\
        \frac{1}{\mu} \frac{d\mu}{d\tau} &=&-\frac{1}{\eta} \frac{\partial v_S}{\partial \omega}, 
         \label{eq:mu}\\
        \frac{d\omega}{d\tau} &=&
         -\frac{\partial v_S}{\partial \eta}+ \frac{\mu}{\eta} \frac{\partial v_S}{\partial \mu}.
         \label{eq:omega}
    \end{eqnarray}
From these basic equations, the following two conserved quantities are obtained:
    \begin{eqnarray}
        \theta &\equiv& \eta \mu , 
        \label{eq:theta_app}\\
        C_\mathrm{KL} &\equiv&
         \frac{v_S}{12} = (1-\eta^2)\left[ 1-\frac{5}{2}(1-\mu^2)\sin^2\omega \right].~~~~
        \label{eq:C_KL_app}
    \end{eqnarray}
These are the same as the previously introduced 
 two conserved quantities~\eqref{eq:theta} and \eqref{eq:C_KL}.
Due to the existence of two conserved values for three equations, 
we get the following single equation for $\eta$:  
    \begin{equation}
        \frac{d\eta}{d\tau} = -\frac{12\sqrt{2}}{\eta}\sqrt{f(\eta)g(\eta)},
        \label{eq:etaN}
    \end{equation}
where 
    \begin{align}
        &f(\eta) \equiv 1-\eta^2-C_\mathrm{KL}, \\
        &g(\eta) \equiv -5\theta^2 + (5\theta^2 + 3 +2C_\mathrm{KL})\eta^2 -3\eta^4 .
    \end{align}
    
Because of KL-oscillations, the eccentricity $e$ takes the maximum 
or minimum value when 
${d\eta}/{d\tau}$ vanishes.

Since the zero of $f(\eta)$ exists only for $C_\mathrm{KL} \geq 0$, 
we classify the KL-oscillations into two types:\\
(i) \, rotation type : $C_\mathrm{KL} \geq 0$
\\ 
(ii) libration type : $C_\mathrm{KL} \leq 0$.
\\
The zero of $f(\eta)$ is given by 
   \begin{equation}
\eta=\eta_0 \equiv \sqrt{1-C_\mathrm{KL} }
\,,
    \end{equation}
while the zeros of $g(\eta)$ are obtained as
   \begin{equation}
 \eta=  \eta_\pm\equiv
        \sqrt{ 
                \frac{5\theta^2+2C_\mathrm{KL}+3\pm \sqrt{(5\theta^2+2C_\mathrm{KL}-3)^2
                +24C_\mathrm{KL}} }{6} 
              }\,.
   \end{equation}
With the conditions $f(\eta)g(\eta)\geq 0$ and $0\leq \eta\leq 1$,
we find 
\begin{eqnarray*}
&&
\eta_-\leq \eta \leq \eta_0 ~~{\rm for}~~{\rm rotation~type} (C_{\rm KL}>0)
\\
&&
\eta_-\leq \eta \leq \eta_+ ~~{\rm for}~~{\rm libration~type} (C_{\rm KL}<0)\,.
\end{eqnarray*}

This gives
\begin{eqnarray}
e_{\rm min} \leq e \leq e_{\rm max}
\,,
\end{eqnarray}
where  for  rotation type  ($C_{\rm KL}>0$), we obtain
   \begin{align}
&e_{\rm min} = \sqrt{C_\mathrm{KL}}\,,
\\
&e_{\rm max}=\sqrt{\frac{4C_\mathrm{KL}}{5\theta^2 +2C_\mathrm{KL}-3 +\sqrt{(5\theta^2 +2C_\mathrm{KL}-3)^2 +24C_\mathrm{KL}}}}\,,
   \end{align}
while for libration type  ($C_{\rm KL}<0$), we have
   \begin{align}
 &e_\mathrm{min} = \sqrt{\frac{4C_\mathrm{KL}}{5\theta^2 +2C_\mathrm{KL}-3 -\sqrt{(5\theta^2 +2C_\mathrm{KL}-3)^2 +24C_\mathrm{KL}}} }  , \\
&e_\mathrm{max} = \sqrt{\frac{4C_\mathrm{KL}}{5\theta^2 +2C_\mathrm{KL}-3 +\sqrt{(5\theta^2 +2C_\mathrm{KL}-3)^2 +24C_\mathrm{KL}}}}.
    \end{align}

From the condition of $e_\mathrm{min}\leq e_\mathrm{max}$, we 
have the  constraints for $\theta$ and $C_\mathrm{KL}$: 
   \begin{align*}
 &  \theta^2 \leq -C_\mathrm{KL} +1 &~~&{\rm (rotation~type)}, \\
&  \theta^2 \leq \frac{1}{5}(-2C_\mathrm{KL}+3-2\sqrt{-6C_\mathrm{KL}})
&~~&{\rm (libration~type)} .
    \end{align*}

\begin{figure}
        \centering
        \includegraphics[width=7cm]{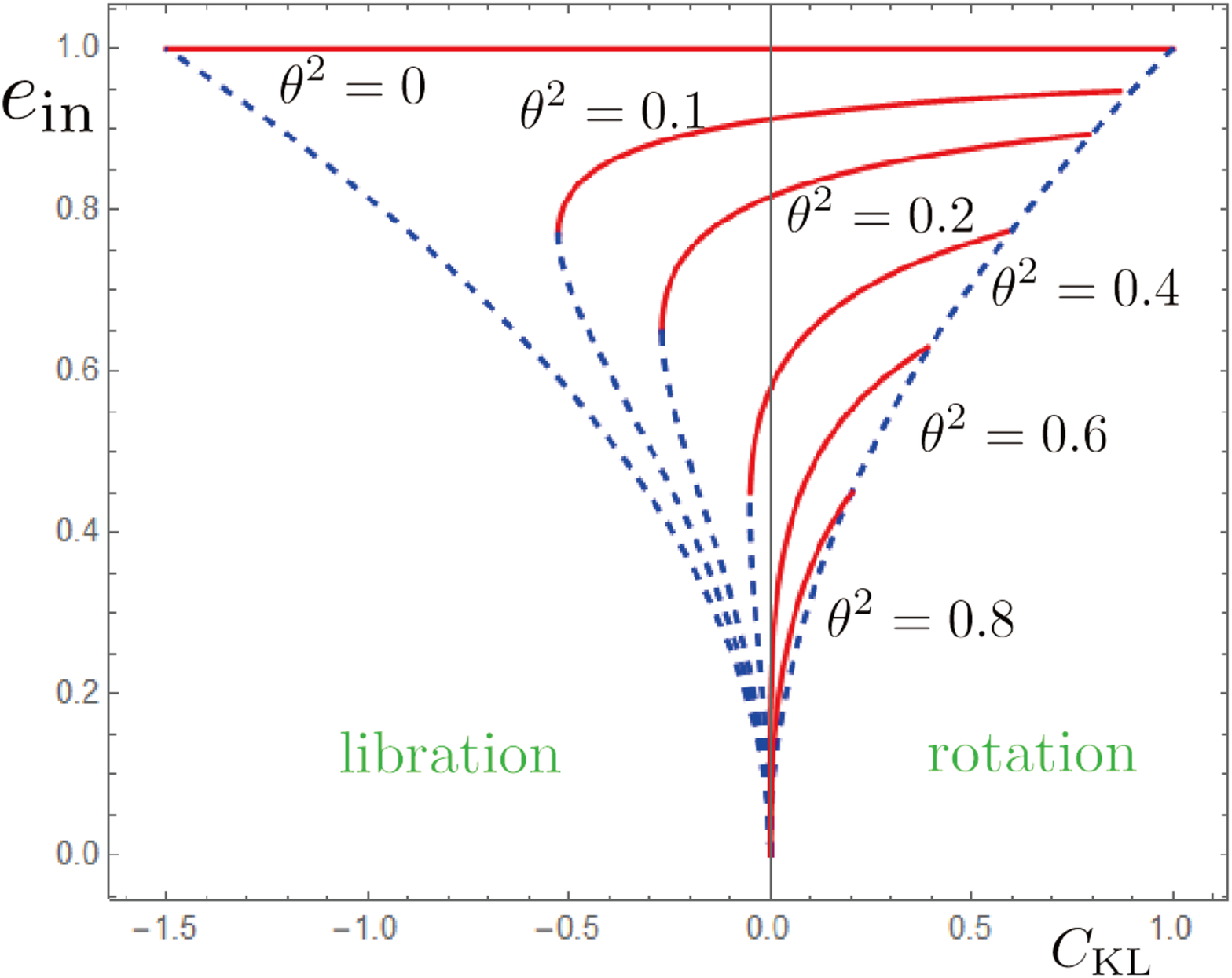}
        \caption{The maximum and minimum values of eccentricity
        in terms of $C_{\rm KL}$. 
        The red solid and blue dotted curves denote the maximum and 
        minimum values of the eccentricity, respectively.
        We choose $\theta^2=0. 01, 0.2, 0.4, 0.6, $ and $0.8$.
        The libration type exists only for $\theta^2<0.6$.
         }
        \label{fig:emin_emax_N}
    \end{figure}

In Fig. \ref{fig:emin_emax_N}, we show 
some examples of $e_{\rm min}$ and $e_{\rm max}$ for four types of KL-oscillations.
We find that the eccentricity oscillates between zero and the maximum value for the initially 
circular types, while it changes between two finite values (finite minimum and finite maximum values). For the libration types, there is no KL-oscillation beyond some critical value of $\theta^2$,
while for rotation types, $\theta^2$ reaches almost unity although the oscillation amplitude becomes smaller for larger $\theta^2$.

\begin{figure}
        \centering
        \includegraphics[width=8cm]{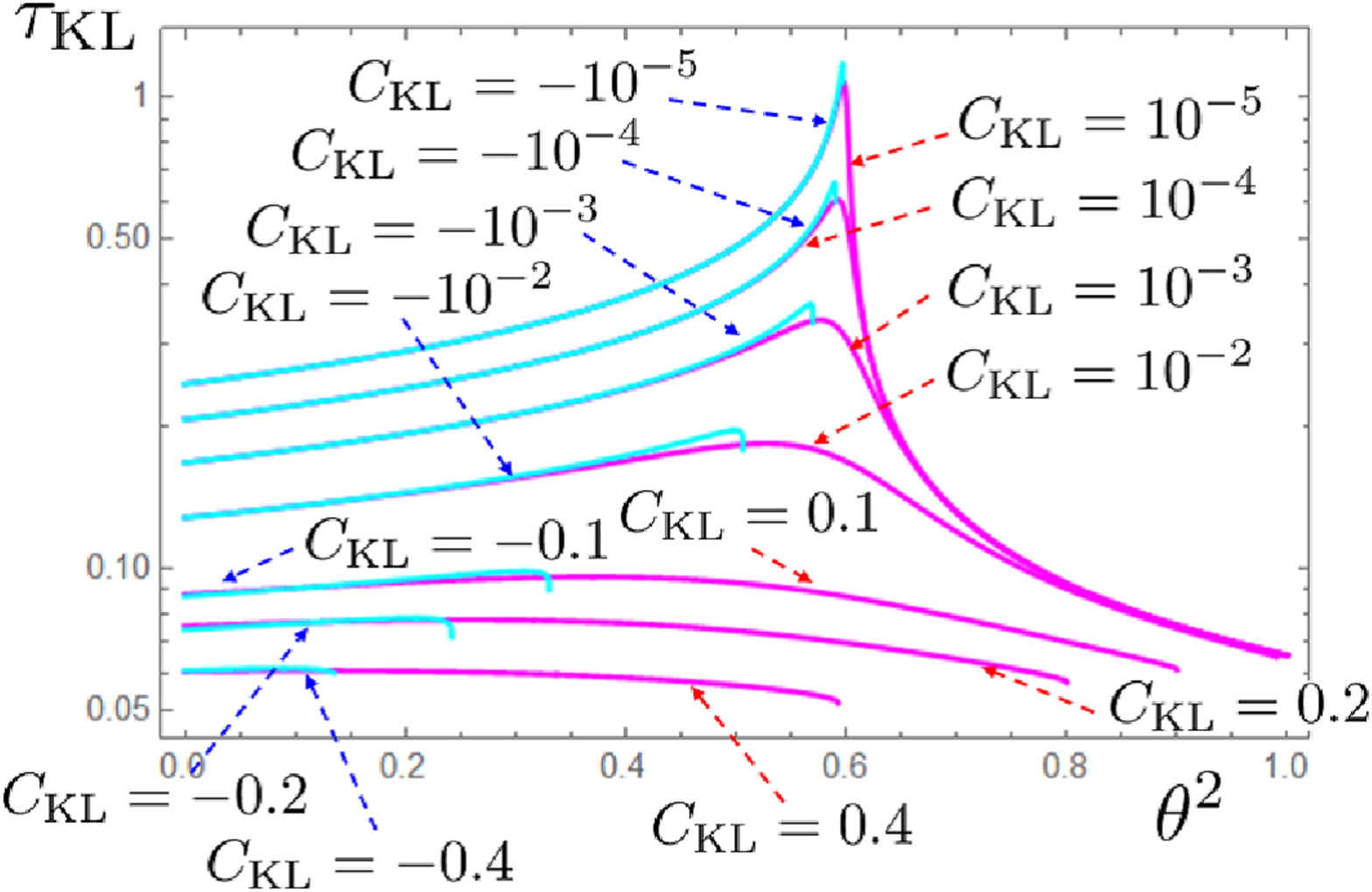}
        \caption{
                    Normalized KL-oscillation period $\tau_{\rm KL}$ 
                    in terms of $\theta^2$. 
                    The cyan and magenta curves denote $\tau_{\rm KL}$ for 
                    the libration and rotation types, respectively.
                }
        \label{fig:tauKL_N}
    \end{figure}

The exact half-period of the 
KL-oscillation $T_\mathrm{KL}$ is defined by the time such that
the eccentricity changes from the minimum value to the maximum value \citep{Antognini15}.
It is evaluated as 
    \begin{equation}
         T_\mathrm{KL} = \frac{na^2}{V_0} \tau_\mathrm{KL},
    \end{equation}
where 
    \begin{equation}
        \tau_\mathrm{KL} = \int_{\eta_\mathrm{min}}^{\eta_\mathrm{max}} \left( \frac{d\eta}{d\tau} \right)^{-1} d\eta .
        \label{eq:tau_KL}
    \end{equation}
Since $\tau_\mathrm{KL}$  has order of unity, the dimensionful factor $na^2/V_0$ 
is used for rough estimation of the KL-timescale, which corresponds to Eq.~\eqref{eq:roughT_KL}.
We find
    \begin{align*}
    \tau_{\rm KL}=\left\{
    \begin{array}{l}
 {1\over 12\sqrt{6(\eta_0^2-\eta_-^2)}}K\left(\sqrt{\eta_+^2-\eta_-^2\over \eta_0^2-\eta_-^2}\right)~~~{\rm for ~libration}
    \\
{1\over 12\sqrt{6(\eta_+^2-\eta_-^2)}}K\left(\sqrt{\eta_0^2-\eta_-^2\over \eta_+^2-\eta_-^2}\right)
~~~{\rm for ~rotation}
      \,,
\end{array}
\right.
  \end{align*}
  where
  $K(k)$ is the complete elliptic integral of the first kind with the modulus $k$.
In Fig. 
  \ref{fig:tauKL_N}, we show $\tau_{\rm KL}$.
  
\subsection{Post-Newtonian Correction}
\label{subsec:GR_app}
  
In the restricted triple system, the first order post-Newtonian (1PN) GR correction 
can be included by adding the correction term to the interaction potential, that is, 
    \begin{equation}
        V_S \rightarrow V_S^{\rm (GR)}=V_S+V^\mathrm{(1PN)},
        \label{eq:PNpotential}
    \end{equation}
where 
    \begin{equation}
        V^\mathrm{(1PN)} = \frac{3G^2m^2}{c^2a^2\sqrt{1-e^2}}.
    \end{equation}
This correction term is derived by double-averaging 
the 1PN Hamiltonian of two-body relative motion
(See e.g. \citet{Migaszewski11}. The original Hamiltonian is obtained in \cite{Richardson88}). 
When the corrected potential $V_S^{\rm (GR)}$ is used instead of $V_S$, 
dimensionless potential $v_S$ is also replaced by 
    \begin{equation}
    v_S^{(\mathrm{GR})} = v_S+12\frac{\epsilon^{(\mathrm{1PN})}}{\eta},
    \end{equation}
where $\epsilon^{(\mathrm{1PN})}$ is the dimensionless constant that describes the 1PN 
GR correction defined by Eq.~\eqref{eq:epsilon}.

The basic equations for the orbital elements
 are the same as Eqs.\eqref{eq:eta}, \eqref{eq:mu} and \eqref{eq:omega} 
 by replacing the potential $v_S$ with $v_S^{(\mathrm{GR})}$.
 Hence we find two conserved quantities again:
    \begin{align}
        &\theta = \eta \mu, \\
        &C_\mathrm{KL}^{(\mathrm{GR})} = C_\mathrm{KL}(\eta,\mu,\omega)+\epsilon^{(\mathrm{1PN})}\left(\frac{1-\eta}{\eta}\right).
    \end{align}
    Note that $C_\mathrm{KL}(\eta,\mu,\omega)$ is 
    not conserved in this case because of 1PN corrections.
    $C_\mathrm{KL}^{(\mathrm{GR})} $ coincides with the Newtonian value 
    $C_\mathrm{KL} $
    if the orbit is circular ($\eta=1$).
    
From three basic equations with two conserved quantities, we obtain 
one single equation for $\eta$ as
    \begin{equation}
        \frac{d\eta}{d\tau} = -{12 \sqrt{2}\over \eta} \sqrt{
                                       f^{(\mathrm{GR})}(\eta) 
                                         g^{(\mathrm{GR})} (\eta)
                                        },
        \label{eq:etaGR}
    \end{equation}
where 
    \begin{align*}
     f^{(\mathrm{GR})}=&1-\eta^2-C_{\rm KL}^{\rm (GR)}+\epsilon^{(\rm 1PN)}
     \left(\frac{1-\eta}{\eta}\right)
     \\
    g^{(\mathrm{GR})}=&-5\theta^2 + \left(5\theta^2 + 3 +2C_\mathrm{KL}^{\rm (GR)}\right)\eta^2 -3\eta^4-2\epsilon^{\rm (1PN)}\eta\left(1-\eta\right) 
    \end{align*}
    
 

In order to find the maximum and minimum values of eccentricity, 
we look for the zeros of $f^{(\mathrm{GR})}(\eta) $ and $g^{(\mathrm{GR})}(\eta) $
under the conditions of $f^{(\mathrm{GR})}g^{(\mathrm{GR})}\geq 0$ with $0\leq \eta\leq 1$.
Hence we solve the cubic equation $\eta f^{(\mathrm{GR})}(\eta) =0$ and 
the quartic equation $g^{(\mathrm{GR})}(\eta) =0$.
There is one root for $\eta f^{(\mathrm{GR})}(\eta) =0$ only if 
 $C_\mathrm{KL}^{(\mathrm{GR})} \geq 0$.
 As a result, just as the Newtonian case, the KL-oscillation is classified into two types: \\
(i) rotation type with $C_\mathrm{KL}^{(\mathrm{GR})} \geq 0$ 
\\
 (ii) libration type with $C_\mathrm{KL}^{(\mathrm{GR})} \leq 0$.

\begin{figure}
        \centering
        \includegraphics[width=8cm]{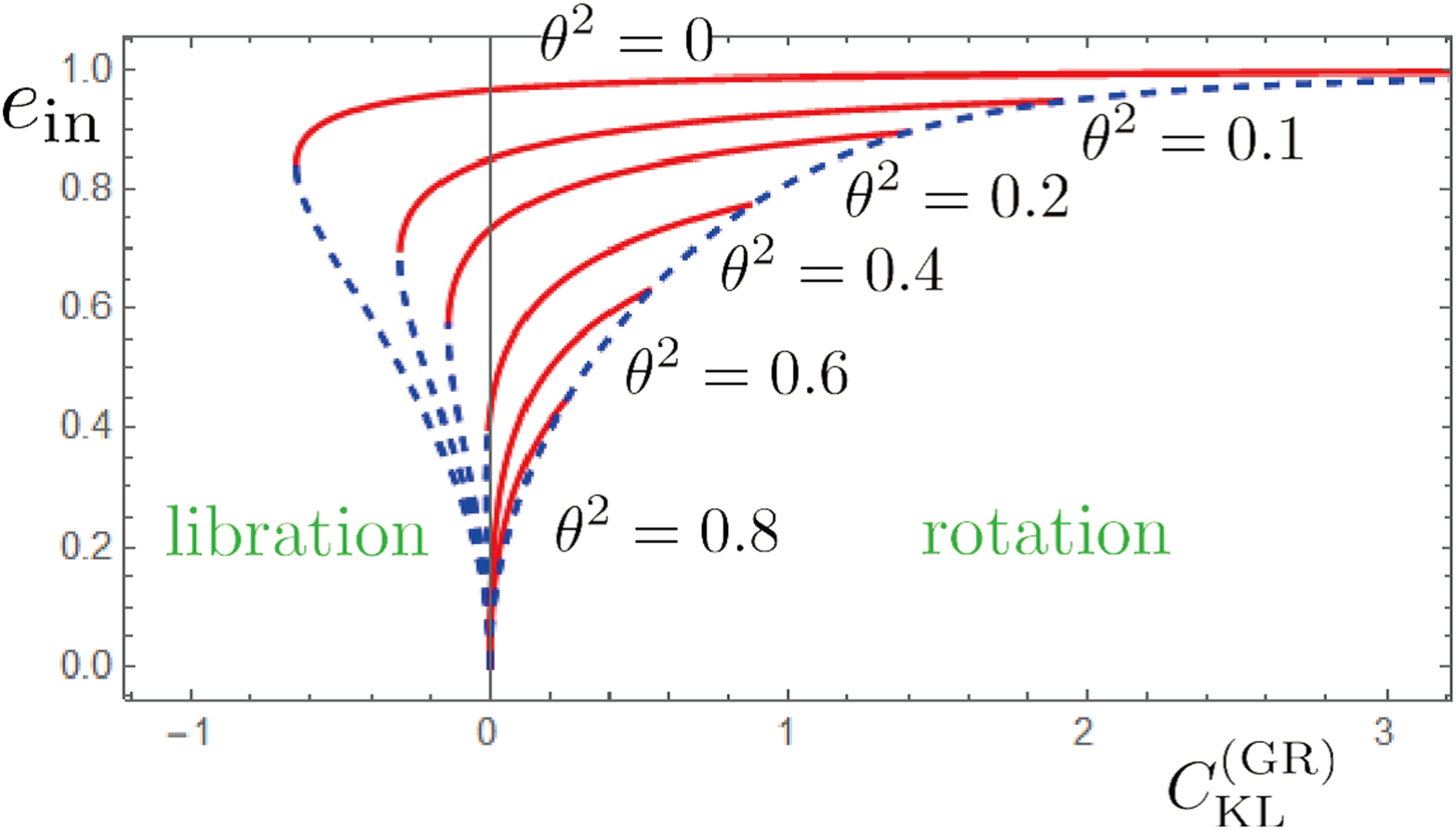}
        \caption{The same figure as Fig. \ref{fig:emin_emax_N} 
      for the PNIB model with the post-Newtonian corrections.
      The libration type exists only for $\theta^2<(3-\epsilon^{\rm (1PN)})/5$.
      We choose $\epsilon^{\rm (1PN)}=0.484$.
            }
        \label{fig:emin_emax_PN}
    \end{figure}

In Fig. ~\ref{fig:emin_emax_PN}, we show  
the maximum and minimum values of the eccentricity in terms of $C_{\rm KL}^{\rm (GR)}$.
Since $C_{\rm KL}^{\rm (GR)}$ is conserved, fixing its value we obtain 
the maximum and minimum values of the eccentricity.
Here we choose $\epsilon^{\rm (1PN)}=0.484$, 
which is the value for  the PNIB model in Table  \ref{tab:ini_GR}. 
The behaviours  are similar to those in Newtonian case
 (Fig. \ref{fig:emin_emax_N}), 
 but the parameter region of $\theta^2$ and $C_{\rm KL}^{\rm (GR)}$ for the KL oscillation is modified.
 
\begin{figure}[h]
        \centering
        \includegraphics[width=7cm]{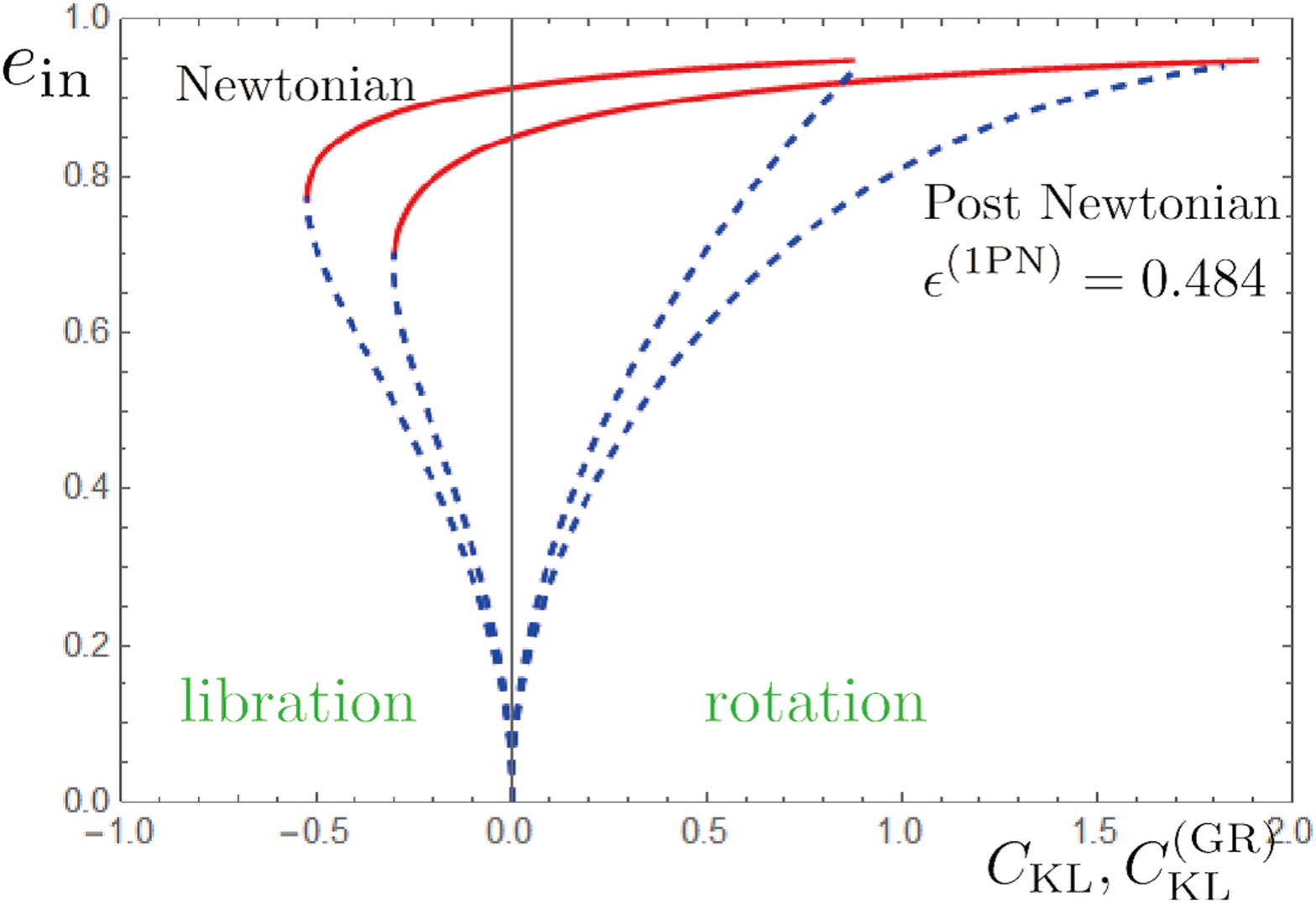}
        \caption{Comparison of the post-Newtonian result with the Newtonian one
      for the PNIB model with $\epsilon^{\rm (1PN)}=0.484$.
      We choose $\theta^2=0.1$.
      For the libration type, for given $C_{\rm KL}$, 
      the oscillation amplitude between $e_{\rm max}$ and $e_{\rm min}$
      in the P-NS case becomes smaller than the Newtonian one, 
      which shows the suppression of the KL-oscillation by the relativistic effect.
      On the other hand, for rotation type, the maximum value decreases, but the 
      oscillation amplitude can increase depending on the parameters.      
            }
        \label{fig:emin_emax_NvsPN}
    \end{figure}

In order to see the relativistic effect, we compare this result with the Newtonian case. 
As an example, in Fig. \ref{fig:emin_emax_NvsPN}, we plot both results for 
$\theta^2=0.1$.
In the libration type, the relativistic effect suppresses the KL oscillation mechanism. The parameter region of $C_{\rm KL}^{\rm (GR)}$ 
where the KL oscillation exists is reduced and the oscillation amplitude of the eccentricity becomes smaller for given value of $C_{\rm KL}^{\rm (GR)}$.
On the other hand, for the rotation type, 
the parameter region of $C_{\rm KL}^{\rm (GR)}$ increases, and 
the oscillation amplitude of the eccentricity is not always reduced
for given value of $C_{\rm KL}^{\rm (GR)}$.
    
\begin{table*}
\begin{tabular}{c|c|c|c|c|c|c|c|c|c}
\hline
Model&Type & $\epsilon^{\rm (GR)}$ & $C_{\rm KL}^{\rm (GR)}$  &   $\theta^2$      &
$e_{\rm min}$&  $e_{\rm max}$  & $\Delta e$&  $\tau_{\rm KL}$  &$T_{\rm KL}$[yrs]  \\
				\hline
PNIB& ICL & $0$  & $-1.64\times10^{-5}$ & $0.25$& $0.0043293$ &   $0.76376$ &$0.75943$ &  $0.27817$  &$ 37.027$  \\
&Libration &  $0.484$  &  $-1.64\times10^{-5}$ & $0.25$ &  $0.0050901$ &   $0.66754$  & $0.66245$  &$0.28302$ &$37.673$  \\
\cline{2-10}
& ICR& $0$  & $7.73\times10^{-5}$ & $0.25$& $0.0087920$ &   $0.76379$ & $0.75500$ &$0.24363$ &$32.430$   \\
&Rotation & $0.484$  &  $7.73\times10^{-5}$ & $0.25$ &  $0.0078891$ &   $0.66760$  & $0.65971$ & $0.24657$  &$32.820 $ \\
\cline{2-10}
&IEL  & $0$  & $-0.0931$ & $0.232$& $0.331241$ &   $0.752117$ &$0.420876$ &  $0.0865099$  &$ 11.515$  \\
&Libration &  $0.484$  &  $-0.0931$ & $0.232$ &  $0.427721$ &   $0.599244$  & $0.171523$  &$0.0837164$ &$11.144$  \\
\cline{2-10}
&IER & $0$  & $0.143$ & $0.32$& $0.378153$ &   $0.738886$ &$0.360732$ & $0.0764598$  &$10.178$  \\
&Rotation&  $0.484$  &  $0.143$ & $0.32$ &  $0.336263$ &   $0.668577$  &  $0.332314$ &$0.0684457$  &$9.111$ \\
\hline
	\end{tabular}
		\caption{The maximum and minimum eccentricities for the PNIB models.
		$\Delta e=e_{\rm max}-e_{\rm min}$ gives the oscillation amplitude.
		$\tau_{\rm KL}$ and  $T_{\rm KL}$ are the reduced KL oscillation timescale and the real period,
		respectively, which are calculated based on the double-averaging method. 
		The first row gives the Newtonian result, while the second row shows the 
		result with post-Newtonian correction.}
			\label{tab:KLtimescale}
\end{table*}		

The timescale of KL-oscillation is evaluated in the same way as the Newtonian case.
With post-Newtonian correction, Eq.~\eqref{eq:etaGR} is substituted into Eq.\eqref{eq:tau_KL} instead of Eq.\eqref{eq:etaN}.
We give the result in Table. \ref{tab:KLtimescale}. 
We also show the Newtonian case as reference.
We find that both timescales are almost the same although the relativistic correction 
changes their values slightly. 
We can conclude that the relativistic effect changes 
the parameter region of $\theta^2$ and $C_{\rm KL}^{\rm (GR)}$ for the KL oscillation, 
and the normalized KL-oscillation period
$\tau_{\rm KL}$ depends mostly on those two conserved quantities.

\section{ Conversion between Orbital Elements and Cartesian Coordinates}
\label{sec:convergence}

\subsection{Initial Condition}
\label{subsec:initial}
We employ six orbital elements to set up initial configurations: 
semi-major axis $a$, eccentricity $e$, inclination $i$, argument of periastron $\omega$, longitude of ascending node $\Omega$, and mean anomaly ${\cal M}$.
These orbital elements should be transformed to the Cartesian coordinates of the constituent bodies to provide the initial conditions for our equations of motion. 
We first calculate the eccentric anomaly $u$ from the mean anomaly ${\cal M}$, solving the following equation with the Newton-Raphson method: 
    \begin{equation}
        {\cal M} = u-e \sin u .
    \end{equation}
We then transform it to the true anomaly $f$ according to the following relation,
    \begin{equation}
        f = \arctan \left\{ \frac{(\sin u) \sqrt{1 - e ^{2}}}{\cos u - e} \right\} .
    \end{equation}
Then the polar coordinates of a body on the orbit are given in terms of the true anomaly
 and other orbital elements as
    \begin{eqnarray}
         r &=& \frac{a(1-e^{2})}{1-e \cos f} ,\\
         \psi &=& \Omega +\arctan\{ \tan (\omega + f) \cos i) \} ,\\
         \theta &=& \arccos\{ \sin(\omega + f) \sin i \} .  
    \end{eqnarray}
These coordinates describe the positions of an orbiting object measured from its companion; 
for the inner binary of the hierarchical triple system, the origin is put at the position of $m_1$ and the orbiting object is $m_2$;
for outer orbit, we set its origin at the position of $m_3$ and orbiting object is the centre of mass of the inner binary.
The velocity of an orbiting body in these coordinates is obtained as
    \begin{eqnarray}
        \dot{r}&=&g_{r} \dot{f} ,\\
        \dot{\theta}&=&g_{\theta} \dot{f} , \\
        \dot{\psi}&=&g_{\psi} \dot{f} ,
    \end{eqnarray}
 where $g_{r}$, $g_{\theta}$, $g_{\psi}$, and $\dot{f}$ are given by 
    \begin{eqnarray}
        g_{r} &=& \frac{a(1- e^{2}) e \sin f}{(1+ e \cos f) ^{2}} ,\\
        g_{\theta}&=& - \frac{1}{\sin\theta} \cos{(\omega + f)} \sin i ,\\
        g_{\psi}&=& \cos^{2}(\psi-\Omega ) \frac{\cos i}{\cos^{2}(\omega + f )} ,
    \end{eqnarray}
    \begin{equation}
        \dot{f} = \sqrt{G m' \left(\frac{2}{r}-\frac{1}{a} \right)
                 \frac{1}{g_{r}^{2} +(r g_{\theta})^{2} +(r \sin \theta g_{\psi})^{2}}} \,,
    \end{equation}
    where $m'$ is total mass of the binary.
We then change the polar coordinates to the Cartesian coordinates and shift the origins 
so that the centre of mass of the entire system coincides with the origin of the coordinates.
The numerical integration of the EIH equations is performed on these Cartesian coordinates.

\subsection{Post-process}
\label{subsec:post}

The computational results described with Cartesian coordinates are transformed back 
to the orbital elements in order to interpret our results.
The semi-major axis $a$ is obtained from the following relation,
    \begin{equation}
        a=-\frac{G m'}{2 E} .          
        \label{semi-major}
    \end{equation}
In this expression, 
$m'$ is defined as $m'=m_{1}+m_{2}$ and $m'=m_{1}+m_{2}+m_{3}$ for the inner 
and outer orbits, respectively.
$E$ is the orbital energy per unit mass given as
    \begin{equation}
        E=\frac{1}{2}v^{2}-\frac{Gm'}{r} ,    
    \end{equation}
in which $v$ is orbital velocity
and $r$ is the separation between the orbiting object and the companion.
The inclination $i$, eccentricity $e$, and longitude of the ascending node $\Omega$ are derived from the following equations:
    \begin{equation}
        i = \arccos \left( \frac{(\bm{r} \times \bm{v})_{z}}{|\bm{r} \times \bm{v}|} \right) ,
    \end{equation}
    \begin{equation}
        e = \sqrt{1-\frac{|\bm{r} \times \bm{v}|^{2}}{a G m'}}   _,
    \end{equation}
    \begin{equation}
        \Omega = \arccos \left( \frac{(\bm{n} \times ( \bm{r} \times\bm{v} ))_{x}}
                                     {|\bm{n} \times ( \bm{r} \times \bm{v} )|} \right),
    \end{equation}
where the subscripts stand for the components of vectors; $\bm{n}$ is 
the unit vector normal to the $xy$ plane.
The argument of periastron $ \omega $ is obtained as follows: 
first, the true anomaly $f$ is given as
    \begin{equation}
        f = \arccos \left( \frac{a(1-e^{2})-r}{e r} \right) ;
    \end{equation}
secondly, the angle $\theta$ of the planet from the ascending node is given as
    \begin{equation}
        \theta = \arccos \left( \frac{x\cos{\Omega} + y\sin{\Omega}}{|\bm{r}|} \right) ,
    \end{equation} 
the argument of periastron is finally obtained as the difference of these arguments,
    \begin{equation}
        \omega = \theta - f .
    \end{equation} 

\section{Direct integration v.s. Double-averaging method with octupole-order expansion}
\label{sec:octupole}

{\KM In order to see the effect of the octupole-order terms, we 
integrate the double-averaging equations with octupole expansion \citep{Ford20, Naoz13a, Naoz13b}
and compare the results with those given by the quadupole-order equations as well as those 
obtained by the direct integration.
Since the octupole-order term is proportional to the difference of the inner binary masses, 
we shall analyze  the models with different-mass inner binary.
Here we show the result for the PBB model. }

\begin{figure}
        \centering
        \begin{minipage}{6.5cm}
            \includegraphics[width=6.5cm]{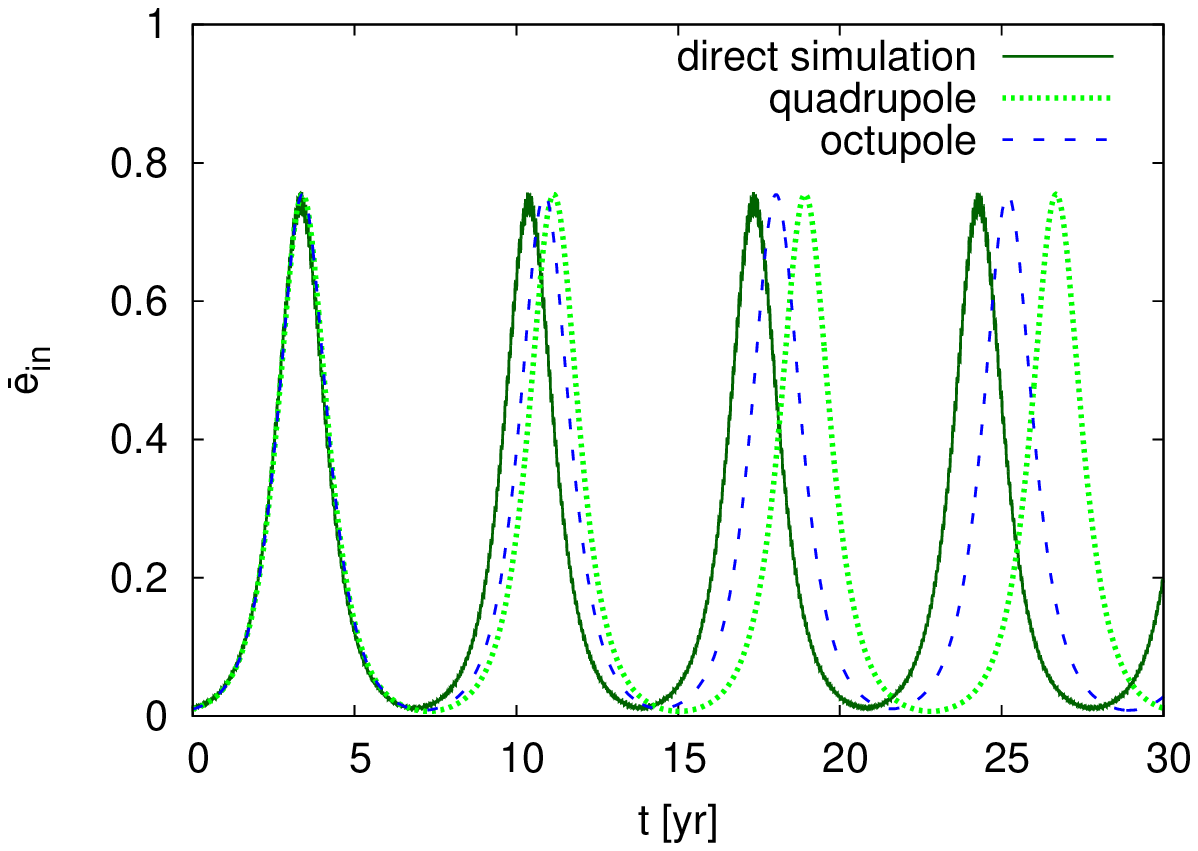}
        \end{minipage}\\
        \begin{minipage}{6.5cm}
            \includegraphics[width=6.5cm]{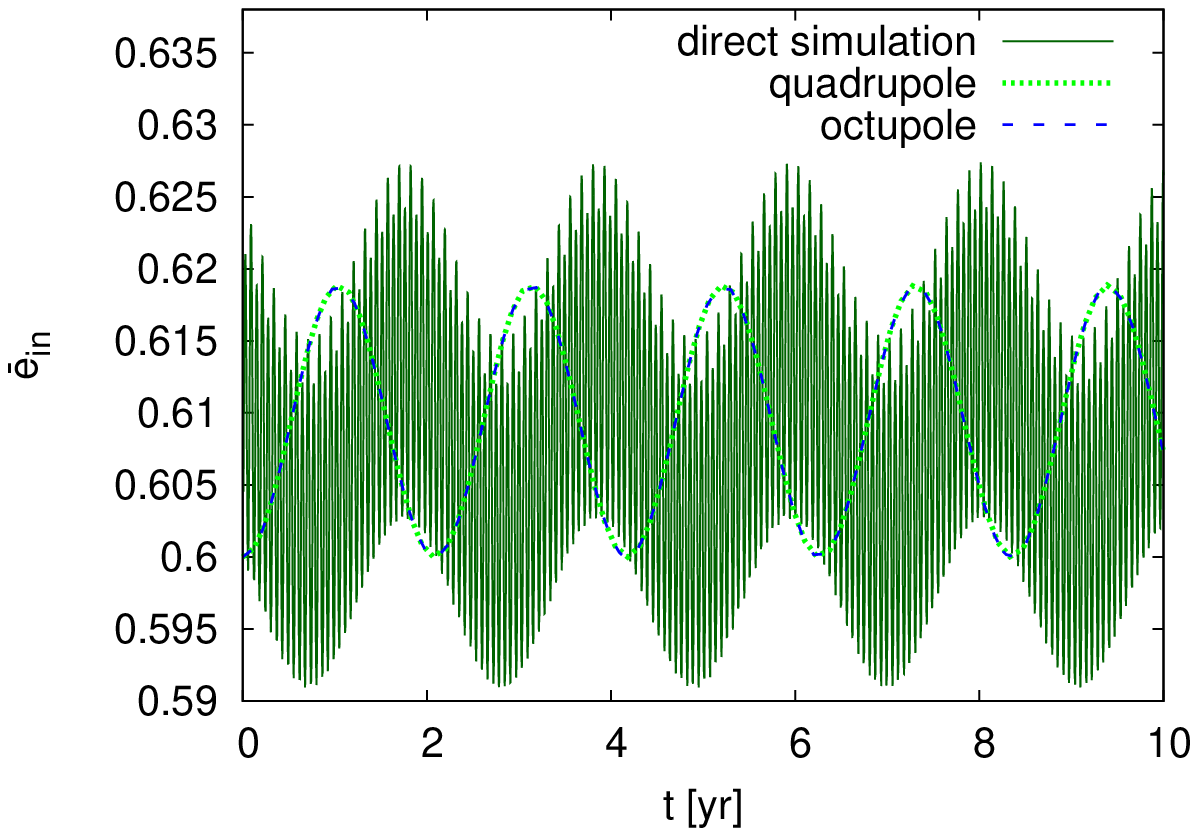}
        \end{minipage} 
        \caption{
                    Comparison between three evolution lines of the averaged inner eccentricity $\bar{e}_\mathrm{in}$ 
                    for the libration type of KL oscillations in the PBB model.
                    Top and bottom panels show the results of ICL and IEL types, respectively.
                    The dark-green solid line describes the evolution obtained from direct simulation while the two dashed lines denote the result obtained by double-averaged calculation: the light-green line is the result 
                    of quadrupole-order equations and dark-blue line is that of octupole-order ones.
                    {\KM In the bottom panel, the oscillation curve of the direct integration becomes 
                     broad. It is because the KL oscillation amplitude becomes small and 
                    it is almost the same as the amplitude of modulation caused by the outer orbit.}
                 }
        \label{fig:PBB_e_l}
    \end{figure}

\begin{figure}
        \centering
               \begin{minipage}{6.5cm}
            \includegraphics[width=6.5cm]{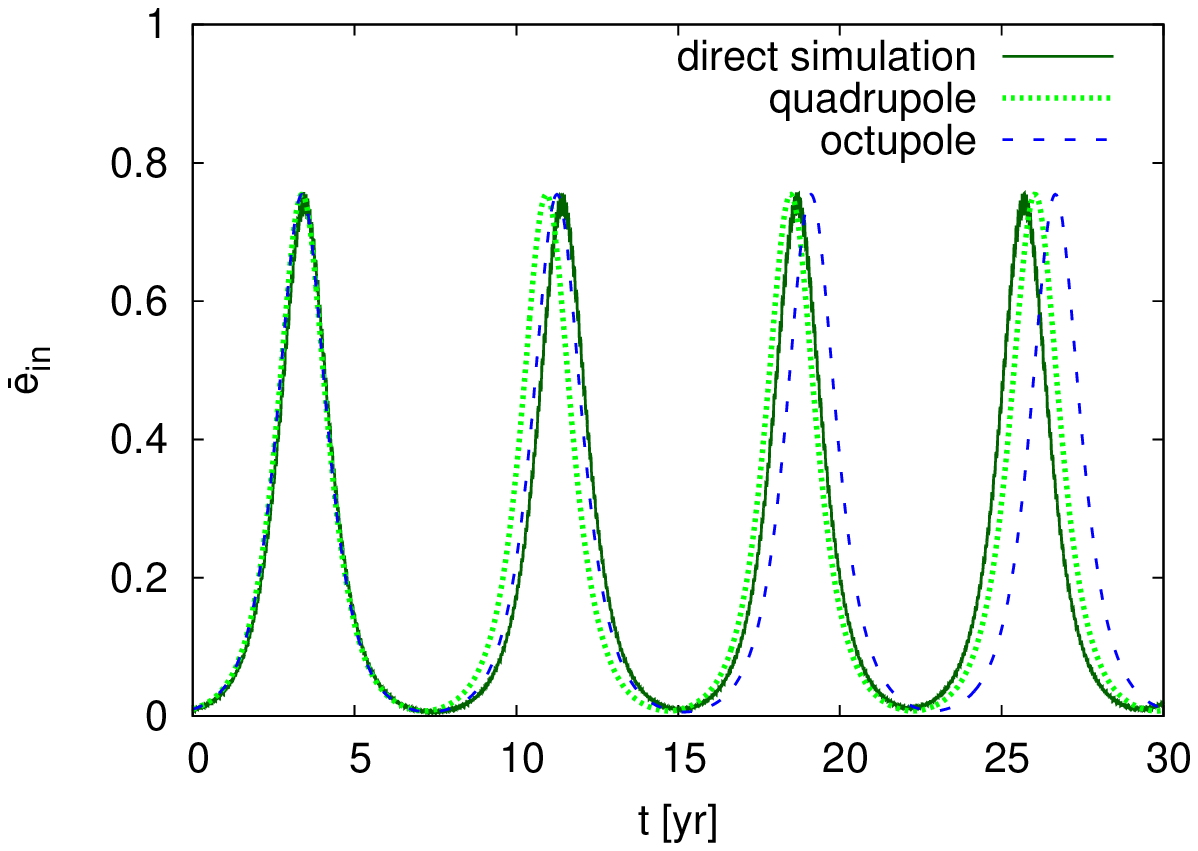}
        \end{minipage}\\
        \begin{minipage}{6.5cm}
            \includegraphics[width=6.5cm]{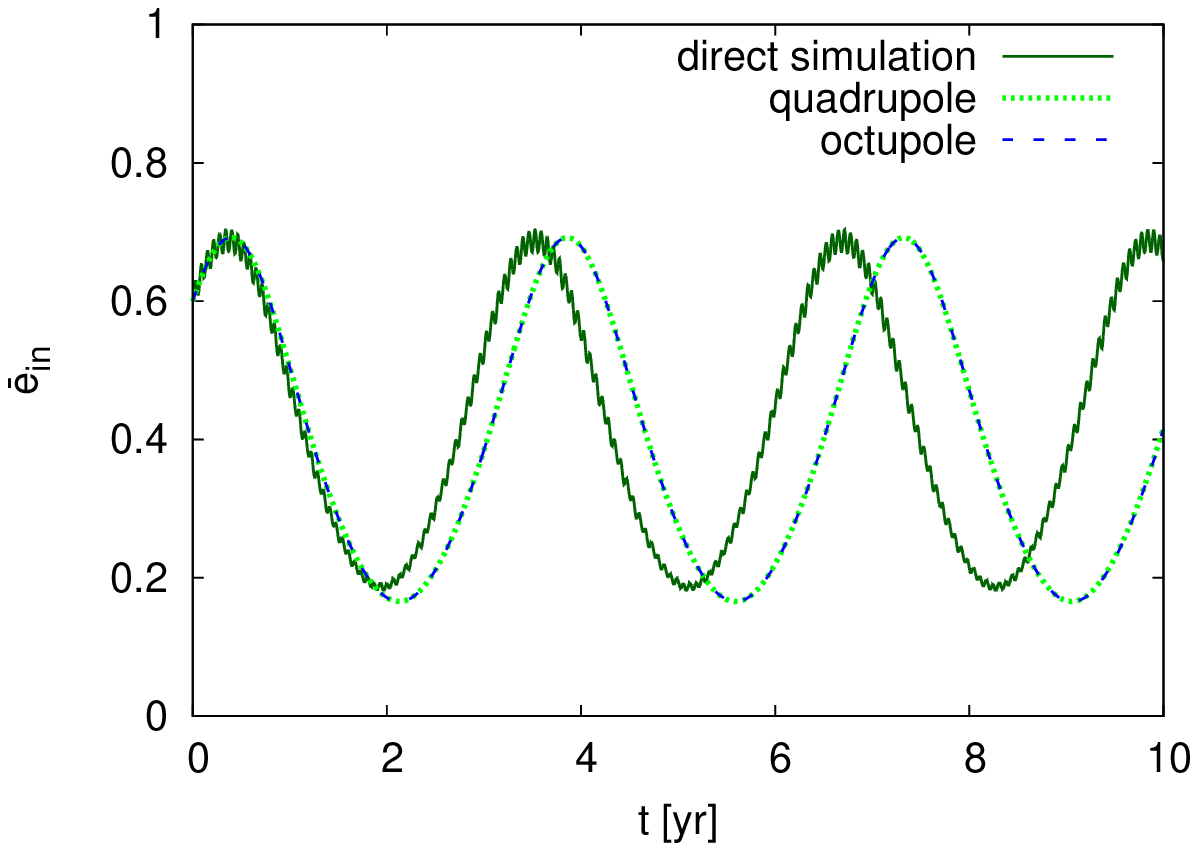}
        \end{minipage} 
        \caption{
                    The same figure as Fig.~\ref{fig:PBB_e_l} for the ``rotation'' type KL-oscillations in PBB model.
                    The top and bottom panels show the results of ICR and IER types, respectively.
                }
        \label{fig:PBB_e_r}
    \end{figure}

{\HS 
Figs.~\ref{fig:PBB_e_l} and \ref{fig:PBB_e_r} are the evolution of inner eccentricity. 
These figures show the evolution curves obtained by direct integration (dark-green solid line), 
by double-averaging method with quadrupole-order term (light-green dashed line),
and by the octupole-order expansion (dark-blue dashed line).  
In the bottom panels of Figs.~\ref{fig:PBB_e_l} and \ref{fig:PBB_e_r}, 
the quadrupole- and octupole-order lines are almost the same, 
but are different from the result of the direct-integration. 
In the top panels of Figs.~\ref{fig:PBB_e_l} and \ref{fig:PBB_e_r}, on the other hand, 
we find the difference between the results of quadrupole- and octupole-order expansion.}

{\KM In the top figure of Fig.~\ref{fig:PBB_e_l}, the octupole-order expansion gives 
better result compared with the quadrupole one. 
While the top figure of Fig.~\ref{fig:PBB_e_r} seems to show the opposite result
when we look at the fourth period.
However if we look at the second period, the result of the octupole-order 
expansion is closer to the direct one.
This is possible because the ICR type in the PBB model shows the irregular period
as discussed in \S \ref{irregular_period}.
Although we have not confirmed that the octupole-order expansion 
improves the calculation, 
we conclude that
the double-averaged calculations not only in  quadrupole-order 
expansion but also in octupole-order one
show clear deviation from the results by direct integration. }

\bsp	
\label{lastpage}
\end{document}